\documentclass[11pt,a4paper]{article}
\usepackage{graphicx}
\usepackage{subcaption}
\usepackage{amsmath,amsfonts,amssymb,bm,tensor}
\usepackage{mathrsfs}
\usepackage{xcolor}
\usepackage{physics}
\usepackage{booktabs}
\usepackage{jcappub}
\usepackage{xcolor}

\newcommand{\m}{m_{\rm P}}

\DeclareMathAlphabet\mathbfcal{OMS}{cmsy}{b}{n}
\allowdisplaybreaks
\graphicspath{{Figures/}}

\title{\begin{center}
Axion Inflation with a \\ Massive Abelian Gauge Field
\end{center}} 

\author[a,b]{Ricardo~Z.~Ferreira,}
\author[c]{Alessio Notari,}%
\author[b]{and Jos\'e Jaime Terente D\'iaz}

\affiliation[a]{Centro de F\'isica das Universidades do Minho e do Porto (CF-UM-UP),\\
Universidade do Minho, P-4710-057 Braga, Portugal}
\affiliation[b]{Faculdade de Ci\^{e}ncias e Tecnologia and CFisUC, Departamento de F\'isica,\\ 
Universidade de Coimbra, Rua Larga P-3004-516 Coimbra, Portugal}
\affiliation[c]{Dipartimento di Fisica, Sapienza University of Rome and INFN,\\
Piazzale Aldo Moro 2, I-00185, Italy}%

\emailAdd{rzferreira@fisica.uminho.pt}
\emailAdd{alessio.notari@uniroma1.it}
\emailAdd{jterente@uc.pt}

\makeatletter
\gdef\@fpheader{}
\makeatother

\abstract{An axial coupling between an inflaton and an Abelian gauge field can trigger the tachyonic amplification of one gauge-field helicity. For a massless vector, modes with physical momentum $k/a\sim |\xi|H$ are enhanced by approximately $\exp(\pi|\xi|)$, and sufficiently efficient production can provide substantial friction for the homogeneous inflaton. We extend this mechanism to a vector of mass $m$. The instability is present only for $|\xi|>\bar m\equiv m/H$, and in the heavy regime the mode amplitude scales as $\exp[\pi(|\xi|-\bar m)]$. Because the amplified modes remain well inside the Hubble radius when $\bar m\gg1$, their contribution to long-wavelength curvature perturbations $\zeta$ is power-law suppressed at fixed background backreaction. In the weak-backreaction regime we obtain a spectrum ${\cal P}^{\rm id}_{\zeta} \propto \bar m^{-w}$, with $w \sim 2$, while including the gauge-induced friction of scalar perturbations gives the scaling ${\cal P}^{\rm id}_{\zeta}\propto \bar{m}^{-s}$, with $s$ between three and four. These estimates indicate that ${\cal P}^{\rm id}_{\zeta}\lesssim 10^{-9}$ on CMB scales should be compatible with gauge field backreaction for $\bar{m}$ larger than order a few hundred. We test the analytical mode functions and backreaction estimates with the first lattice simulations based on a massive-vector extension of the \texttt{Pencil Code}, including simulations in the strongly backreacting regime.} 

\begin{document}
\maketitle

\section{\label{sec:intro}Introduction}
Axion-like particles are pseudo-scalars endowed with an approximate continuous shift symmetry that is broken only by non-perturbative effects \cite{tHooft:1976rip,tHooft:1976snw,Coleman:1985rnk}. Such fields arise ubiquitously in extensions of the Standard Model of particle physics, and are particularly well-suited to early-Universe model building, since the shift symmetry protects the potential of light scalars from radiative corrections and Planck-suppressed operators \cite{Peccei:2006as,Kawasaki:2013ae,Baumann:2014nda,Marsh:2015xka}. A generic and largely model-independent consequence of the shift symmetry is that the leading interaction of an axion-like field with a gauge sector occurs through parity-violating couplings \cite{Carroll:1989vb,Carroll:1998zi,Andrianov:1998ay}. During inflation, the rolling axion background renders this interaction time dependent and can drive the tachyonic amplification of one gauge-field helicity. In the massless Abelian case \cite{Anber:2009ua}, this mechanism has become one of the most extensively studied sources of particle production during inflation, owing to the broad and distinctive phenomenology it generates, namely, chiral gravitational waves,\footnote{A stochastic gravitational-wave background sourced by the Chern-Simons coupling is generically chiral and may be blue-tilted, potentially placing it within the reach of interferometers and pulsar-timing arrays~\cite{Bastero-Gil:2022fme,Niu:2023bsr,Barbon:2025wjl}.} non-Gaussianity,\footnote{See Ref.~\cite{Caravano:2024xsb} for lattice simulations of a system with a spectator axion-gauge sector. The authors argue that, although non-Gaussianity is suppressed in the strong-backreaction regime, it remains larger than in the minimal model in which the inflaton itself is the axion.} primordial magnetic fields,\footnote{According to Refs.~\cite{Domcke:2018eki,Sobol:2019xls,Fujita:2022fwc,Iarygina:2025ncl}, however, the unstable gauge mode also drives non-perturbative pair production of charged particles via the Schwinger effect, significantly suppressing the resulting electromagnetic spectrum. Lattice simulations \cite{Iarygina:2025ncl} show that this suppression undermines the prospects for primordial magnetogenesis in high-scale axion inflation, unless heavy charged fermions or, equivalently, low-scale inflation are invoked. See also Ref.~\cite{VicenteGarcia-Consuegra:2026vaf} for a recent stochastic treatment of Schwinger production in non-stationary gauge backgrounds.} and primordial black holes, amongst others (see Refs.~\cite{Garretson:1992vt,Field:1998hi,Anber:2006xt,Barnaby:2011vw,Cook:2011hg,Linde:2012bt,Caprini:2014mja,Ferreira:2014zia,Ferreira:2015omg,Adshead:2016iae,Notari:2016npn,Ferreira:2017lnd,Ferreira:2017wlx,Domcke:2018eki,Caravano:2021bfn,Caravano:2022epk} and the more recent~\cite{Figueroa:2023oxc,vonEckardstein:2023gwk,Sharma:2024nfu,Figueroa:2024rkr,vonEckardstein:2025oic,Jamieson:2025ngu,Sobol:2026nfh,Franciolini:2026cps} for entry points into this extensive literature).   

The same gauge-field particle production back-reacts on the homogeneous evolution of the inflaton. The amplified modes source a non-zero gauge-field pseudo-scalar density, which enters the inflaton equation of motion as an additional, velocity-dependent friction term \cite{Anber:2009ua,Notari:2016npn}. This is especially appealing in the context of axion inflation. Its minimal realisation, Natural Inflation \cite{Freese:1990rb}, is by now in significant tension with CMB data \cite{Stein:2021uge,AtacamaCosmologyTelescope:2025nti}: reproducing the observed scalar spectral index requires a super-Planckian decay constant, which is difficult to reconcile with general expectations from high-energy completions~\cite{Banks:2003sx,Svrcek:2006yi,Palti:2019pca}; moreover, it predicts a tensor-to-scalar ratio in excess of the latest Planck/BICEP/Keck upper bound \cite{Planck:2018jri,BICEP:2021xfz}. Strong gauge-field friction offers, in principle, a means of alleviating this tension by sustaining slow roll on steeper potentials, or over shorter effective field ranges, than would otherwise be possible. Considerable recent effort has gone into characterising this backreacting regime, both analytically and through lattice simulations \cite{Caravano:2021bfn,Caravano:2022epk,Figueroa:2023oxc,Figueroa:2024rkr,Sharma:2024nfu,Jamieson:2025ngu}, and the friction-dominated phase is now substantially better understood, although a number of questions remain open \cite{vonEckardstein:2023gwk,Barbon:2025wjl,Sobol:2026nfh}. It is, in any case, well established that the regime in which the gauge fields appreciably affect the background is closely tied to the one in which the sourced scalar power spectrum becomes large and strongly non-Gaussian~\cite{Barnaby:2011vw}, in some cases, to the point of jeopardising the perturbative description of the system~\cite{Ferreira:2015omg}.

The aim of the present work is to understand how this picture is modified when the Abelian gauge field is massive. Massive vectors arise naturally in a variety of settings \cite{Ruegg:2003ps,Graham:2015rva,Ferreira:2017lnd,Wang:2020ioa,Kolb:2020fwh,Bastero-Gil:2021wsf,Lysenko:2025sdo}: through the Higgs or St\"uckelberg mechanisms, from symmetry breaking at high scales, or as effective descriptions of gauge fields propagating in a medium (we return to several well-motivated UV realisations at the end of this introduction). Introducing a mass alters the tachyonic instability; in particular, for sufficiently large masses, the band of amplified modes is shifted towards shorter physical wavelengths \cite{Dimopoulos:2012av,Lu:2019tjj}. This opens up the intriguing possibility that gauge fields may continue to contribute efficiently to local background quantities (and hence to the friction), while becoming progressively less capable of sourcing the long-wavelength curvature perturbations probed by the CMB.

Our analysis combines analytical estimates with lattice simulations. Analytically, we characterise the massive gauge-field instability, compute the expectation values governing the background backreaction, and estimate the inverse-decay contribution to the scalar power spectrum. Numerically, we extend the open-source \texttt{Pencil Code}~\cite{PencilCode:2020eyn} to accommodate a gauge-field mass and use it both to validate the analytical results and to probe the backreacting regime directly. The simulations confirm the analytical expectation that increasing the vector mass suppresses the sourced scalar perturbations at fixed gauge-field backreaction. We thereby go beyond earlier studies of massive gauge fields~\cite{Niu:2022fki,Niu:2022quw,Niu:2023bsr}, which evaluated the relevant loop integrals numerically in order to obtain the non-Gaussianity and gravitational-wave signals, by providing an analytical description of the mode functions, the power spectrum, and the backreaction quantities, together with the first lattice simulations of this scenario. Both ingredients are necessary for a quantitative assessment of the viability of the friction-dominated regime.

Before turning to our main results, it is useful to illustrate the range of settings in which a massive Abelian gauge field of the required type may arise.\footnote{Note that a relic population of such massive vectors after inflation has been proposed also as a dark matter candidate~\cite{Graham:2015rva,Bastero-Gil:2021wsf}.} We highlight three representative possibilities, ranging from the known Standard Model to heavy relic gauge bosons and medium-induced effective masses:

\paragraph{Standard Model $Z$ Boson~~--} A possible connection with the neutral electroweak (EW) gauge sector may be envisaged, although a direct identification with the $Z$ boson would require accounting for its non-Abelian EW interactions. Nevertheless, we expect the same mechanism of exponential growth to remain operative whenever these non-Abelian interactions can be neglected. The viability of this scenario depends crucially on whether the Hubble scale during inflation, $H_I$, exceeds the EW scale $v_{\textrm{EW}}$. If $H_I \gg v_{\textrm{EW}}$, the Standard Model Higgs is light during inflation and is driven towards its stochastic equilibrium attractor, with a typical expectation value $v \sim \mathcal{O}\bigl(H_I/\lambda^{1/4}\bigr)$~\cite{Starobinsky:1994bd,Gorghetto:2023vqu}. In this case, the $Z$-boson mass remains relatively small, unless, for instance, a sizeable non-minimal coupling of the Higgs to gravity is admitted. If, on the other hand, $H_I \ll v_{\textrm{EW}}$, the Higgs is heavy during inflation, sits at its standard EW vacuum expectation value, and the $Z$-boson mass takes its laboratory value, $m_Z \simeq 91.2~\textrm{GeV}$. This second regime lies precisely within the massive-gauge-field range of interest, and an axial coupling between the inflaton and the $Z$ boson is expected, similarly to the QCD axion case~\cite{Alonso-Alvarez:2017hsz}.

\paragraph{GUT-Scale Gauge Boson~~--} A second, equally simple possibility is that an Abelian gauge field gets a mass at the GUT scale, with a mass of the same order, $m_A \sim 10^{15}\text{--}10^{16}~\textrm{GeV}$ (see, \textit{e.g.}, Ref.~\cite{Langacker:2008yv,Maleknejad:2020yys} for further motivation). As we show below, such a field can drive a friction-dominated backreaction regime for $H_I \simeq 10^{13}~\textrm{GeV}$.

\paragraph{Thermal Mass~~--} A third possibility, discussed in Ref.~\cite{Ferreira:2017lnd}, is that the gauge field acquires an effective mass through thermal or non-linear effects, either via interactions with other particles in the plasma that can be generated by the gauge fields themselves or, in the non-Abelian case, through self-interactions.

\subsection{Outline and Conventions}
This work is organised as follows. Section~\ref{sec2-massive-eqs} introduces the model and its field equations. Sections~\ref{sec3-weakback} and \ref{sec4-AnalyticalSolutionsforWeakBackreaction} analyse the gauge-field modes in the weak-backreaction regime. Section~\ref{sec5-scalarpowerspectrum-fullsec} derives the inverse-decay contribution to the scalar spectrum, and Sec.~\ref{sec:Backreaction-FULL} identifies the weak, mild, and strong backreaction regimes. The lattice results are presented in Sec.~\ref{sec:pencil-code-lattice-simulations}, and Sec.~\ref{sec:outlook-conclusions} summarises our conclusions. For many of the asymptotic approximations used below, we draw on Ref.~\cite{NIST:DLMF}. 

Throughout this work, natural units ($\hbar = c = 1$) are adopted, with
\begin{equation}
\m \equiv (8\pi G)^{-1/2} \simeq 2.44\times10^{18}\,\mathrm{GeV}
\end{equation}
denoting the reduced Planck mass and $G$ Newton's gravitational constant. We employ the `mostly plus' metric signature, $(-,+,+,+)$. Finally, $\epsilon^{\mu\nu\alpha\beta}$ denotes the totally antisymmetric Levi--Civita tensor (rather than the alternating symbol), with convention $\epsilon^{0123}=1/\sqrt{-g}$, where $g$ is the determinant of the metric.

\section{\label{sec2-massive-eqs}Massive Abelian Gauge Field Axially Coupled to the Inflaton}
\subsection{Action and Field Equations}
We consider the evolution of a massive Abelian gauge field axially coupled to a scalar field. The system is described by the following action:
\begin{align}
    \label{eq:action}S = \int \textrm{d}^4x \sqrt{-g} \left[\frac{\m^2}{2}R -\frac{1}{2}\partial_{\mu} \phi \partial^{\mu} \phi -V(\phi) -\frac{1}{4} F_{\mu\nu}F^{\mu\nu}-\frac{m^2}{2} A_{\mu} A^{\mu}-\frac{\alpha}{4f}\phi F_{\mu\nu} \tilde{F}^{\mu\nu}\right],
\end{align}
where $R$ is the Ricci scalar, $\phi$ a scalar field with potential $V(\phi)$, and $A_{\mu}$ the Abelian gauge field with mass $m$ and field-strength tensor $F_{\mu\nu}\equiv \partial_{\mu} A_{\nu}-\partial_{\nu} A_{\mu}$, whose dual is $\tilde{F}^{\mu\nu} \equiv \frac{1}{2}\epsilon^{\mu\nu\alpha\beta}F_{\alpha\beta}$. Although the mass term explicitly breaks the gauge invariance associated with $A_\mu$, the situations we have in mind are those in which the symmetry is restored through the Higgs or St\"uckelberg mechanisms, or in which the mass term effectively describes a thermal mass \cite{Ferreira:2017lnd}. The scalar field is coupled to $A_{\mu}$ through an axial interaction involving a dimensionless parameter $\alpha$ and the decay constant $f$, as is typical for an axion. 

To derive the evolution equations, we consider the spatially flat Friedmann-Lema\^{i}tre- Robertson-Walker (FLRW) metric, with line element
\begin{equation}
    \label{eq:flrw-metric}\textrm{d}s^2 = -a^2(\eta) \left(\textrm{d}\eta^2-\delta_{ij} \textrm{d}x^{i} \textrm{d}x^{j}\right).
\end{equation}
$a(\eta)$ is the scale factor and $\eta$ the conformal time. The covariant components of the gauge field are defined as $A_{\mu}(\eta,\mathbf{x})  \equiv \left(A_0(\eta,\mathbf{x}),\mathbf{A}(\eta,\mathbf{x})\right)$. One may then introduce the conventional electric and magnetic fields, 
\begin{equation}
    \label{eq:def-electric-magnetic-gauge-invariant-definitions}\mathbf{E} \equiv -\mathbf{A}^{'}+\nabla A_0~, \ \textrm{and} \ \ \mathbf{B} \equiv \nabla \cross \mathbf{A}~,
\end{equation}
respectively, where primes indicate derivatives with respect to conformal time, and `$\cross$' denotes the three-dimensional cross product. Using standard vector calculus identities, we readily find Gauss's law for magnetism and Faraday's law of induction: 
\begin{equation}
   \nabla\cdot\mathbf{B} = 0~, \qquad 
   \nabla \cross \mathbf{E} = -\mathbf{B}^{'}~.
\end{equation}
The presence of the axial coupling and the gauge-field mass term in Eq.~\eqref{eq:action} modifies the remaining two Maxwell equations. By varying the action with respect to $\phi$ and $A_{\mu}$, we obtain the set of field equations
\begin{align} 
    \label{eq:eq-sf-full}&\phi^{''} - \nabla^2 \phi +2\mathcal{H} \phi^{'} +a^2 V_{,\phi} = \frac{\alpha}{a^2 f} \mathbf{E}\cdot \mathbf{B}~,\\
    \label{eq:nablaE}&\nabla \cdot \mathbf{E} = a^2 m^2 A_0 -\frac{\alpha}{f} \nabla \phi \cdot \mathbf{B}~,\\
    \label{eq:crossB}&\nabla \cross \mathbf{B} = \mathbf{E}^{'} -a^2 m^2 \mathbf{A} + \frac{\alpha}{f} \left(\phi^{'} \mathbf{B}+\nabla \phi \cross \mathbf{E}\right).
\end{align}
$\mathcal{H}(\eta) \equiv a^{'}/a$ is the comoving Hubble parameter, $V_{,\phi}\equiv \textrm{d}V/\textrm{d}\phi$, and $\nabla^2$ denotes the Laplacian operator in flat space. For completeness, we show the covariant form of these equations, and of those of the metric field tensor, in Appendix~\ref{sec:cov-fe}. In terms of the explicit gauge-field components, Eq.~\eqref{eq:crossB} can be rewritten using the constraint equation therein,
\begin{equation}
    \label{eq:constraint-incoord}A^{'}_0 + 2\mathcal{H} A_0 = \nabla \cdot \mathbf{A}~.
\end{equation}
This constraint reflects the presence of three (rather than four) physical degrees of freedom of the spin-1 field. One then finds \cite{Agrawal:2018vin}
\begin{equation}
    \label{eq:eq-vecA-afterconst}\mathbf{A}^{''}-\left(\nabla^2-a^2m^2\right)\mathbf{A}+2\mathcal{H}\nabla A_0 =\frac{\alpha}{f} \left[\phi^{'} \left(\nabla \cross \mathbf{A}\right) -\nabla \phi \cross \left(\mathbf{A}^{'}-\nabla A_0\right)\right].
\end{equation}

Finally, for the system under discussion, the Friedmann equation generally reads
\begin{equation}
    \label{eq:Friedmann-eq}3\mathcal{H}^2 \m^2 = a^{-2} \langle\rho\rangle~,
\end{equation}
with `$\langle...\rangle$' denoting spatial average over a cubic domain of a certain size. $\rho \equiv \rho_A + \rho_\phi$ is the total comoving energy density, while $\rho_A$ is the comoving energy density of the massive gauge field $A_{\mu}$, given by 
\begin{equation}
    \label{eq:A-comov-energy}\rho_A = \frac{1}{2}\left[|\mathbf{E}|^2+|\mathbf{B}|^2+a^2m^2 \left(A_0^2+|\mathbf{A}|^2\right)\right],
\end{equation}
and $\rho_\phi$ is the comoving energy density of the scalar field $\phi$, which reads 
\begin{equation}
    \label{eq:phi-comov-energy}\rho_\phi = \frac{1}{2} a^2 \left[\phi^{'2}+(\nabla \phi)^2\right]+a^4V~.
\end{equation} 

\section{\label{sec3-weakback}Weak Backreaction Regime}
Having derived the general equations of motion for the massive Abelian gauge field axially coupled to a scalar field, we proceed to study their solutions during cosmic inflation, with particular emphasis on the growth of one of the gauge-field polarisations. Throughout, we consider the regime in which backreaction of the gauge field on the scalar field equation of motion is small. In particular, we now assume that the scalar field (the inflaton) is approximately homogeneous, allowing us to neglect gradients of $\phi$, such that $\phi=\phi(\eta)$. In Sec.~\ref{sec:Backreaction}, an estimate of when this approximation breaks down is provided.

\subsection{Equations for the Transverse and Longitudinal Modes}
We begin by Fourier transforming $A_{\mu}$ as
\begin{equation}    
    \label{eq:Fourier-expansion-Amu}A_{\mu}(\eta,\mathbf{x}) = \int \frac{\textrm{d}^3\mathbf{k}}{(2\pi)^{3/2}}\mathcal{A}_{\mu}(\eta,\mathbf{k}) e^{i\mathbf{k}\cdot \mathbf{x}}~,
\end{equation}
and by decomposing its Fourier components into longitudinal ($L$) and transverse ($T$) modes as
\begin{align}
    \label{eq:def-longitudinal and transverse-components}\mathbfcal{A}_L \equiv \frac{\mathbf{k}\left(\mathbf{k}\cdot \mathbfcal{A}\right)}{k^2}~, \qquad \mathbfcal{A}_T \equiv \mathbfcal{A}-\mathbfcal{A}_L~,
\end{align}
$\mathbfcal{A}$ being the Fourier-transformed spatial vector with components $\mathcal{A}_i$. From Eq.~\eqref{eq:constraint-eq-A0} in Fourier space it then follows that 
\begin{equation}
    \label{eq:constraint-A0}\mathcal{A}_0 =-\frac{i\mathbf{k}\cdot \mathbfcal{A}_L^{'}}{k^2+a^2m^2}~,
\end{equation}
where, due to the approximate homogeneity of the scalar field, $\mathcal{A}_0$ is determined by the longitudinal component solely. The polarisation vector of the longitudinal mode satisfies $\mathbf{k}\cdot \bm{\epsilon}_L(\mathbf{k}) = k$, whereas the transverse modes are more conveniently decomposed in a basis of circular polarisation vectors satisfying $\mathbf{k}\cross \bm{\epsilon}_{\pm}(\mathbf{k}) = \mp ik\bm{\epsilon}_{\pm}(\mathbf{k})$. In all cases, irrespective of the basis, $|\bm{\epsilon}_\lambda|^2\equiv \bm{\epsilon}_{\lambda}^{*}(\mathbf{k})\cdot \bm{\epsilon}_{\lambda}(\mathbf{k}) = 1$. 

The set of equations of motion for those three modes $\mathcal{A}_{\lambda}(\eta,k)$, with $\lambda = L, \pm$, is then obtained to be\footnote{In Appendix~\ref{App:Decomposition into Transverse and Longitudinal Modes}, we present the full equations of motion in Fourier space when gradients of $\phi$ are included.}  
\begin{eqnarray}
    \label{eq:longitudinal-mode-homogeneous}\mathcal{A}_L^{''} +2\mathcal{H}\frac{k^2}{k^2+a^2m^2}\mathcal{A}_L^{'}+\left(k^2+a^2m^2\right)\mathcal{A}_L &=& 0~,\\
    \label{eq:oscillator-freq}\mathcal{A}_{\pm}^{''} +\omega_{\pm}^2(\eta,k)  \mathcal{A}_{\pm}&=&0~,
\end{eqnarray}
$\omega_{\pm}$ being the effective frequencies for the $\mathcal{A}_{\pm}$ modes, with dispersion relations given by 
\begin{equation}
    \label{eq:oscillator-freq-wplusminus}\omega_{\pm}^2(\eta,k) \equiv k^2+a^2m^2 \pm 2\mathcal{H}k\xi~,
\end{equation}
and the parameter $\xi(\eta)$ is defined as
\begin{equation}
    \label{eq:def-xi}\xi \equiv -\frac{\alpha\phi^{'}}{2\mathcal{H}f}~.
\end{equation}
As is evident, the longitudinal mode does not undergo any instability-induced enhancement because it does not couple to the scalar field through the axial interaction. Consequently, as discussed below, the phenomenology is similar to that of the massless gauge field, where the main effects arise from the enhancement of one of the transverse polarisation states.

\subsection{\label{sec:instability-band-discussion}Instability Band}
A quasi-de Sitter expansion is assumed now, with $H\equiv \mathcal{H}/a$ being the Hubble parameter, approximately constant in time, and with the conformal one given by $\mathcal{H} \simeq -1/\eta$, where $\eta<0$. The scale factor is $a\simeq -1/(\eta H)$. In this case, $\xi$ can be written in terms of the slow-roll parameter $\epsilon \equiv (\phi^{'}/\mathcal{H})^2/(2\m^2) $ if $\phi$ is the inflaton, such that $\xi \simeq -(\alpha\m/f)\sqrt{\epsilon/2}$.

In the quasi-de Sitter background described above, the dispersion relations in Eq.~\eqref{eq:oscillator-freq-wplusminus} take the form
\begin{equation}
    \label{eq:oscillator-freq-wplusminus-DESITTER}\omega^2_{\pm} \simeq k^2 +\frac{\bar{m}^2}{\eta^2}\mp 2k\frac{\xi}{\eta}~,
\end{equation}
where $\bar{m}\equiv m/H$ is the gauge-field mass in units of the constant Hubble parameter. $\omega_{+}$ admits zeros and can be imaginary whenever $|\xi| > \bar{m}$. In what follows, this condition is assumed to hold, together with a positive scalar-field velocity, $\phi^{'} >0$, so that $\xi <0$ and $\xi/\eta>0$, and thus the `$+$' polarisation is unstable. The instability band is then determined by the two roots of $\omega_+^2$, which occur at the physical momenta $\tilde{k}_{1,2}\equiv k_{1,2}/a$, given by
\begin{equation} \label{eq:Instability Band}
    \tilde{k}_{1,2} =  \left( |\xi|\pm \sqrt{\xi^2-\bar{m}^2}\right)H~.
\end{equation}
Between $\tilde{k}_{1}$ and $\tilde{k}_{2}$, $\omega_+$ becomes imaginary, and one therefore expects tachyonic enhancement of the corresponding transverse mode. The main effect of the mass is to shorten the instability band. For $|\xi|=\bar{m}$, the two roots coincide, and the dispersion relation $\omega_+^2$ remains non-negative.

\subsection{Electric and Magnetic Power Spectra and Gauge-Field Densities}
Before closing this section, it is useful to express the corresponding electric and magnetic fields, defined in Eq.~\eqref{eq:def-electric-magnetic-gauge-invariant-definitions}, as\footnote{\label{footnote:zeromass-electric}Notice that if $m=0$, $\mathbf{E}$ would only depend, as expected, on the sum of $\lambda = \pm$ (same as $\mathbf{B}$) because $\bm{\epsilon}_L(\mathbf{k}) = \mathbf{k}/k$.}
\begin{align}
    &\label{eq:electric-field-decom}\mathbf{E}(\eta,\mathbf{x}) = \int \frac{\textrm{d}^3\mathbf{k}}{(2\pi)^{3/2}}e^{i\mathbf{k}\cdot \mathbf{x}}\left[\frac{k\mathbf{k}}{k^2+a^2m^2}\mathcal{A}^{'}_L(\eta,\mathbf{k})-\sum_{\lambda=L,\pm}\bm{\epsilon}_{\lambda}(\mathbf{k})\mathcal{A}_{\lambda}^{'}(\eta,\mathbf{k})\right],
    \\
    &\label{eq:magnetic-field-decom}\mathbf{B}(\eta,\mathbf{x}) =\sum_{\lambda = \pm} \lambda \int \frac{\textrm{d}^3\mathbf{k}}{(2\pi)^{3/2}}k\bm{\epsilon}_{\lambda}(\mathbf{k})e^{i\mathbf{k}\cdot \mathbf{x}} \mathcal{A}_{\lambda}(\eta,\mathbf{k})~.
\end{align}
As for the electric and magnetic power spectra, $\mathcal{P}_E(k)$ and $\mathcal{P}_B(k)$, respectively, we begin by promoting the gauge-field components to quantum operators satisfying the standard commutation relations, as described in Appendix~\ref{App:Fields quantization}. Hatted symbols will be used to denote the associated operators. It then follows that
\begin{align}
    \label{eq:PE-power}&\mathcal{P}_E(k) \equiv \frac{k^3}{2\pi^2} \left[\left(1-\frac{k^2}{k^2+a^2m^2}\right)^2|\mathcal{A}_L^{'}|^2+|\mathcal{A}_{+}^{'}|^2+|\mathcal{A}_{-}^{'}|^2\right],\\
    \label{eq:PB-power}&\mathcal{P}_B(k) \equiv   \frac{k^5}{2\pi^2}\left[|\mathcal{A}_{+}|^2+|\mathcal{A}_{-}|^2\right],
\end{align}
where $|\mathcal{A}_{\lambda}^{'}|^2 = \mathcal{A}_{\lambda}^{'}\mathcal{A}^{'*}_{\lambda}$ with $\lambda = L,\pm$. As can be seen, the magnetic-field spectrum remains oblivious to the mass of the gauge field at the explicit level, since no $\hat{A}_0$, and hence no longitudinal component, enters its definition. The massive and massless cases nevertheless differ because the mode functions satisfy different equations of motion. In the massless limit, the longitudinal contribution to the electric-field spectrum goes away, as expected. 

For later convenience, the comoving gauge-field energy density, $\rho_A$, is written as
\begin{align}
\rho_A={}&\frac{1}{4\pi^2}\int \textrm{d}\ln k\,k^3\left\{
\frac{a^2m^2}{k^2+a^2m^2}|\mathcal{A}'_L|^2
+|\mathcal{A}'_+|^2+|\mathcal{A}'_-|^2
+a^2m^2|\mathcal{A}_L|^2 \right.\nonumber\\
&\left.\hspace{4.7cm} +(k^2+a^2m^2)\left(|\mathcal{A}_+|^2+|\mathcal{A}_-|^2\right)\right\}.
\label{eq:total-gauge-energy-density}
\end{align}
The operator expression in Eq.~\eqref{eq:total-gauge-energy-density} contains the usual zero-point divergence. Throughout the analytical discussion, $\rho_A$ and $\langle\mathbf E\cdot\mathbf B\rangle$ denote the produced, vacuum-subtracted contributions. We again observe that, in the massless limit, the longitudinal mode drops out of $\rho_A$, which then reduces to the standard electromagnetic energy density, $\frac{1}{2}\left(\langle \hat{\mathbf{E}}^2\rangle + \langle \hat{\mathbf{B}}^2\rangle \right)$, with $\langle \hat{\mathbf{E}}^2\rangle$ and $\langle \hat{\mathbf{B}}^2\rangle$ given in the general case by Eq.~\eqref{eq:pieces-of-energy-density-gauge-fields-rhoA}.
 
Finally, the last quantity relevant for the remainder of this article is the pseudo-scalar density $\langle \hat{\mathbf{E}}\cdot \hat{\mathbf{B}}\rangle$, which governs the backreaction of the gauge field on the inflaton equation of motion \eqref{eq:eq-sf-full}:\footnote{\label{footnote:the1/2}The $1/2$ in $\langle \hat{\mathbf{E}}\cdot \hat{\mathbf{B}}\rangle$ arises from taking the symmetrised product of the two field operators (see Appendix~\ref{App:Fields quantization}).}
\begin{equation}
    \label{eq:backreacting-term-EB}\langle \hat{\mathbf{E}}\cdot \hat{\mathbf{B}} \rangle = -\frac{1}{4\pi^2}\sum_{\lambda = \pm} \lambda \int \textrm{d}\ln k~k^4\frac{\partial}{\partial \eta}\left[|\mathcal{A}_{\lambda}(\eta,k)|^2\right].
\end{equation}
Note that the latter expression is identical to that of the massless gauge-field case. This is because the mass term only affects the longitudinal contribution to the electric field, which vanishes since $\mathbf{k}\cdot \bm{\epsilon}_{\pm}(\mathbf{k})=0$. 

\section{\label{sec4-AnalyticalSolutionsforWeakBackreaction}Analytical Solutions for Weak Backreaction}
In the previous section, the general equations for the gauge field were derived, along with a number of useful definitions. Attention now turns to obtaining an analytical solution for the unstable `$+$' polarisation mode in Eq.~\eqref{eq:oscillator-freq}. The analysis focuses primarily on the regime $|\xi| > \bar{m} \gg 1$, although most formulas also apply to light and massless gauge fields. Further details on all these approximations are provided in Appendix~\ref{sec:app-asymp}, including the superhorizon solutions presented in Sec.~\ref{sec:superhorizon-expressions}. For completeness, the dynamics of the stable `$-$' mode are examined as well.

\subsection{Whittaker Equation}
In the slow-roll regime of inflation, $\xi$ is approximately constant, and Eq.~\eqref{eq:oscillator-freq} takes on the form of the Whittaker's equation (see Chap.~13 of Ref.~\cite{NIST:DLMF}):
\begin{equation}
    \label{eq:Whittaker-eq-Aplusminus-z}\frac{\textrm{d}^2 \mathcal{A}_{\pm}}{\textrm{d}z^2}+\left(-\frac{1}{4}\pm \frac{i\xi}{z}+\frac{\bar{m}^2}{z^2}\right)\mathcal{A}_{\pm} = 0~,
\end{equation}
where $z\equiv 2ik\eta$. The general solution is expressed in terms of the so-called Whittaker functions $W_{\kappa,\mu}$; here $\kappa$ is a generic first Whittaker index, while $\mu$ is the second index (and not a spacetime label). This avoids confusing the Whittaker index with the matching parameter $\kappa$ introduced below:
\begin{equation}
    \label{eq:sol-whittaker-eq}\mathcal{A}_{\pm}(\eta) = \alpha_1 W_{\pm i\xi,\mu}(z) + \alpha_2 W_{\mp i\xi,\mu}(-z)~,
\end{equation}
with $\alpha_{1,2}$ denoting integration constants. Here $\mu\equiv \sqrt{1/4-\bar{m}^2}$ is either real and non-negative when the massive gauge field is light, $0\leq\bar{m} \leq 1/2$, or purely imaginary when it is heavy, $\bar{m}>1/2$. Imposing the Bunch-Davies mode function on sufficiently short-wavelength scales, namely $\mathcal{A}_{\pm}(-k\eta\to \infty) \sim e^{-ik\eta}/\sqrt{2k}$,  along with the normalisation condition ensuring that the annihilation and creation operators satisfy the canonical commutation relations (see Eq.~\eqref{eq:comm-rel-ann-cr}), leads to $\alpha_2=0$ and $\alpha_1=e^{\mp\xi\pi/2}/\sqrt{2k}$, and hence\footnote{\label{footnotesub}In the subhorizon limit, assuming that $2k|\eta| \gg |1/2+\mu\mp i \xi||1/2-\mu\mp i \xi|$, one has $W_{\pm i\xi,\mu}(2ik\eta) \simeq e^{-ik\eta}(2ik\eta)^{\pm i \xi}$, which is independent of $\mu$ (the gauge-field mass), so that 
\begin{equation}
    \mathcal{A}_{\pm}(\eta) \simeq \alpha_1 \exp\left\{-i[k\eta \mp \xi \ln(-2k\eta)]\pm \xi \pi/2\right\}+\alpha_2 \exp\left\{i[k\eta \mp \xi \ln(-2k\eta)]\pm \xi \pi/2\right\}.
\end{equation}
As can be seen, logarithmic terms appear in the exponentials, $\pm i\xi \ln(-2k\eta)$. However, deep inside the subhorizon regime, $k|\eta| \gg |\xi|$, they contribute only a subleading correction to the phase relative to the dominant oscillatory term $k\eta$, since their contribution to the local frequency is proportional to $\xi/\eta$. Consequently, the asymptotic behaviour remains that of plane waves, up to a slowly varying logarithmic phase.}
\begin{equation}
    \label{eq:solution-to-Whittaker-eq-Aplusminus}\mathcal{A}_{\pm}(\eta)= \frac{e^{\mp\xi\pi/2}}{\sqrt{2k}}W_{\pm i\xi,\mu}(2ik\eta)~.
\end{equation}

\subsection{\label{app: Matching procedure}Approximate Solution via Matching}
The Whittaker function is, however, not well suited for analytical estimates of the effects of the gauge field. We therefore provide an approximate solution for the massive gauge-field mode functions in the regime $|\xi| > \bar{m}\gg 1$ by means of a matching procedure. Specifically, we consider two limiting forms of Eq.~\eqref{eq:oscillator-freq}: the early-time limit, $|k^2\mp 2 \xi k/\eta| \gg \bar{m}^2/\eta^2$, and the late-time limit, $|\bar{m}^2/\eta^2\mp 2 \xi k/\eta| \gg k^2$,\footnote{We refer the reader to Sec.~\ref{sec:asymp-Tricomi-various} for an alternative derivation based directly on the Whittaker solution.} and match the respective solutions in the common regime of validity: 
\begin{equation}\label{eq:Instability Band 2}
    |\xi|+\sqrt{\xi^2 -\bar{m}^2} >-k\eta > |\xi|-\sqrt{\xi^2 -\bar{m}^2}~,
\end{equation}
which corresponds precisely to the instability band in Eq.~\eqref{eq:Instability Band}.

At early times (`$\textrm{E}$'), the mass term in Eq.~\eqref{eq:Whittaker-eq-Aplusminus-z} is negligible, and the equation therefore reduces to the massless gauge-field case, $\mu=1/2$. The solution satisfying vacuum initial conditions, $\mathcal{A}^{\textrm{E}}_{\pm}(\eta)$, can be expressed in terms of the irregular Coulomb functions $H^{\pm}_0$, which are related to Whittaker functions through \cite{Anber:2009ua}
\begin{equation}
    W_{\pm i \xi,1/2}(2ik\eta) = H^{\pm}_0(-\xi,\mp k \eta) e^{\xi \pi/2\mp i \sigma_0(-\xi)}~,
\end{equation}
with $\sigma_0(-\xi) \equiv \textrm{ph}[\Gamma(1-i\xi)]$ being the so-called `Coulomb phase shift', and $\textrm{ph}[\Gamma(1-i\xi)]$ denotes the principal value of the phase (argument) of the complex-valued gamma function $\Gamma(1-i\xi)$ (see footnote~\ref{footnote:indication-logarithm-phase}). For large $|\xi|$ while $\xi k \eta$ is held fixed, 
and $2|\xi|\gg -k\eta \gg (8|\xi|)^{-1}$, 
we arrive at 
\begin{eqnarray}
    \label{eq:A+-massless-regime1}|\mathcal{A}^{\textrm{E}}_{+}(\eta)|&\simeq& \frac{1}{\sqrt{2k}}\left(\frac{k\eta}{2\xi}\right)^{1/4}e^{|\xi|\pi-2\sqrt{2\xi k \eta}}~,\\
    \label{eq:A--massless-regime1}|\mathcal{A}^{\textrm{E}}_{-}(\eta)|&\simeq& \frac{1}{\sqrt{2k}}\left(\frac{k\eta}{2\xi}\right)^{1/4}~,
\end{eqnarray}
which are the amplitudes of the transverse modes in the massless case \cite{Anber:2009ua}. At late times (`$\textrm{L}$'), however, $|\bar{m}^2/\eta^2 \mp 2\xi k/\eta| \gg k^2$ as $\eta \rightarrow 0^{-}$. Introducing the variables 
\begin{equation}
    x \equiv 2\sqrt{\pm 2\xi k \eta}~, \ \ \ \mathcal{A}_{\pm}(\eta) \equiv \sqrt{-\eta}\,f_{\pm}(x)~,
\end{equation}
Eq.~\eqref{eq:Whittaker-eq-Aplusminus-z} reduces to the standard modified Bessel equation,
\begin{equation}
    x^2\frac{\textrm{d}^2f_{\pm}}{\textrm{d}x^2}
    +x\frac{\textrm{d}f_{\pm}}{\textrm{d}x}
    -\left(x^2+4\mu^2\right)f_{\pm}=0~,
\end{equation}
whose general solution is
\begin{equation}
    \label{eq: matching procedure second regime}\mathcal{A}_{\pm}^{\textrm{L}}(\eta) = \sqrt{-\eta}\left[c_1^{\pm}K_{2\mu}\left(2\sqrt{\pm 2\xi k \eta}\right)+c_2^{\pm}I_{2\mu}\left(2\sqrt{\pm 2\xi k \eta}\right)\right],
\end{equation}
where $I_{\nu}(z)$ and $K_{\nu}(z)$ are the modified Bessel functions of the first and second kind, respectively. By matching the early-time and late-time solutions for $\mathcal{A}_{\pm}$ in the overlapping regime of applicability, we determine the constants of integration $c_1^{\pm}$ and $c_2^{\pm}$. For the sole purpose of this matching procedure, we further expand Eq.~\eqref{eq: matching procedure second regime} in the limit $|\sqrt{\pm 2\xi k \eta}| \gg (1,|\mu|)$, (see Sec.~10.40 of Ref.~\cite{NIST:DLMF}), and retain only the leading-order term. The resulting expression is then compared with Eqs.~\eqref{eq:A+-massless-regime1} and \eqref{eq:A--massless-regime1}, and leads to $c_2^{\pm}=0$, $c_1^{+} = \sqrt{2/\pi}e^{|\xi|\pi}$, and $c_1^{-}=\sqrt{2/\pi}$. 

Below, the Bessel expressions provide an effective ultraviolet cutoff of the late-time approximation. They suppress short-wavelength modes outside the instability band while retaining the tachyonic enhancement. Nevertheless, although the Bessel approximation provides a good description at late times, the matching procedure becomes less accurate as $\bar m \to |\xi|$. In this regime, we find that the Bessel solution can underestimate the peak amplitude of the mode function by a factor of a few. This discrepancy originates from the fact that, in deriving the Bessel approximation, the $k^2$ term in the mode equation is neglected. As a result, the lower edge of the instability band in Eq.~\eqref{eq:Instability Band 2} is effectively replaced by $-k \eta = \bar m^2/(2|\xi|)$. These two scales coincide in the small-mass limit, but differ by a factor of two as $\bar m \to |\xi|$.

\begin{figure}
    \centering
\includegraphics[width=1.00\linewidth]{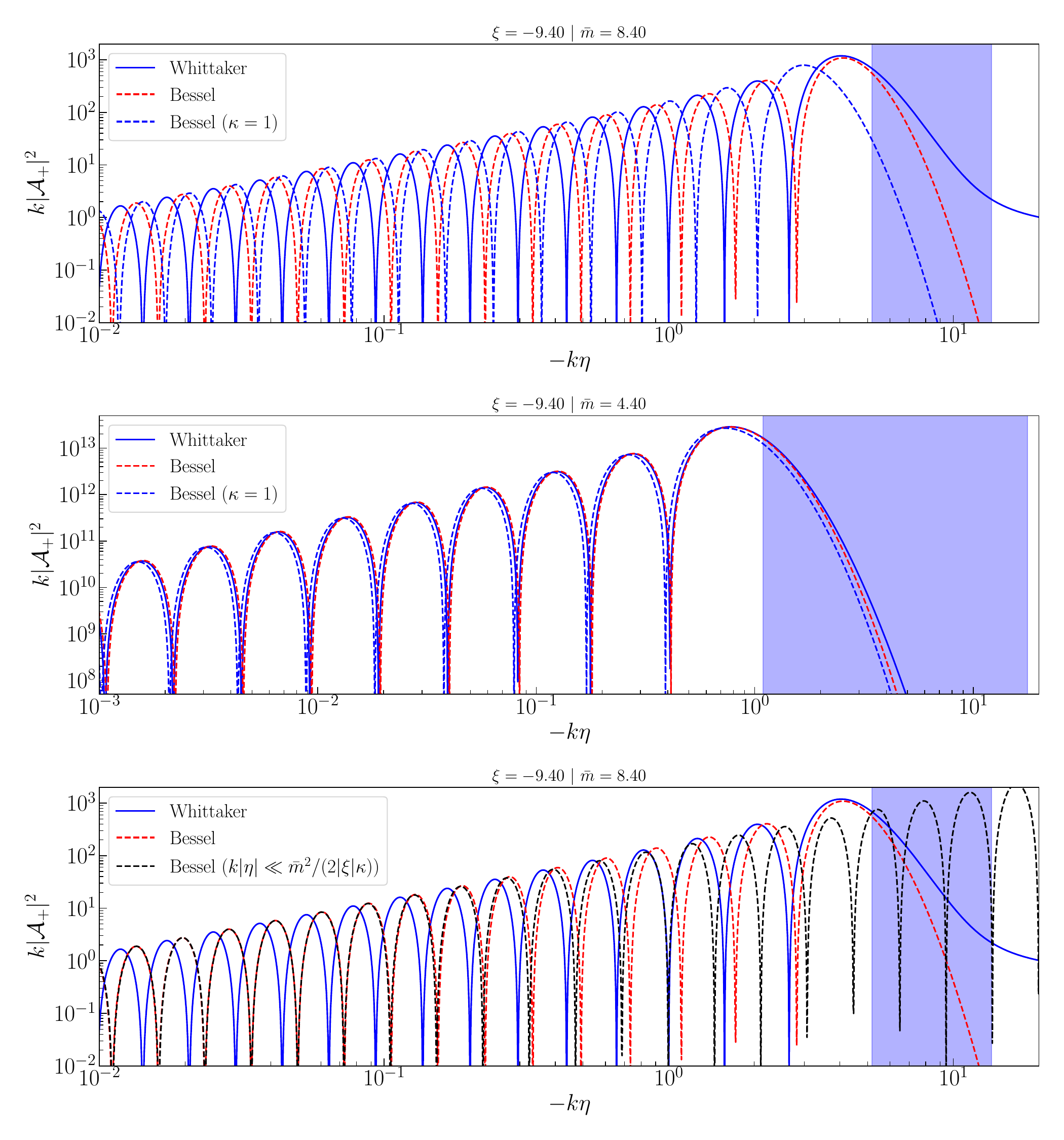}
    \caption{Comparison of $k|\mathcal{A}_{+}|^2$ as a function of $-k\eta$ obtained from the full Whittaker solution, Eq.~\eqref{eq:solution-to-Whittaker-eq-Aplusminus} (solid blue), and its Bessel approximation, Eq.~\eqref{eq:mod-sq-Aplus-largemass-kappa} (dashed red and blue for $\kappa\neq 1$ and $\kappa =1 $, respectively), for two representative values of the mass (first and second panels). The blue band indicates the instability region defined by Eq.~\eqref{eq:Instability Band 2}. The third panel additionally displays the large-$\bar{m}$ limit of the Bessel approximation, Eq.~\eqref{eq:heavy-bessel-envelope} (black dashed), for the same values of ($\xi$,$\bar{m}$) as in the first panel.}
    \label{fig:kA2-Bessel-Whittaker}
\end{figure}

To account for this mismatch, we rescale the Bessel argument, equivalently replacing $k$ by $\kappa k$. We choose $\kappa$ so that the Bessel function peaks at the lower edge of the instability band, $|\xi|-\sqrt{\xi^2-\bar{m}^2}$, and is given by
\begin{equation}
    \kappa
    \equiv
    \frac{\bar m^2}{2|\xi|}
    \frac{1}{|\xi|-\sqrt{\xi^2-\bar m^2}}=
    \frac{1}{2}\left(1+\sqrt{1-\frac{\bar m^2}{\xi^2}}\right). \label{eq:kappa-definition}
\end{equation}
Thus, $\kappa$ varies from $\kappa\simeq 1$ for $\bar m\ll |\xi|$ to $\kappa\simeq 1/2$ in the narrow-instability limit $\bar m\simeq |\xi|$. The corrected mode functions take the form (see footnote~\ref{footnote-clarification-argument-stable} for clarification regarding the argument of the Bessel function of the stable mode)
\begin{eqnarray}
    \label{eq:mod-sq-Aplus-largemass-kappa}
    \mathcal{A}_{+}(\eta)
    &\simeq&
    \sqrt{\frac{-2\eta}{\pi}}\,
    e^{|\xi|\pi}
    K_{2i\tilde{\mu}}\left(2\sqrt{2\xi \kappa k\eta}\right),
    \\
    \label{eq:mod-sq-Aminus-largemass-kappa}
    \mathcal{A}_{-}(\eta)
    &\simeq&
    \sqrt{\frac{-2\eta}{\pi}}\,
    K_{2i\tilde{\mu}}\left(-2i\sqrt{2\xi \kappa k \eta}\right).
\end{eqnarray}
We defined $\tilde{\mu} \equiv -i\mu$, making $\tilde{\mu}$ real when $\bar{m}> 1/2$, and we remind the reader that $\xi k\eta>0$. Fig.~\ref{fig:kA2-Bessel-Whittaker} shows the comparison of the full Whittaker solution with the Bessel approximations, both with $\kappa \neq 1$ and $\kappa=1$. The $\kappa$-corrected solution is seen to provide a more accurate description of the mode function near its peak, particularly when $|\xi|\sim \bar m$.

\subsection{\label{sec:gauge-field-rho-EB-helicity}Gauge-Field Energy and Pseudo-Scalar Densities}
With the aim of performing the integrals in Eqs.~\eqref{eq:total-gauge-energy-density} and \eqref{eq:backreacting-term-EB}, it is useful to employ some identities involving derivatives of the modified Bessel function $K_{2i\tilde{\mu}}(z)$. Our interest lies in the unstable mode $\mathcal{A}_{+}$. The time derivative of $|\mathcal{A}_{+}|^2$ in Eq.~\eqref{eq:mod-sq-Aplus-largemass-kappa} can be written as (see Sec.~10.29 of Ref.~\cite{NIST:DLMF}) 
\begin{equation}
    \label{eq:partialeta-A+ms-approx}\frac{\partial}{\partial\eta}[\left|\mathcal{A}_{+}(\eta,k)\right|^2] \simeq -\frac{2}{\pi}e^{2|\xi|\pi}K_{2i\tilde{\mu}}(x)\left\{K_{2i\tilde{\mu}}(x)-\frac{x}{2}\left[K_{1+2i\tilde{\mu}}(x)+K_{1-2i\tilde{\mu}}(x)\right]\right\},
\end{equation}
where $x\equiv 2\sqrt{2\xi\kappa k \eta} \in \mathbb{R}^{+}$. The right-hand side is real in spite of the fact that the last two modified Bessel functions have complex orders, $1\pm 2i\tilde{\mu}$, since $(K_{1+2i\tilde{\mu}} + K_{1-2i\tilde{\mu}})^{*} = K_{1+2i\tilde{\mu}}+K_{1-2i\tilde{\mu}}$.\footnote{In fact, according to Sec.~10.34 of Ref.~\cite{NIST:DLMF}, because the argument of the modified Bessel function in Eq.~\eqref{eq:mod-sq-Aplus-largemass-kappa} is real and the order is purely imaginary, together with the property $K_{-\nu}(z) = K_{\nu}(z)$, we have that $K_{2i\tilde{\mu}}(x)$ is real-valued, with $x\in\mathbb{R}^{+}$.} On the other hand, $|\mathcal{A}_{+}^{'}|^2 = \mathcal{A}_{+}^{'}\mathcal{A}_{+}^{'*}$ is instead given by 
\begin{equation}
    \label{eq:mspartialeta-A+-approx}|\mathcal{A}_{+}^{'}|^2 \simeq \frac{e^{2|\xi|\pi}}{-2\pi \eta}\left\{K_{2i\tilde{\mu}}(x)-\frac{x}{2}\left[K_{1+2i\tilde{\mu}}(x)+K_{1-2i\tilde{\mu}}(x)\right]\right\}^2.
\end{equation}
Owing to the instability, the amplitude of the `$+$' polarisation mode is much larger than those of the other two modes, $|\mathcal{A}_{+}|\gg |\mathcal{A}_{-}|, |\mathcal{A}_L|$. The same hierarchy applies to the respective amplitudes of their time derivatives, as $\mathcal{A}_{+}$ grows exponentially. Equations~\eqref{eq:total-gauge-energy-density} and \eqref{eq:backreacting-term-EB} can then be approximated as
\begin{eqnarray}
    \label{eq:rhoA-approx-Aplus}\rho_A &\simeq& \frac{1}{4\pi^2}\int \textrm{d}\ln k~k^3 \left[|\mathcal{A}_{+}^{'}|^2+\left(k^2+\frac{\bar{m}^2}{\eta^2}\right)|\mathcal{A}_{+}|^2\right],\\
    \label{eq:EB-approx-Aplus}\langle \hat{\mathbf{E}}\cdot \hat{\mathbf{B}}\rangle &\simeq& -\frac{1}{4\pi^2}\int \textrm{d}\ln k~k^4 \frac{\partial}{\partial \eta}[|\mathcal{A}_{+}|^2]~,
\end{eqnarray}
respectively. 

\begin{figure}
    \centering
\includegraphics[width=1.00\linewidth]{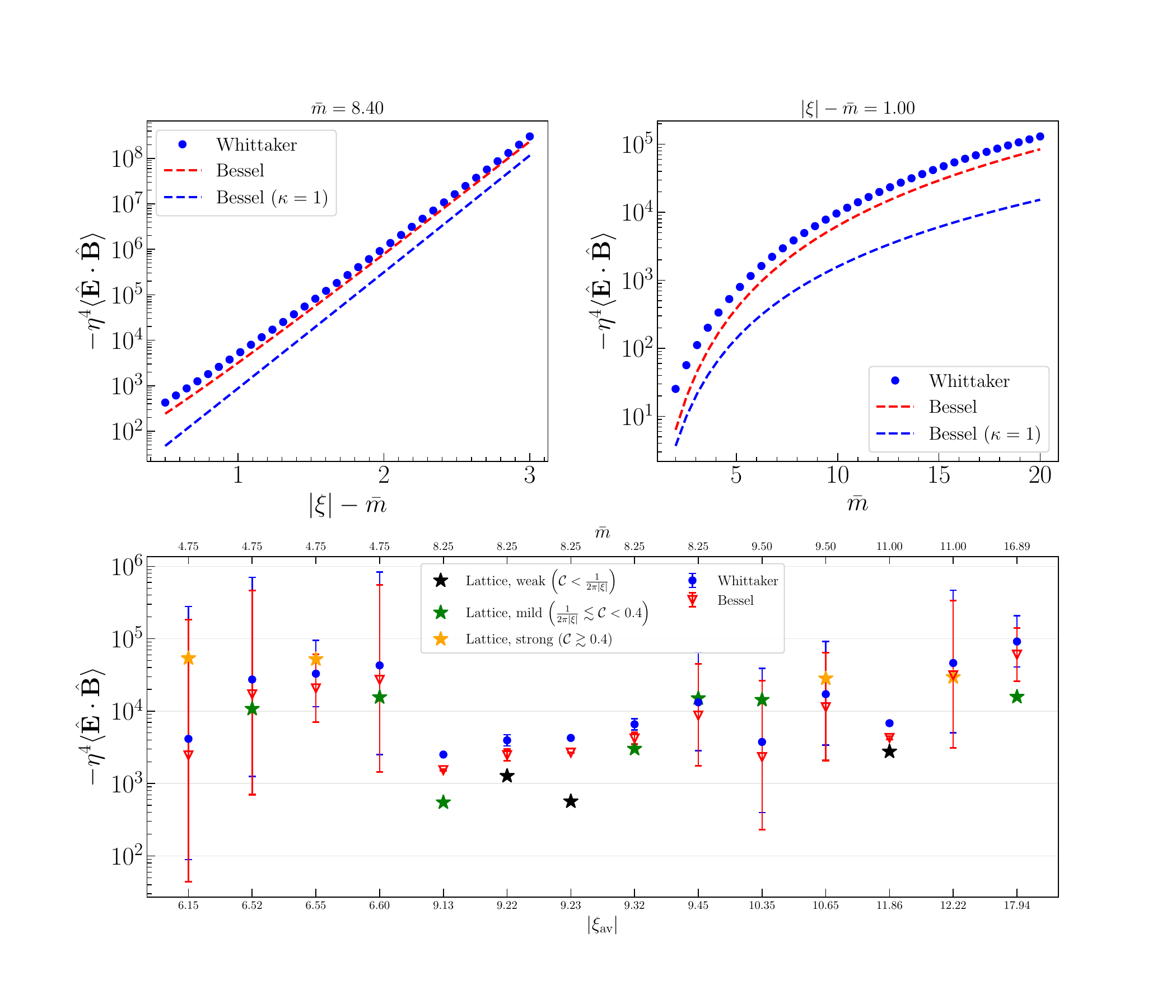}
    \caption{Comparison of $-\eta^4\langle\hat{\mathbf{E}}\cdot\hat{\mathbf{B}}\rangle$ computed from the full Whittaker solution, Eq.~\eqref{eq:solution-to-Whittaker-eq-Aplusminus} (blue circles), the $\kappa$-corrected Bessel approximation, Eq.~\eqref{eq:mod-sq-Aplus-largemass-kappa} (red dashed), and the case of $\kappa=1$ (blue dashed). The upper-left panel shows the dependence on $|\xi|-\bar m$ for fixed $\bar m=8.4$, while the upper-right panel displays the dependence on $\bar m$ along $|\xi|-\bar m=1$. The lower panel compares the Whittaker and Bessel results with lattice simulations for different values of $(|\xi|,\bar m)$ (see Table~\ref{tab:parameter_points}). Each simulation samples a range of values for $|\xi|$, hence the average value in the lower horizontal axis and the error bars indicating the interval. Green and orange stars correspond to simulations satisfying $(2\pi|\xi|)^{-1}\lesssim \mathcal{C}<0.4$ and $\mathcal{C}\gtrsim0.4$, respectively, while black stars denote simulations with $\mathcal{C}<(2\pi|\xi|)^{-1}$ (see Eq.~\eqref{eq:def-curly-C-ratio}). Overall, the Bessel approximation reproduces the Whittaker prediction well throughout the parameter range considered, and both are in reasonable agreement, within a factor of few for the runs with narrow $|\xi|-\bar{m}$, with the lattice results.}
    \label{fig:EB-Bessel-Whittaker}
\end{figure}

We note that the approximate expressions above, written in terms of the modified Bessel functions, are valid deep in the IR. After changing the integration variable to $x\equiv 2\sqrt{2\xi \kappa k \eta}$, the lower integration limit can therefore be taken to be zero. Moreover, the upper limit may be extended to infinity, since the Bessel-based expression is exponentially suppressed for large $x$, rendering the contribution from this region comparatively negligible. The integrals evaluate to (see Sec.~\ref{sec:app-detailed-EBandrho} for further details)
\begin{equation}
     \label{eq:main-result-rho-EB-very-heavy}
     \rho_A \simeq \frac{4\bar{m}^7}{105\pi^2}\frac{e^{2(|\xi|-\bar{m})\pi}}{\kappa^3|\xi|^3\eta^4}\left(1+\frac{\bar{m}^2}{6\kappa^2\xi^2}\right), \qquad
     \langle \hat{\mathbf{E}}\cdot \hat{\mathbf{B}}\rangle \simeq -\frac{3\bar{m}^7}{70\pi^2}\frac{e^{2(|\xi|-\bar{m})\pi}}{\left(\kappa\xi\eta\right)^4}~,
\end{equation}
where we assumed $\tilde{\mu}\simeq \bar{m}\gg 1$. The term $\bar{m}^2/(6\kappa^2 \xi^2)$ can be neglected when $\kappa\simeq 1$ (and therefore $\bar{m}\ll |\xi|$), but should be retained otherwise. The accuracy of the Bessel approximation in reproducing the pseudo-scalar density $\langle \hat{\mathbf{E}}\cdot \hat{\mathbf{B}}\rangle$ is illustrated in Fig.~\ref{fig:EB-Bessel-Whittaker}. The lower panel further compares both analytical predictions with lattice simulations, which are generally found to be in reasonable agreement, within a factor of few, with the Whittaker and Bessel results. Further details on the classification of the different backreaction regimes in terms of $\cal C$, defined in Eq.~\eqref{eq:def-curly-C-ratio}, are presented and discussed in detail in Sec.~\ref{sec:Backreaction-FULL}, while the corresponding simulation results and parameter values are summarised in Table~\ref{tab:parameter_points}. 

It is instructive to compare these results with those obtained in the case of a massless gauge field, for which $\mu = \frac{1}{2}$ \cite{Barnaby:2011vw}.\footnote{\label{footnote:whataboutthecaseofrealmu}The massless and light gauge field cases, corresponding to $\mu=1/2$ and $\mu\in [0,1/2)$, respectively, yield the same expressions as in Eqs.~\eqref{eq:heavy-gauge-field-rhoA-analytical} and \eqref{eq:heavy-gauge-field-EB-analytical}, with the real-valued $\mu$ appearing in place of $i\tilde{\mu}$ (and therefore $-\tilde{\mu}^2 \rightarrow \mu^2$; cf.~Eqs.~\eqref{eq:rhorealmu-amust} and \eqref{eq:EBrealmu-amust}).} We find
\begin{eqnarray}
    \label{eq:large-mass-limit-comparison-masslessrhoA}\frac{\rho_A}{\rho_{A}|_{m=0}} &\simeq& \frac{256\pi}{27\kappa^3}\left(\frac{m}{H}\right)^7 e^{-2\pi m/H}\left(1+\frac{m^2}{6\kappa^2H^2\xi^2}\right)~,\\
   \label{eq:large-mass-limit-comparison-masslessEB}\frac{\langle \hat{\mathbf{E}}\cdot \hat{\mathbf{B}} \rangle}{\langle \hat{\mathbf{E}} \cdot \hat{\mathbf{B}}\rangle|_{m=0}} &\simeq& \frac{16\pi}{3\kappa^4}\left(\frac{m}{H}\right)^7 e^{-2\pi m/H} ~.
\end{eqnarray}
The massive analogue exhibits a power-law enhancement proportional to $\bar{m}^7$, and an exponential suppression $e^{-2 m\pi/H}$ signalling the fact that the instability disappears when the mass overcomes $|\xi|$. We further note that, because the very same enhancement affects both quantities, their ratio remains essentially unchanged relative to the massless case except for the presence of the $\kappa$ and $\bar{m}^2/(6\kappa^2 \xi^2)$ terms:
\begin{equation}
    \label{eq:ratio-EB-rho-analytic-kappa}\frac{|\langle \hat{\mathbf{E}}\cdot \hat{\mathbf{B}}\rangle|}{\rho_A}\simeq \frac{27}{4}\frac{\kappa |\xi|}{6\kappa^2 \xi^2+\bar{m}^2}~.
\end{equation}

For completeness, in the light-field regime $0\leq\mu\leq1/2$, where $\kappa=1+\mathcal O(\bar m^2/\xi^2)$, the corresponding expressions are
\begin{align}
\label{eq:EBrealmu-amust}
\langle \hat{\mathbf E}\cdot\hat{\mathbf B}\rangle
&\simeq-\frac{3}{35}\Gamma(4+2\mu)\Gamma(4-2\mu)
\frac{e^{2\pi|\xi|}}{2^9\pi^3(\xi\eta)^4}~,\\
\label{eq:rhorealmu-amust}
\rho_A
&\simeq\frac{\Gamma(3+2\mu)\Gamma(3-2\mu)}{5}
\left[\frac{11-8\mu^2}{21}
+\frac{(4-\mu^2)(9-4\mu^2)}{63\xi^2}\right]
\frac{e^{2\pi|\xi|}}{2^7\pi^3|\xi|^3\eta^4}~.
\end{align}

\section{\label{sec5-scalarpowerspectrum-fullsec}Scalar Power Spectrum}
In this section, we first derive an integral form for the scalar perturbations sourced by the gauge fields. We then study separately the cases of massless/light and heavy gauge fields. Owing to the technical nature of the analysis, we focus here on the main steps of the derivation, while several intermediate calculations are deferred to Appendix~\ref{app:detailed-calculation-inverse-decay}. Readers primarily interested in the final results may proceed directly to Eqs.~\eqref{eq:PidPvacratio-light-semianalytical} and \eqref{eq:loop-contribution-weak-backreaction}, where the analytical expressions for the sourced scalar power spectrum are presented.

\subsection{\label{sec:scalarpowerspectrum-weak}Integral Form of the Sourced Scalar Spectrum}
The inverse-decay process, whereby two amplified gauge quanta combine to produce an inhomogeneous inflaton mode, supplies a stochastic source for the scalar perturbation. In the spatially flat gauge, and in the standard slow-roll approximation in which metric-induced terms are neglected, the first-order field equation is (see Eq.~\eqref{eq:eq-sf-full})
\begin{equation}
\label{eq:sf-eq-pert-linear}
\delta\phi''-\nabla^2\delta\phi+2\mathcal H\delta\phi'
+a^2V_{,\phi\phi}(\phi)\delta\phi
=\frac{\alpha}{a^2f}\,\delta(\mathbf E\cdot\mathbf B)~,
\end{equation}
where $\phi$ is the homogeneous background, $\delta\phi$ is its perturbation, and
\begin{equation}
\delta(\mathbf E\cdot\mathbf B)
\equiv \mathbf E\cdot\mathbf B-\langle\mathbf E\cdot\mathbf B\rangle
\end{equation}
is the connected fluctuation of the gauge-field source. The omitted scalar-metric contributions are slow-roll suppressed in the regime considered here \cite{Barnaby:2011vw,Ferreira:2014zia}. Defining the canonical variable $\delta\varphi\equiv a\,\delta\phi$, the Fourier-space equation becomes \cite{Anber:2009ua,Barnaby:2011vw}
\begin{equation}
\label{eq:sf-eq-pert-linear-Fourier}
\delta\varphi_{\mathbf k}''+
\left(k^2+a^2m_\phi^2-\frac{a''}{a}\right)\delta\varphi_{\mathbf k}
=J_{\mathbf k}~,
\end{equation}
where $m_\phi^2\equiv V_{,\phi\phi}$ and
\begin{equation}
\label{eq:source-Fouriermodes-sf-evolution}
J(\eta,\mathbf k)=\frac{\alpha}{a(\eta)f}
\int\frac{\textrm d^3\mathbf x}{(2\pi)^{3/2}}
\delta(\mathbf E\cdot\mathbf B)(\eta,\mathbf x)e^{-i\mathbf k\cdot\mathbf x}~.
\end{equation}

It is convenient to express the result in terms of the gauge-invariant curvature perturbation, which in spatially flat gauge takes the form~\cite{Malik:2008im}
\begin{equation}
    \zeta \equiv -\frac{\mathcal{H}}{\phi^{'}}\delta \phi~.
\end{equation}
The statistical properties of the scalar perturbations are then encoded in the two-point correlation function of $\zeta$. Starting from the Fourier-space correlator, $\langle \hat{\zeta}(\mathbf{k})\hat{\zeta}(\mathbf{k}')\rangle$, we define the dimensionless power spectrum $\mathcal{P}_{\zeta}$ as
\begin{equation}
    \label{eq:def-normalised-power-spectrum-total}\langle \hat{\zeta}(\mathbf{k})\hat{\zeta}(\mathbf{k}')\rangle\equiv\frac{2\pi^2}{k^3}\mathcal{P}_{\zeta}(k)\delta^{(3)}(\mathbf{k}+\mathbf{k}')~,
\end{equation}
which can be written as  
\begin{equation}
    \label{eq:relation-total-pzeta-pvac-pid}\mathcal{P}_{\zeta}(k) \equiv \mathcal{P}_{\zeta}^{\textrm{vac}}(k)+\mathcal{P}_{\zeta}^{\textrm{id}}(k)~.
\end{equation}
The superscripts `vac' and `id' stand for the vacuum and inverse-decay contributions, respectively. As we show in Sec.~\ref{app:detailed-calculation-inverse-decay-POWER-SPECTRUM} the inverse-decay contribution is given by
\begin{align}
    \nonumber&\frac{\mathcal{P}^{\textrm{id}}_{\zeta}}{\mathcal{P}^{\textrm{vac}}_{\zeta}}\simeq \frac{\mathcal{P}_{\zeta}^{\textrm{vac}}}{4\pi^2}\xi^2e^{4|\xi|\pi}\int^{\infty}_0 \frac{\textrm{d}q_*}{q_*}\int^{1+q_*}_{|1-q_*|}\frac{\textrm{d}p_*}{p_*}\left[(q_*+p_*)^2-1\right]^2 \\
    \nonumber&\times \left\{\int^{\infty}_{0}\textrm{d}x\left[\sin(x)-x\cos(x)\right]\left\{q_*K_{2\mu}(2\sqrt{2 |\xi| \kappa p_*x})K_{2\mu}(2\sqrt{2 |\xi| \kappa q_*x})-q_*\sqrt{2|\xi|\kappa p_* x}\right.\right.\\
    \label{eq:the-ratio-semianalytical-many-bessels1}&\left.\left.\times K_{2\mu}(2\sqrt{2|\xi|\kappa q_*x})\left[K_{1-2\mu}(2\sqrt{2|\xi|\kappa p_*x})+K_{1+2\mu}(2\sqrt{2|\xi|\kappa p_*x})\right]+(q_* \leftrightarrow p_*)\right\}\right\}^2,
\end{align}
with $q_*\equiv q/k$ and $p_*\equiv p/k$, so that $q\equiv |\mathbf{q}|$ and $p\equiv |\mathbf{k}-\mathbf{q}|$ denote the momenta of the two gauge quanta sourcing the scalar perturbation through the inverse-decay process. We employed the definition of $\mathcal{P}_{\zeta}^{\textrm{vac}}$ in terms of $\xi$ given in Eq.~\eqref{eq:def-As-intermsof-xi}.

The inner integral in Eq.~\eqref{eq:the-ratio-semianalytical-many-bessels1} can be evaluated analytically in the massless case when $|\xi|$ is large \cite{Anber:2009ua,Barnaby:2011vw}. However, the presence of the gauge-field mass adds further complications. In the following subsections, we discuss the appropriate treatment of the integrals, beginning with the massless and light gauge-field regimes before turning to the heavy-field case, which constitutes the primary focus of this work.

\subsection{The Case of Massless and Light Gauge Fields}
For a massless or light vector, $0\leq\bar m\leq 1/2$, $\mu=\sqrt{1/4-\bar m^2}$ is real. In the range relevant for the onset of inverse decay, $|\xi|\simeq3$, the matching parameter of Eq.~\eqref{eq:kappa-definition} obeys $\kappa=1+\mathcal{O}(\bar m^2/\xi^2)$. We therefore set $\kappa=1$ in this subsection, consistently with the accuracy of the large-$|\xi|$ expansion.

As we show in Appendix~\ref{app:detailed-calculation-inverse-decay}, for massless and light fields, the integral in Eq.~\eqref{eq:the-ratio-semianalytical-many-bessels1} further simplifies to 
\begin{align}
  \frac{\mathcal{P}^{\textrm{id}}_{\zeta}}{\mathcal{P}^{\textrm{vac}}_{\zeta}}
  &\simeq\frac{2\kappa}{\pi^{2}}\mathcal{P}^{\textrm{vac}}_{\zeta}|\xi|^{3}e^{4\pi|\xi|}
   \int_{0}^{\infty}\!\textrm{d}q_{*}
   \int_{|1-q_{*}|}^{1+q_{*}}\!\textrm{d}p_{*}\,
   \big[(q_{*}+p_{*})^{2}-1\big]^{2}\big(\sqrt{q_{*}}+\sqrt{p_{*}}\big)^{2}
   \nonumber\\
  &\quad\times\Bigg\{\int_{0}^{\infty}\!\textrm{d}x\,\sqrt{x}\,
   [\sin x-x\cos x]\,
   K_{2\mu}\!\left(2\sqrt{2|\xi|\kappa q_{*}x}\right)
   K_{2\mu}\!\left(2\sqrt{2|\xi|\kappa p_{*}x}\right)\Bigg\}^{2}~.
  \label{eq:the-ratio-semianalytical}
\end{align}
The dominant part of the time integral in Eq.~\eqref{eq:the-ratio-semianalytical} is captured by retaining  the first correction term in the large-argument expansion of the modified Bessel functions in Eq. \eqref{eq:large-argument-K-heavy} 
\begin{equation}
K_{2\mu}(y)\simeq \sqrt{\frac{\pi}{2y}}e^{-y}
\left[1-\frac{1/4-4\mu^2}{2y}\right].
\label{eq:light-bessel-large-argument}
\end{equation}
The complete asymptotic series is given in Eq.~\eqref{eq:large-argument-K-heavy}. The time integrations are dominated by the region with $x \lesssim 1$. Thus we can expand $\sin x-x\cos x\simeq x^3/3$, and the inner integral becomes
\begin{align}
\mathcal I\simeq{}\frac{\pi(q_*p_*)^{-1/4}}{12\sqrt{2|\xi|}}
\int_0^\infty\!\textrm d x\,x^3e^{-2\sqrt{2|\xi|x}(\sqrt{q_*}+\sqrt{p_*})}\left(1-\frac{1/4-4\mu^2}{4\sqrt{2|\xi|q_*x}}\right)
\left(1-\frac{1/4-4\mu^2}{4\sqrt{2|\xi|p_*x}}\right).
\end{align}
The $x$ integral is elementary and gives
\begin{equation}
\mathcal I\simeq
\frac{15\pi}{12288\sqrt{2}\,|\xi|^{9/2}}
\frac{(q_*p_*)^{-1/4}}{(\sqrt{q_*}+\sqrt{p_*})^6}
\left[
\frac{168}{(\sqrt{q_*}+\sqrt{p_*})^2}
-\frac{(1/4-4\mu^2)(47/4+4\mu^2)}{\sqrt{q_*p_*}}
\right].
\label{eq:light-time-integral}
\end{equation}
The remaining momentum integrals can also be evaluated analytically. The result is
\begin{equation}
\frac{\mathcal P^{\rm id}_\zeta}{\mathcal P^{\rm vac}_\zeta}
\simeq \frac{25\mathcal{P}_{\zeta}^{\textrm{vac}}}{16777216}\frac{e^{4\pi|\xi|}}{|\xi|^6}\,\mathcal F(\mu)~,
\label{eq:PidPvacratio-light-semianalytical}
\end{equation}
where $\mathcal P^{\rm vac}_\zeta=H^4/(4\pi^2\dot\phi^2)$ at leading slow-roll order and
\begin{align}
\mathcal F(\mu)\equiv{}&\int_0^\infty\frac{\textrm d q_*}{\sqrt{q_*}}
\int_{|1-q_*|}^{1+q_*}\frac{\textrm d p_*}{\sqrt{p_*}}
\frac{[(q_*+p_*)^2-1]^2}{(\sqrt{q_*}+\sqrt{p_*})^{10}} \left[
\frac{168}{(\sqrt{q_*}+\sqrt{p_*})^2}
-\frac{f(\mu)}{\sqrt{q_*p_*}}
\right]^2\nonumber\\
={}&\frac{7168}{143}-\frac{256}{99}f(\mu)+\frac{32}{945}f(\mu)^2~,
\qquad
f(\mu)\equiv(1/4-4\mu^2)(47/4+4\mu^2)~.
\label{eq:curlyF-ratio-powerspectrum}
\end{align}

The condition $\mathcal P^{\rm id}_\zeta=\mathcal P^{\rm vac}_\zeta$ is solved by
\begin{equation}
|\xi|\simeq-\frac{3}{2\pi}W_{-1}\!\left[-\frac{2\pi}{3}
\left(\frac{25\mathcal{P}_{\zeta}^{\textrm{vac}}\mathcal F(\mu)}{16777216}\right)^{1/6}\right],
\label{eq:xi-from-perturbation-backreaction-onset}
\end{equation}
where $W_{-1}$ is the lower real branch of the Lambert function, defined on $[-e^{-1},0)$. Taking $\mathcal{P}_{\zeta}^{\textrm{vac}}=2.1\times10^{-9}$ as a benchmark gives $|\xi|\simeq2.80$ in the massless case consistently with previous literature. 

\subsection{\label{sec:scalarpowerspectrum-weak-CASE-HEAVY-FIELDS}The Case of Heavy Gauge Fields}
In the case of heavy gauge fields, instead of performing a large argument expansion, it is more convenient to perform the opposite expansion.
In the limit where $\sqrt{2 \kappa |\xi|q_{*}x}$, $\sqrt{2\kappa|\xi|p_{*}x} \ll \tilde{\mu}$,
the oscillatory Bessel function has the envelope \cite{Dunster:1990hlc}
\begin{equation}
K_{2i\tilde\mu}(y)\simeq-\sqrt{\frac{\pi}{\tilde\mu}}e^{-\pi\tilde\mu}
\sin\!\left[2\tilde\mu\ln(y/2)-\arg\Gamma(1+2i\tilde\mu)\right].
\label{eq:heavy-bessel-envelope}
\end{equation}
We impose the validity of this approximation through the momentum-dependent cutoff
$\Lambda(u)=\tilde\mu^2/(c\kappa|\xi|u)$, where $u=\max(q_*,p_*)$ and $c$ is an order-one matching constant. The detailed reduction of the momentum and time integrals is given in Sec.~\ref{app:heavy-weak-spectrum}. The leading result is
\begin{equation}
\frac{\mathcal P^{\rm id}_\zeta}{\mathcal P^{\rm vac}_\zeta}
\simeq \frac{32c_\ell}{c^6}\mathcal{P}_{\zeta}^{\textrm{vac}}
\frac{\tilde\mu^{10}}{\kappa^5|\xi|^3}
\exp[4\pi(|\xi|-\tilde\mu)]~,
\label{eq:loop-contribution-weak-backreaction}
\end{equation}
with $c_\ell \simeq0.04$. 
In arriving to this result, we have used the approximated form of the loop integral given in Eq.~\eqref{eq:the-ratio-semianalytical}. Although, when comparing with the direct numerical evaluation of Eq.~\eqref{eq:the-ratio-semianalytical-many-bessels1} we find that a correction factor of $g(\tilde \mu) = (\tilde \mu^{-2}+0.06\tilde \mu^{-1})$ with $c\simeq 1.27$ improves the agreement between the two. Therefore, henceforth we will use
\begin{equation}
\frac{\mathcal P^{\rm id}_\zeta}{\mathcal P^{\rm vac}_\zeta}
\simeq \frac{32c_\ell}{c^6}\mathcal{P}_{\zeta}^{\textrm{vac}} g(\tilde \mu)
\frac{\tilde \mu^{10}}{\kappa^5|\xi|^3}
\exp[4\pi(|\xi|-\tilde \mu)]~
\label{eq:loop-contribution-weak-backreaction-correction} \, ,
\end{equation}
in the remainder of the paper.\footnote{
However, we emphasise that the function $g(\mu)$ was introduced phenomenologically and calibrated against the numerical evaluation of the loop integrals. We also find that $\mathcal{O}(1)$ variations in the powers of $\tilde{\mu}$ and $|\xi|$ in Eq.~\eqref{eq:loop-contribution-weak-backreaction} describe the data equally well. For example, a comparably good fit is obtained with $\mathcal{P}_{\zeta}^{\rm id}/\mathcal{P}_{\zeta}^{\textrm{vac}}= 0.12 ~\mathcal{P}_{\zeta}^{\textrm{vac}}\tilde{\mu}^{9.5}\exp\left[4\pi\left(|\xi|-\tilde{\mu}\right)\right]/(\kappa^{6}|\xi|^{4}).$ At this stage, the exponents in the analytical parametrisation should therefore be understood as having $\mathcal{O}(1)$ uncertainties. Despite this fact, we verified that, within the range of $\mu$ plotted, the predictions in Fig.~\ref{fig:Summary-plot} using either of the expressions almost overlap, thus showing the robustness of the conclusions.}

\begin{figure}[t]
\centering
\includegraphics[width=0.82\linewidth]{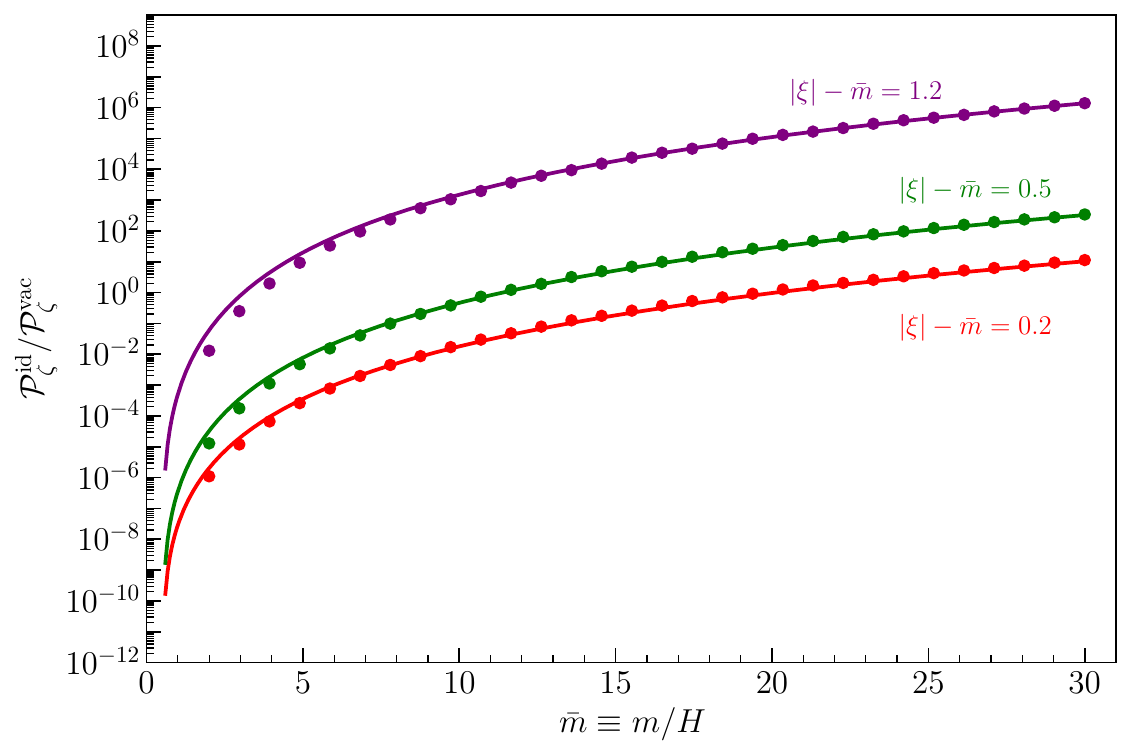}
\caption{Ratio $\mathcal P^{\rm id}_\zeta/\mathcal P^{\rm vac}_\zeta$ as a
function of $\tilde \mu \sim \bar m=m/H$ at fixed $|\xi|-\bar m=0.2$ (red), $0.5$ (green), and
$1.2$ (purple), for $\mathcal P^{\rm vac}_\zeta=2.1\times10^{-9}$. Dots show the direct
numerical evaluation of the full-bracket expression,
Eq.~\eqref{eq:the-ratio-semianalytical-many-bessels1}; solid lines show the
corrected heavy closed form, Eq.~\eqref{eq:loop-contribution-weak-backreaction-correction}. The closed form reproduces the numerical result across
more than fifteen orders of magnitude, the residual mismatch being largest at
the lightest masses, where the heavy-field expansion is least accurate.}
\label{fig:PidrPvac_heavy_fixedxim}
\end{figure}

\section{Estimate of the Onset of Mild and Strong Backreaction \label{sec:Backreaction-FULL}}
We distinguish three regimes. In the \emph{weak backreaction} regime the produced gauge field acts as a source but does not appreciably induce friction effects,  neither on the homogeneous inflaton nor on its linear perturbations. In the \emph{mild backreaction} regime the velocity dependence of gauge field production adds substantial friction to the scalar perturbations while the homogeneous trajectory remains close to ordinary slow roll. In the \emph{strong backreaction} regime the gauge-field friction also changes the background evolution. This ordering is important because the threshold for friction on perturbations  is parametrically lower than the threshold for the one on the background.

\subsection{\label{sec:Backreaction}Strong Gauge-Field Backreaction}
The equation of motion for the inflaton zero-mode is given in cosmic time by
\begin{equation}
\ddot\phi+3H\dot\phi+V_{,\phi}
=\frac{\alpha}{a^4f}\langle\hat{\mathbf E}\cdot\hat{\mathbf B}\rangle~.
\label{eq:sf-eq-backreaction}
\end{equation}
Here and below $\langle\hat{\mathbf E}\cdot\hat{\mathbf B}\rangle$ is the comoving quantity of Eq.~\eqref{eq:backreacting-term-EB}; the corresponding physical density contains the factor $a^{-4}$. We quantify the amount of gauge field backreaction by
\begin{equation}
\mathcal C\equiv
\frac{\alpha\,|\langle\hat{\mathbf E}\cdot\hat{\mathbf B}\rangle|}
{fa^4 (3H|\dot\phi|)}~ \qquad \textrm{(Backreaction Parameter)}~.
\label{eq:def-curly-C-ratio}
\end{equation}
The standard slow-roll solution is reliable for $\mathcal C\ll 1$. We observe in simulations, as discussed in the next section, that the background trajectory begins to change at $\mathcal C\gtrsim 0.4$, and a friction-dominated attractor can be reached when
\begin{equation}
\frac{\alpha}{fa^4}|\langle\hat{\mathbf E}\cdot\hat{\mathbf B}\rangle|
\simeq |V_{,\phi}|~.
\label{eq: strong backreaction}
\end{equation}
Although the weak backreaction formulas break down in the strong backreaction regime, we nevertheless use the weak-regime expressions to estimate the onset of this transition and then test the result on the lattice.

\paragraph{Massless and Light Gauge Fields:}
For massless and light fields, $0\leq\mu\leq1/2$, Eqs.~\eqref{eq:EBrealmu-amust} and \eqref{eq:rhorealmu-amust}, together with $\xi=-\alpha\dot\phi/(2fH)$, give the backreaction parameter
\begin{equation}
\mathcal C\simeq
\frac{\mathcal{P}_{\zeta}^{\textrm{vac}}}{2240\pi}\Gamma(4+2\mu)\Gamma(4-2\mu)
\frac{e^{2\pi|\xi|}}{|\xi|^3}~.
\label{eq:C-light}
\end{equation}
The estimate of the onset of backreaction, $\mathcal C=1$, can therefore be inverted as
\begin{equation}
|\xi|\simeq-\frac{3}{2\pi}W_{-1}\!\left[-\frac{2\pi}{3}
\left(\frac{\mathcal{P}_{\zeta}^{\textrm{vac}}\Gamma(4+2\mu)\Gamma(4-2\mu)}{2240\pi}\right)^{1/3}\right].
\label{eq: backreaction light gauge fields}
\end{equation}

\paragraph{Heavy Gauge Fields:}
Using the heavy-field pseudo-scalar density in Eq.~\eqref{eq:main-result-rho-EB-very-heavy}, we find
\begin{align}
\mathcal C\simeq{}
\frac{4\mathcal{P}_{\zeta}^{\textrm{vac}}}{35}
\frac{\bar m^7}{\kappa^4|\xi|^3}
\exp[2\pi(|\xi|-\bar m)]~.
\label{eq:C-heavy}
\end{align}
The second line uses $\mathcal{P}_{\zeta}^{\textrm{vac}}=H^4/(4\pi^2\dot\phi^2)$. The onset of strong backreaction on the background, $\mathcal C \sim 1$, then corresponds to
\begin{equation}
|\xi| - \bar{m} \simeq\frac{1}{2\pi}
\ln\!\left(\frac{35\kappa^4|\xi|^3}{4\mathcal{P}_{\zeta}^{\textrm{vac}}\bar m^7}\right).
\label{eq:heavy-background-onset}
\end{equation}

\begin{figure}[t]
\centering
\includegraphics[width=0.88\linewidth]{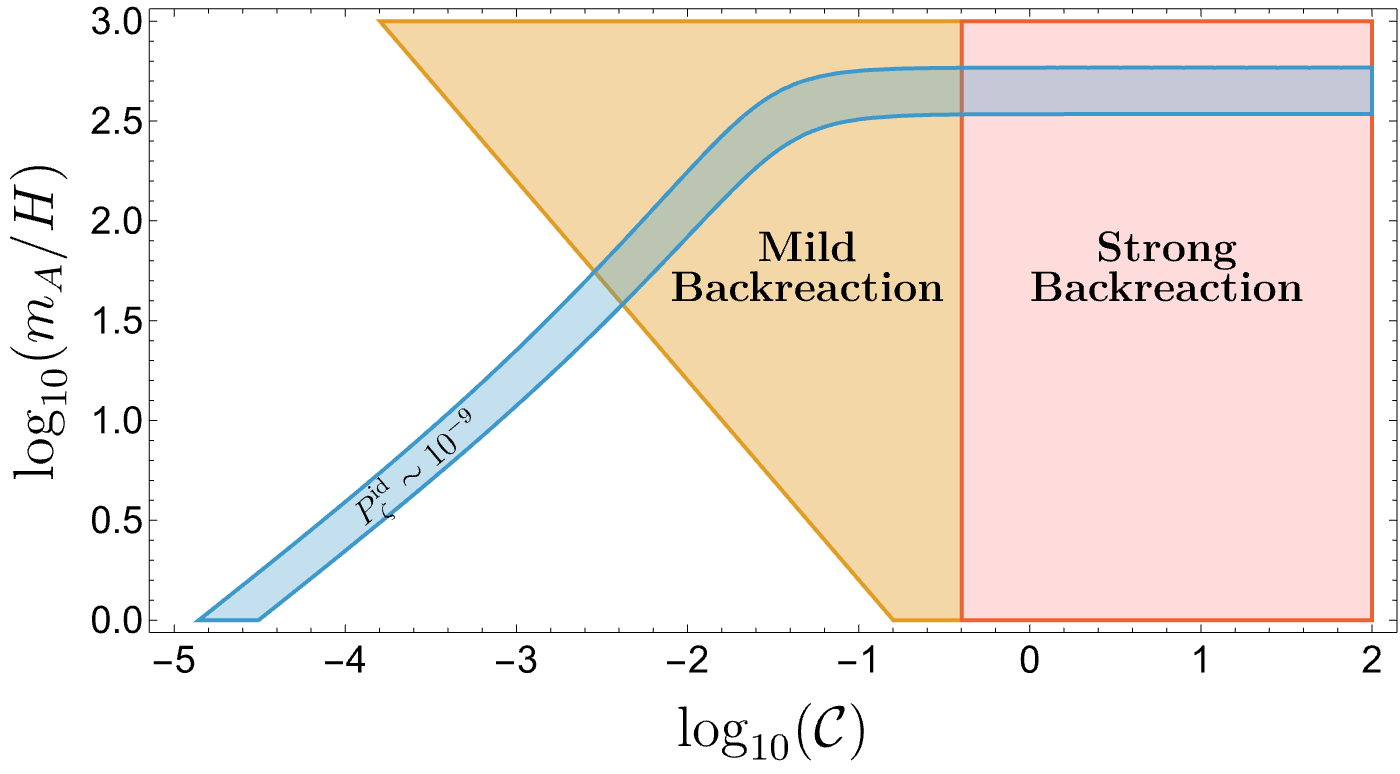}
\caption{Characterisation of the parameter space in the $(\mathcal C,\bar m)$ plane. The orange boundary is $\mathcal C\simeq (2\pi|\xi|)^{-1}$, evaluated with $|\xi|\simeq\bar m$, and separates weak from mild backreaction. The vertical red boundary at $\mathcal C\simeq0.4$ marks the onset of background backreaction. The blue band indicates the vector masses for which the sourced scalar power is of order $10^{-9}$, with its width representing an order-one  uncertainty. Above the band the sourced contribution is smaller than $10^{-9}$.}
\label{fig:Summary-plot}
\end{figure}

\subsection{Mild Gauge-Field Backreaction}
In the previous subsection we estimated the region of the parameter space in which gauge-field production backreacts on the homogeneous inflaton trajectory by providing an additional source of friction. We now ask whether an analogous effect occurs for the scalar perturbations following the arguments in Ref.~\cite{Anber:2009ua}.

The basic observation is that the source term appearing in Eq.~\eqref{eq:sf-eq-pert-linear} depends not only on spacetime, but also on the inflaton velocity $\dot\phi$, through the dependence of the gauge-field mode functions on $\xi$. Expanding this source term around the background value $\dot\phi=\dot\phi_0+\delta\dot\phi$, one obtains
\begin{eqnarray}
	\left< \hat{\mathbf{E}}\cdot \hat{\mathbf{B}}(\mathbf{x},t,\dot\phi) \right>
	=
	\left< \hat{\mathbf{E}}\cdot \hat{\mathbf{B}}(\mathbf{x},t,\dot\phi_0)\right>
	+
	\delta \dot\phi\,
	\frac{\partial}{\partial \dot\phi}
	\left< \hat{\mathbf{E}}\cdot \hat{\mathbf{B}}(\mathbf{x},t,\dot\phi_0)\right>
	+ \cdots .
\end{eqnarray}
The second term is proportional to $\delta\dot\phi$ therefore, once substituted into the equation of motion for $\delta\phi$, it acts as an additional friction term as\footnote{We alert the reader that this equation should be taken as a phenomenological description and not as a first-principle derivation. Indeed, a careful derivation should include the retarded Green's function for the gauge field, see \emph{e.g.}~Ref.~\cite{Laine:2021ego} for an analogous discussion in the case of warm inflation.} 
\begin{equation}
 \delta \ddot\phi_k+3(1+\nu_{\rm f})H\delta\dot\phi_k
+\left(\frac{k^2}{a^2}+V_{,\phi\phi} \right) \delta\phi_k
=\frac{\alpha}{a^4f}\,\delta(\mathbf E\cdot\mathbf B)~,
\label{eq:eom-delta-phi-friction}
\end{equation}
where
\begin{equation}
\nu_{\rm f}\equiv-\frac{\alpha}{3Hfa^4}
\frac{\partial\langle\hat{\mathbf E}\cdot\hat{\mathbf B}\rangle}{\partial\dot\phi}~
\label{eq:nu-definition}
\end{equation}
and we have written the equation of motion for $\delta \phi$ in cosmic time. The coefficient $\nu_{\rm f}$ controls the amount of gauge-induced friction on the scalar perturbations. The pseudo-scalar density depends on $\dot \phi$ mostly through the exponential and so
\begin{equation}
\frac{\partial\langle\hat{\mathbf E}\cdot\hat{\mathbf B}\rangle}{\partial\dot\phi}
\simeq\frac{\alpha\pi}{fH}
\langle\hat{\mathbf E}\cdot\hat{\mathbf B}\rangle~,
\qquad
\nu_{\rm f}\simeq2\pi|\xi|\,\mathcal C~.
\label{eq:nu-C-relation}
\end{equation}
Thus scalar perturbations experience substantial friction already at
\begin{equation}
\mathcal C\gtrsim\frac{1}{2\pi|\xi|}~,
\label{eq:mild-threshold}
\end{equation}
well before the background threshold $\mathcal C\sim 0.4$.

The last step is to understand how the loop induced power spectrum ${\cal P}_\text{id}$ is modified in this mild backreaction regime.\footnote{Note that both ${\cal P}_\text{id}$ and ${\cal P}_\text{vac}$ are modified in this regime. However, in the following, we will assume the inverse decay contribution to dominate over the vacuum piece and so disregard the vacuum contribution.} The homogeneous solutions are $x^rJ_\rho(x)$ and $x^rY_\rho(x)$, and the retarded Green function is
\begin{equation}
G_k(\eta,\eta')=\Theta(\eta-\eta')\frac{\pi}{2k}
 x^r{x'}^{1-r}
\left[J_\rho(x)Y_\rho(x')-Y_\rho(x)J_\rho(x')\right],
\label{eq:green-friction-full}
\end{equation}
where we assumed $\nu_{\rm f}$ to be approximately constant and defined
\begin{equation}
r\equiv\frac{3}{2}(1+\nu_{\rm f})~,
\qquad
\rho\equiv\sqrt{r^2-\frac{V_{,\phi\phi}}{H^2}}~.
\end{equation}
At late time,
\begin{equation}
G_k(0,\eta')=\Theta(\eta-\eta') \frac{2^{\rho-1}\Gamma(\rho)}{k}
\lim_{x\to0}x^{r-\rho}
{x'}^{1-r}J_\rho(x')~.
\label{eq:green-friction}
\end{equation}
In the slow-roll limit $V_{,\phi\phi}/H^2\ll r^2$, one has $\rho\simeq r$. Repeating the heavy-field time integral as in Sec. \ref{sec:scalarpowerspectrum-weak-CASE-HEAVY-FIELDS} with this Green function changes the dimensionless coefficient to
\begin{equation}
c_\ell^{(f)}=\frac{\Gamma(r)^2}{243\,\Gamma(r+1)^2}
=\frac{1}{243r^2}
\simeq\frac{4}{2187\nu_{\rm f}^2}
\qquad(\nu_{\rm f}\gg1)~.
\label{eq:cl-friction}
\end{equation}
Consequently, we estimate the inverse decay contribution to the power spectrum to be\footnote{Note that, unlike the expression in the weak backreaction regime, Eq.~\eqref{eq:loop-contribution-weak-backreaction-correction}, the function $g(\mu)$ does not appear in the friction term. The reason is that, in the weak backreaction regime, we found that including $g(\mu)$ improved the agreement with the numerical evaluation of the full loop integral. By contrast, in the mild and strong backreaction regimes, Eq.~\eqref{eq: Power spectrum, inverse decay, with friction} already provides reasonable agreement with the lattice simulations without the need for this additional factor.}
\begin{equation}
\frac{\mathcal P^{\rm id}_{\zeta,f}}{\mathcal P^{\rm vac}_\zeta}
\simeq\frac{32c_\ell^{(f)}}{c_f^6}\mathcal{P}_{\zeta}^{\textrm{vac}} 
\frac{\tilde\mu^{10}}{\kappa^5|\xi|^3}
 e^{4\pi(|\xi|-\tilde\mu)}~,
\label{eq: Power spectrum, inverse decay, with friction}
\end{equation}
where $c_f\sim 1$ parametrises the same cutoff-matching ambiguity as the parameter $c$ in the weak backreaction case (see Eq.~\eqref{eq:loop-contribution-weak-backreaction-correction}). By comparing the analytical fit with the results with the lattice simulation we find $c_f=0.75$ to be in good agreement. 

\subsection{Power Spectrum for Weak, Mild, and Strong Backreaction}
In the previous subsections we identified the three different backreaction regimes. We can now relate the power spectrum in each of the regimes with the backreaction parameter $\mathcal{C}$. Combining Eqs.~\eqref{eq:loop-contribution-weak-backreaction-correction} and \eqref{eq:C-heavy}, and using $\tilde\mu\simeq\bar m$, gives a particularly transparent result at a fixed amount of backreaction $\mathcal C$. In the weak regime, $\mathcal C\ll(2\pi|\xi|)^{-1}$, 
\begin{equation}
\mathcal P^{\rm id}_\zeta\simeq
\frac{2450c_\ell}{c^6}
\frac{\kappa^3|\xi|^3}{\bar m^4}\,g(\bar m) \mathcal  C^2
\simeq\frac{2.9}{\bar m} g(\bar m)\,\mathcal C^2~,
\label{eq: power spectrum weak backreaction regime}
\end{equation}
where the last expression uses $|\xi|\simeq\bar m$, $\kappa\simeq1/2$, $c=1.27$, and $c_\ell=0.04$. The exponential sensitivity has been absorbed into $\mathcal C$, leaving an inverse power of the vector mass.

In the mild regime, $(2\pi|\xi|)^{-1}\lesssim\mathcal C<0.4$, Eqs.~\eqref{eq:nu-C-relation} and \eqref{eq:cl-friction} cancel the explicit $\mathcal C$ and exponential dependence:
\begin{equation}
\mathcal P^{\rm id}_{\zeta,f} =  {\cal P}_\zeta^\text{id}\times \left(\frac{c}{c_f} \right)^6 \frac{c_\ell^{(f)}}{c_\ell\, g(\bar{m})} 
\simeq \frac{0.64 \kappa^3 \xi}{\bar{m}^4} \simeq \frac{0.08}{\bar{m}^3}~. \label{eq: power spectrum mild backreaction regime}
\end{equation}
In this strong backreaction regime friction continues to act on perturbations and so we expect the perturbations to follow the same behavior as in the mild backreaction regime in Eq.~\eqref{eq: power spectrum mild backreaction regime}.

Based on these estimations on Eqs.~\eqref{eq: power spectrum weak backreaction regime} and \eqref{eq: power spectrum mild backreaction regime} above we can now explore the region of parameters that are compatible with the CMB normalisation of the power spectrum, ${\cal P}_\zeta=2.1 \times 10^{-9}$ \cite{Planck:2018jri}. We show the results in Fig.~\ref{fig:Summary-plot} in terms of the backreaction parameter $\cal C$ and the gauge field mass $m$. We have approximated $|\xi| \simeq m$ which is a good approximation for large masses since the exponential dependence is now stored in $\cal C$. We have also identified the region of where there is backreaction on the perturbations (mild) and on the background (strong).

The blue band represents the region where the gauge field contribution to the power spectrum is of order $10^{-9}$. Above the band, for larger masses, the gauge field contribution would be smaller and the vacuum contribution should instead provide the necessary contribution to the power spectrum if the mechanism is active on CMB scales. 
Requiring strong backreaction and a correct CMB normalisation thus seems to require $\bar m$ of order a few hundred. 

\begin{figure}[t]
\centering
\begin{subfigure}[b]{0.49\textwidth}
\centering
\includegraphics[width=\textwidth]{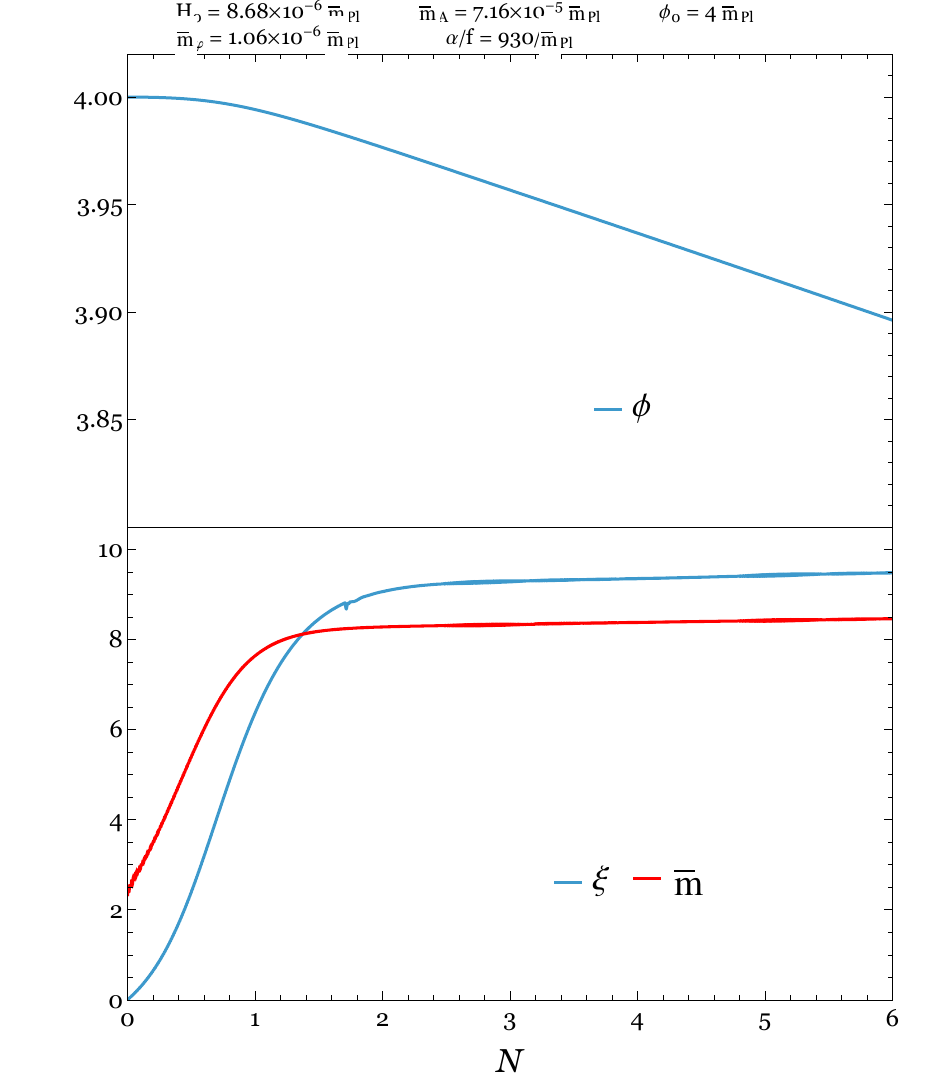}
\caption{Inflaton, $|\xi|$, and $\bar m$.}
\end{subfigure}\hfill
\begin{subfigure}[b]{0.49\textwidth}
\centering
\includegraphics[width=\textwidth]{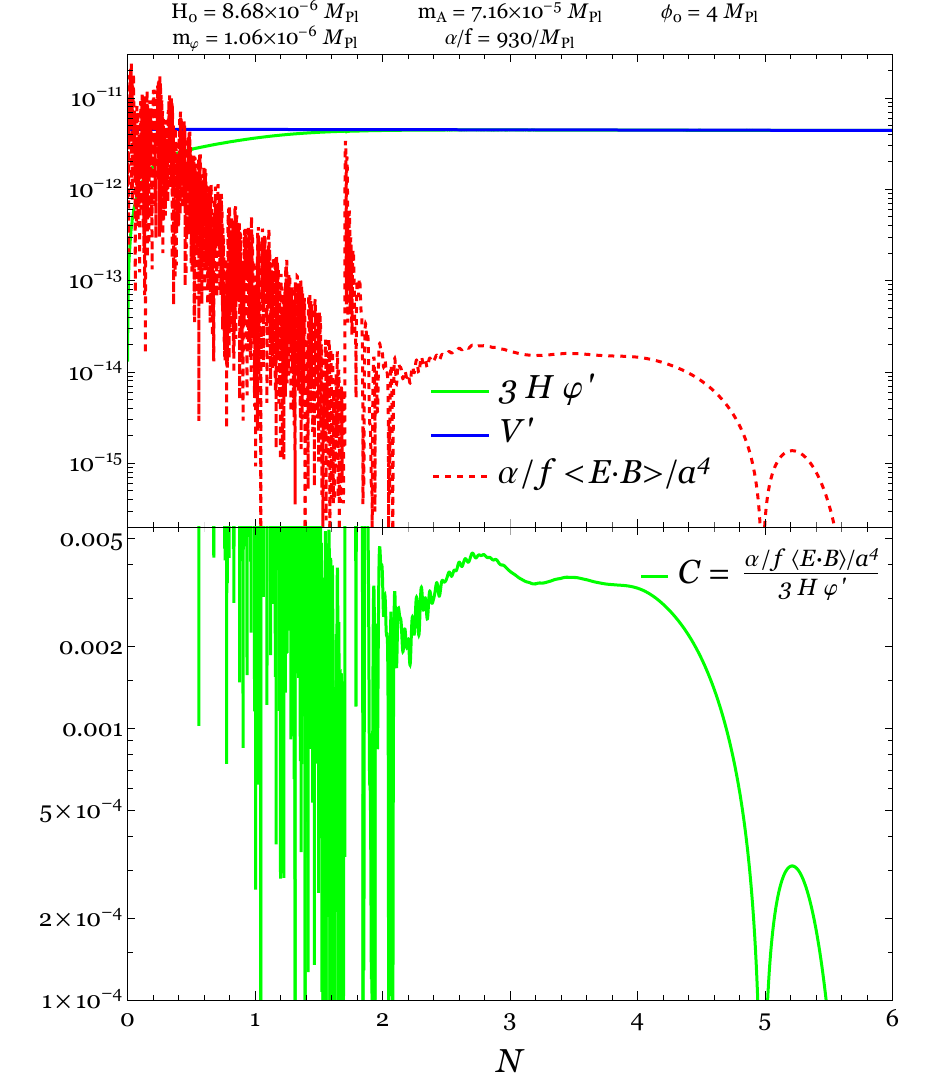}
\caption{Terms in the homogeneous equation and $\mathcal C$.}
\end{subfigure}
\caption{Representative run with weak backreaction, with $\bar m\simeq8.25$. The left panels show the homogeneous inflaton and the evolution of $|\xi|$ and $\bar m=m/H$. The right panels compare $3H\dot\phi$, $V_{,\phi}$, and $\alpha\langle\mathbf E\cdot\mathbf B\rangle/(fa^4)$ and show the ratio $\mathcal C$. After the initial transient, $\mathcal C$ remains well below unity and the background follows the standard slow-roll trajectory in the range $2.5\lesssim N\lesssim 4.5$. Afterwards the relevant window of modes is not covered anymore with our lattice.}
\label{fig:lattice-background-weak}
\end{figure}

\begin{figure}[t]
\centering
\begin{subfigure}[b]{0.92\textwidth}
\centering
\includegraphics[width=\textwidth]{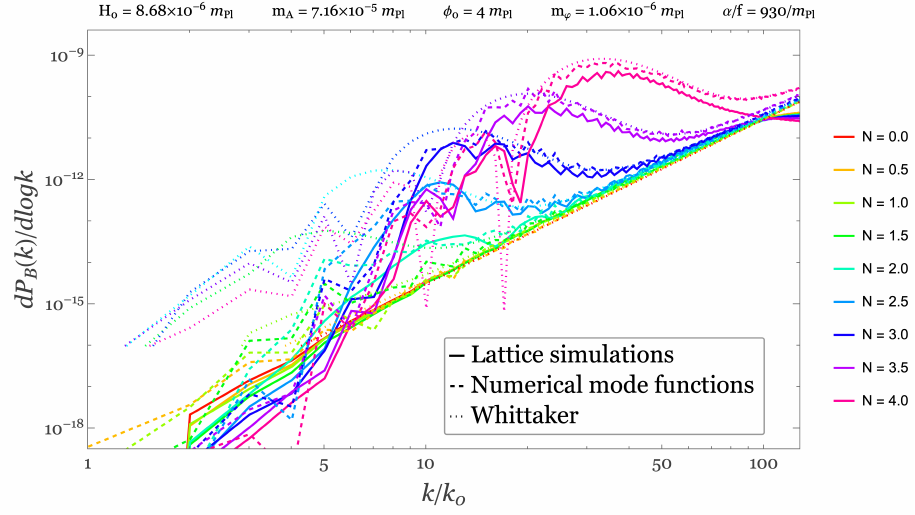}
\caption{Magnetic gauge-field spectrum.}
\end{subfigure}

\vspace{0.3cm}
\begin{subfigure}[b]{0.92\textwidth}
\centering
\includegraphics[width=\textwidth]{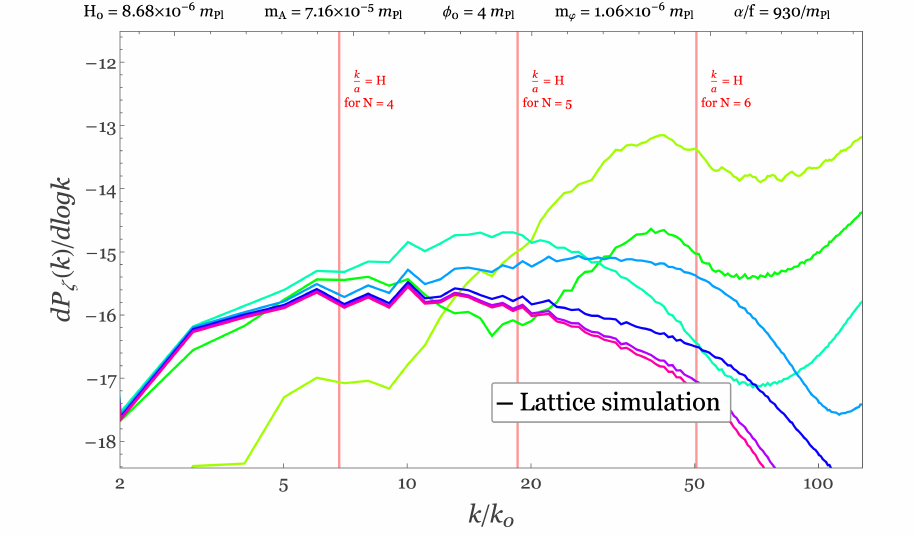}
\caption{Curvature spectrum.}
\end{subfigure}
\caption{Spectral evolution in the weak-backreaction run of Fig.~\ref{fig:lattice-background-weak}. In the upper panel, solid curves are the lattice spectra, dashed curves are obtained by evolving the linear mode equation with the instantaneous background, and dotted curves show the Whittaker approximation. The lower panel shows the lattice curvature spectra; the vertical markers indicate $k/a=H$ at the quoted e-folds. Modes that have crossed the horizon approach an approximately time-independent plateau.}
\label{fig:lattice-spectra-weak}
\end{figure}

\subsection{Effective Field Theory Constraints}
As discussed, the gauge-field mass suppresses particle production unless it is compensated by a larger instability parameter $|\xi|$, which could be achieved, \emph{e.g.} via a stronger axial coupling to the rolling inflaton (large $\alpha/f$). However, this must remain compatible with the cutoff of the effective theory underlying the axial coupling. The precise cutoff is UV dependent but adopting the conventional dimensional-analysis estimate $\Lambda_{\rm EFT}\simeq 4\pi f$, we require the gauge field mass to satisfy $m<4\pi f$.

Near the narrow-instability regime $\kappa\simeq1/2$ in Eq.~\eqref{eq:C-heavy} and
\begin{equation}
\mathcal C\simeq\frac{4\alpha^2m^2}{35\pi^2f^2}e^{2\pi(|\xi|-\bar m)}~,
\qquad
\frac{m}{4\pi f}=\frac{\sqrt{35}}{8\alpha}\sqrt{\mathcal C}\,e^{-\pi\Delta}~.
\label{eq:EFT-heavy-relation}
\end{equation}
The condition $m<4\pi f$ therefore requires
\begin{equation}
\alpha>\frac{\sqrt{35}}{8}\sqrt{\mathcal C}\,e^{-\pi(|\xi|-\bar m)}~.
\label{eq:EFT-alpha-bound}
\end{equation}
For example, $\mathcal C=1$ and $|\xi|-\bar m=0.5$ give $\alpha\gtrsim 0.15$. This is compatible with a perturbative dimensionless coupling, but it removes part of the otherwise viable parameter space.

\begin{figure}[t]
\centering
\begin{subfigure}[b]{0.49\textwidth}
\centering
\includegraphics[width=\textwidth]{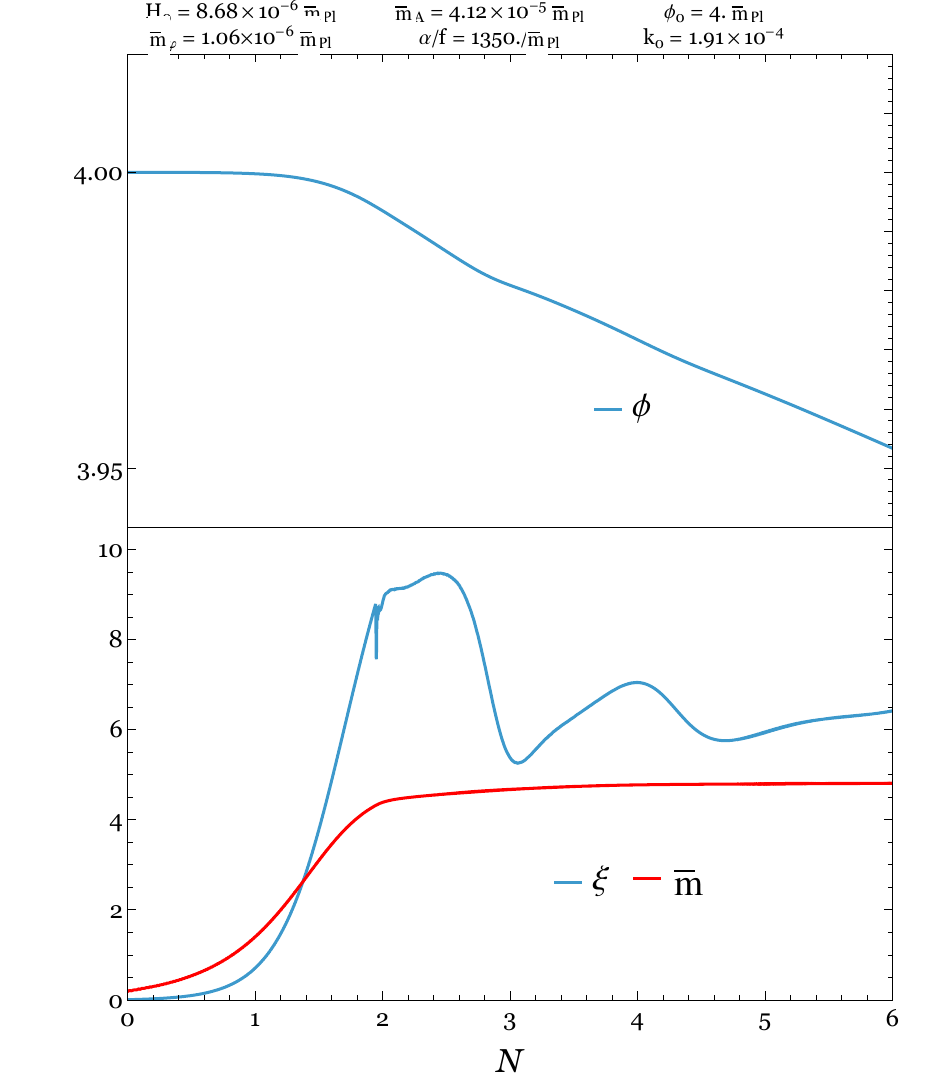}
\caption{$\bar m\simeq4.75$: background variables.}
\end{subfigure}\hfill
\begin{subfigure}[b]{0.49\textwidth}
\centering
\includegraphics[width=\textwidth]{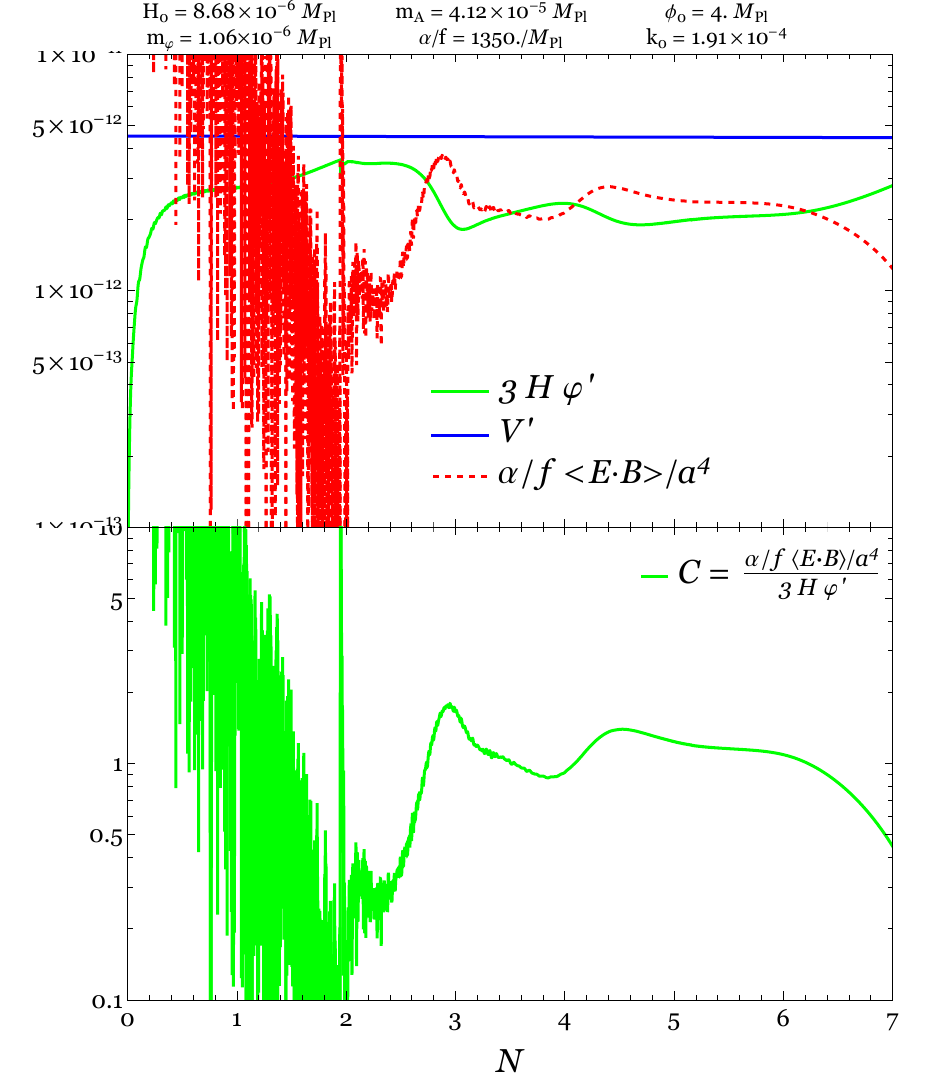}
\caption{$\bar m\simeq4.75$: force balance.}
\end{subfigure}

\vspace{0.3cm}
\begin{subfigure}[b]{0.45\textwidth}
\centering
\includegraphics[width=\textwidth]{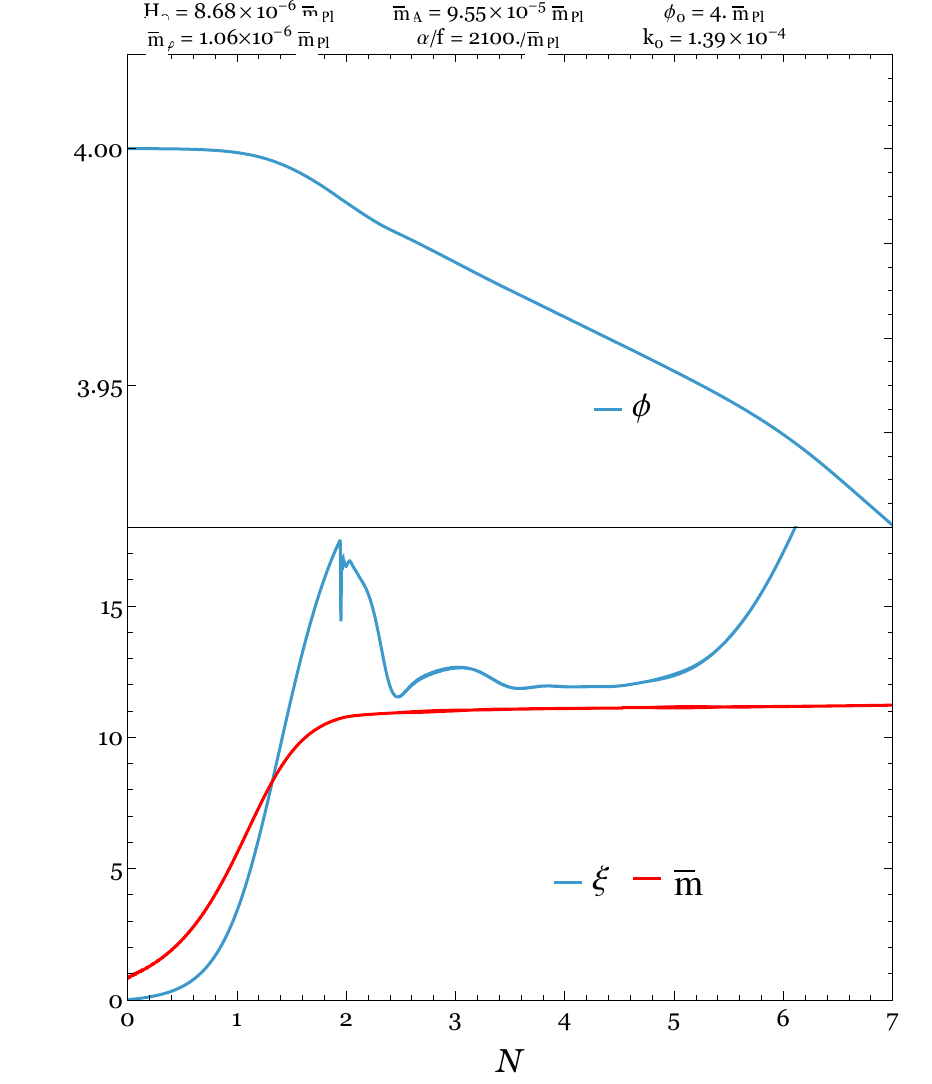}
\caption{$\bar m\simeq11$: background variables.}
\end{subfigure}\hfill
\begin{subfigure}[b]{0.45\textwidth}
\centering
\includegraphics[width=\textwidth]{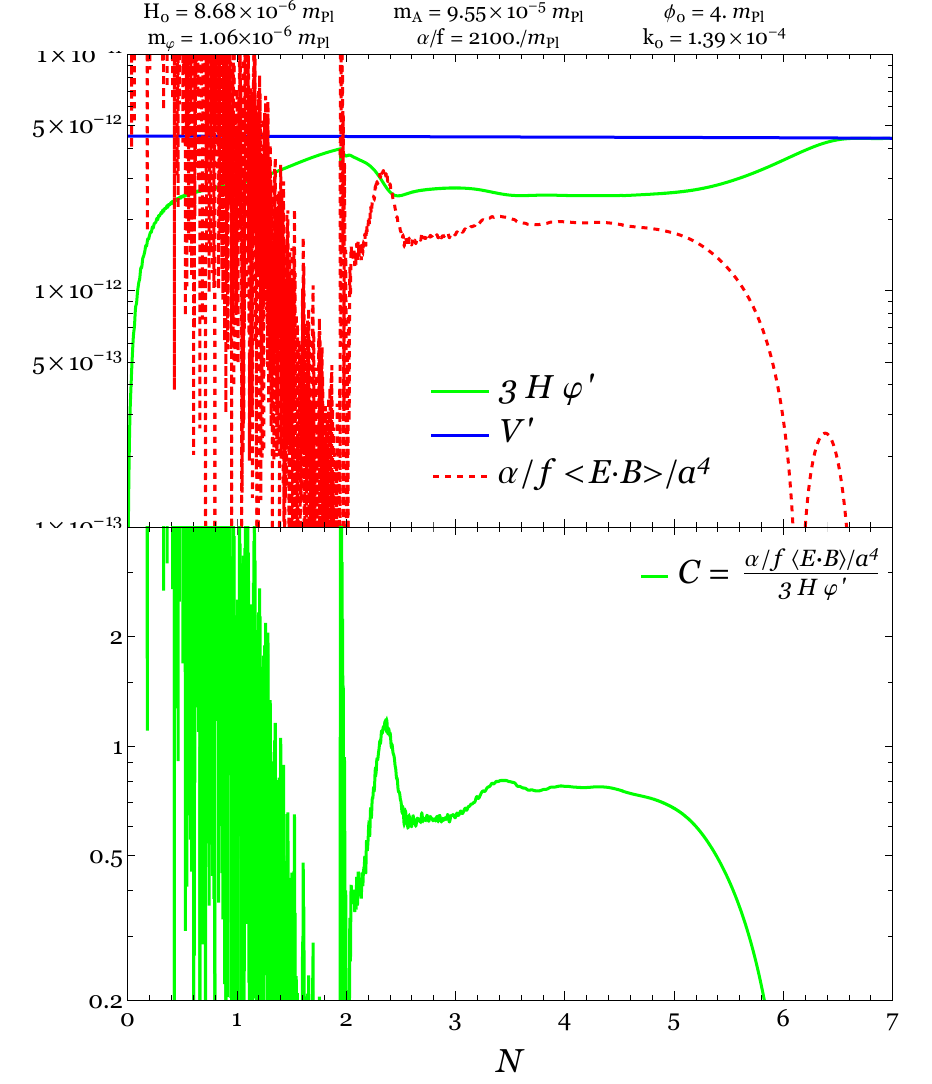}
\caption{$\bar m\simeq11$: force balance.}
\end{subfigure}
\caption{Two runs in which gauge field production substantially affects the homogeneous dynamics. For $\bar m\simeq 4.75$ (upper row), $\mathcal C$ reaches the threshold for strong backreaction $\mathcal C\sim 0.4$ and $|\xi|$ responds non-monotonically as the gauge-field source participates in the force balance. The lower row shows a heavier run, $\bar m\simeq11$, in which the backreaction is significant but the sourced curvature perturbation spectrum is more strongly suppressed. The high-frequency structure before the interaction switch-on is an initial-state transient and is excluded from the physical comparison.}
\label{fig:lattice-background-strong}
\end{figure}

\begin{figure}[t]
\centering

\begin{subfigure}[b]{0.49\textwidth}
\centering
\includegraphics[width=\textwidth]{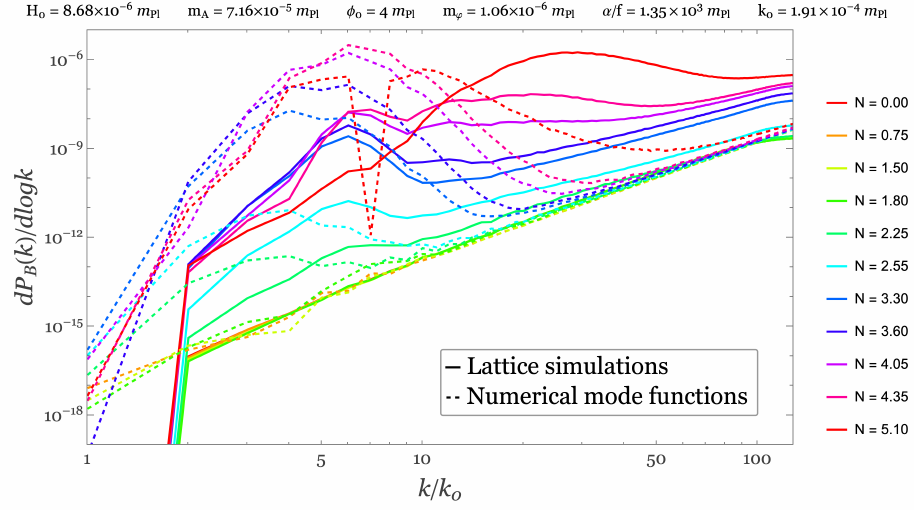}
\caption{Magnetic spectrum, $\bar m\simeq4.75$.}
\end{subfigure}\hfill
\begin{subfigure}[b]{0.49\textwidth}
\centering
\includegraphics[width=\textwidth]{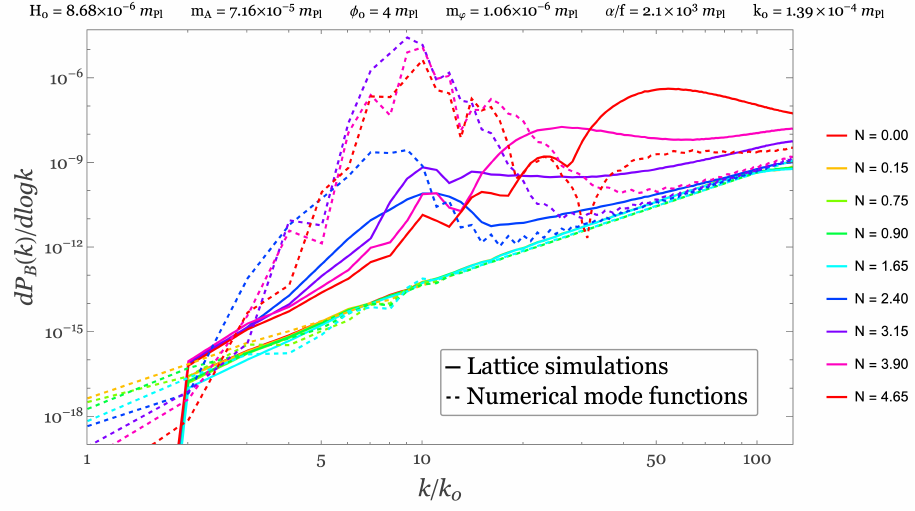}
\caption{Magnetic spectrum, $\bar m\simeq11$.}
\end{subfigure}

\vspace{0.3cm}

\begin{subfigure}[b]{0.49\textwidth}
\centering
\includegraphics[width=\textwidth]{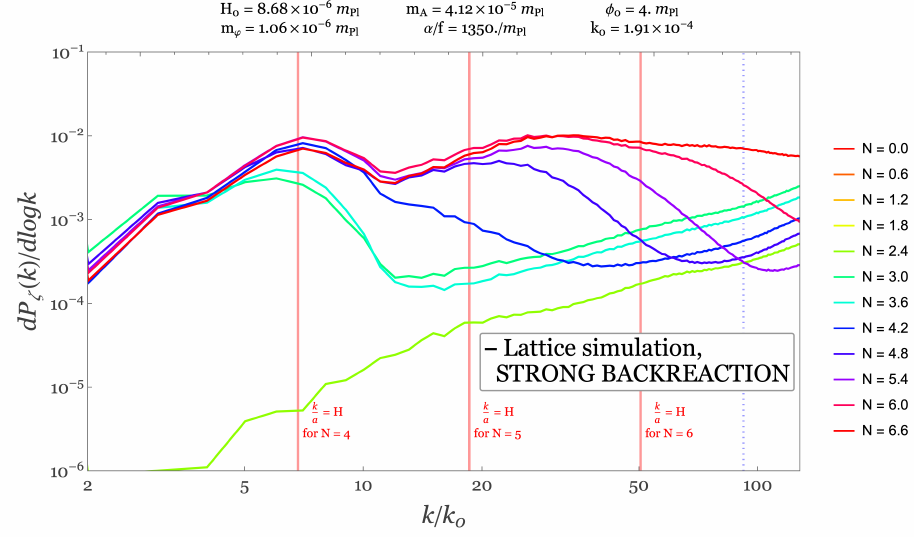}
\caption{Curvature perturbation spectrum, $\bar m\simeq4.75$.}
\end{subfigure}\hfill
\begin{subfigure}[b]{0.49\textwidth}
\centering
\includegraphics[width=\textwidth]{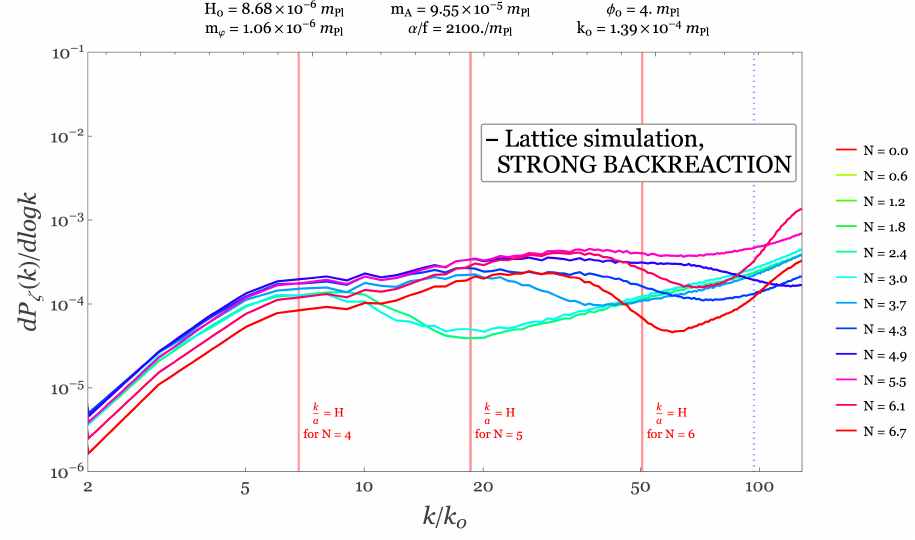}
\caption{Curvature perturbation spectrum, $\bar m\simeq11$.}
\end{subfigure}

\caption{
Magnetic and curvature perturbation spectra in the two backreacting runs of Fig.~\ref{fig:lattice-background-strong}.
The upper panels show the evolution of the magnetic spectrum $\textrm{d} \mathcal{P}_B/\textrm{d}\log k$, comparing the lattice results with the corresponding numerical mode-function evolution.
The lower panels show the curvature perturbation spectra extracted from the full lattice simulations.
In both runs the curvature perturbation spectra develop broad, approximately frozen plateaus behind the horizon-crossing markers. Given the fact that the background has oscillations, it is not surprising that oscillations are visible also in ${\cal P}_\zeta$. 
Increasing the vector mass lowers the plateau amplitude at comparable background relevance, in qualitative agreement with the mass suppression in Eq.~\eqref{eq: power spectrum mild backreaction regime}.
Because these runs are nonlinear, the comparison should be regarded primarily as a test of the mass scaling and of the existence of the strong-backreaction regime, rather than as a precision test of the perturbative normalisation.
}
\label{fig:lattice-spectra-strong}
\end{figure}

\section{\label{sec:pencil-code-lattice-simulations}Results from Pencil Code Simulations}
We simulate the nonlinear system with the \texttt{Pencil Code} \cite{PencilCode:2020eyn}, following the axion--$U(1)$ implementation of Ref.~\cite{Sharma:2024nfu}. The mass term is added to the gauge-field evolution and to the energy/current terms (in the modules \texttt{backreaction.f90} and \texttt{disp\_current.f90}). Apart from these mass-dependent terms, the discretisation and initialisation follow the reference implementation. We use lattices of size $256^3$. For the runs displayed below, the simulation outputs record $m_\phi=1.06\times10^{-6}\,\m$ and we use $N\equiv\ln(a/a_0)$ for the number of e-folds elapsed since initialization. We initialise the gauge field with vacuum fluctuations and the scalar field as a homogeneous background. Since the vacuum is unrenormalised, there is an initial UV-dominated energy density contribution, which however is diluted as $a^{-4}$. To prevent such a contribution from dominating the  Friedmann constraint, the axial interaction is switched on only after this transient has redshifted enough, at approximately $N=1.7$ in the representative runs shown below. This delayed switch-on is a numerical prescription for the initial state and is kept fixed when comparing runs. In all runs with our lattice the relevant window of modes is well covered in the range $2.5\lesssim N\lesssim 4.5$.

The simulations serve three purposes: to test the Whittaker/Bessel mode functions before strong backreaction, to verify the location and motion of the gauge-field spectral peak, and to measure the spectrum of curvature perturbations when the analytical estimate ceases to be controlled. The parameter choices used in the comparison are collected in Table~\ref{tab:parameter_points}.

The mode-function comparison in Fig.~\ref{fig:lattice-spectra-weak} shows that the peak is set by the instability band and that the mode functions and the Whittaker functions agree well with the full numerical results, at least in weak backreaction. The curvature perturbation spectra in Figs.~\ref{fig:lattice-spectra-weak} and \ref{fig:lattice-spectra-strong} also exhibit the expected freeze-out after horizon crossing. In the strongly backreacting runs the produced field changes the background on the same time scale as the instability, producing a characteristic oscillatory behavior already observed in the massless case~\cite{Cheng:2015oqa,Notari:2016npn}. 

\begin{figure}[t]
\centering
\includegraphics[width=0.78\linewidth]{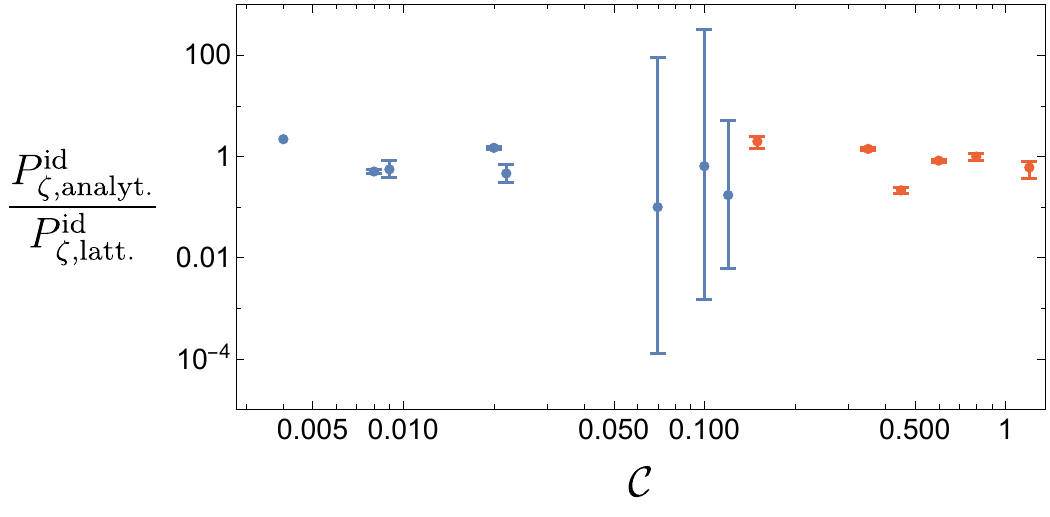}
\caption{Ratio of the analytical inverse-decay estimate to the lattice result as a function of $\mathcal C$. Values near unity indicate agreement in normalisation. We have used the weak backreaction expression in Eq.~\eqref{eq:loop-contribution-weak-backreaction-correction} for smaller $\cal C$ (blue points) and the mild/strong backreaction expression in Eq.~\eqref{eq: power spectrum mild backreaction regime} for larger $\cal C$ (red points). The data points correspond to the midpoints of the $|\xi|-\mu$ intervals reported in Table \ref{tab:parameter_points}. The uncertainty bands span the results obtained at the endpoints of each interval.}
\label{fig:analytical-over-lattice}
\end{figure}

Figure~\ref{fig:analytical-over-lattice} shows the ratio of the analytical
inverse-decay estimates to the lattice results as a function of the
backreaction parameter $\mathcal{C}$, for both the weak- and
strong-backreaction regimes, described by
Eqs.~\eqref{eq:loop-contribution-weak-backreaction-correction} and
\eqref{eq: Power spectrum, inverse decay, with friction}, respectively.
The analytical expression used for the blue points scales as
$\exp[4\pi(|\xi|-\mu)]$ and is therefore exponentially sensitive to
$|\xi|-\mu$. Despite this sensitivity, the estimates agree very well with
the lattice results at small $\mathcal{C}$, while those at intermediate
$\mathcal{C}$ carry substantially larger uncertainties. By contrast, the
orange points are obtained from
Eq.~\eqref{eq: Power spectrum, inverse decay, with friction}, where the
exponential factor is traded by $\mathcal{C}^2$, yielding a much milder
dependence on $\xi$ and $\mu$.

We find that both analytical estimations seem to yield a good agreement to the lattice simulations thus supporting the robust conclusion that inhomogeneities are suppressed at large masses.

\section{\label{sec:outlook-conclusions}Conclusions}
We have analysed axion inflation coupled to a massive Abelian gauge field. The vector mass narrows the tachyonic instability and shifts the amplified modes to subhorizon physical momenta. For $|\xi|>\bar m=m/H\gg1$, the unstable helicity is enhanced by $\exp[\pi(|\xi|-\bar m)]$. The local quantities that control the homogeneous dynamics acquire compensating powers of $\bar m$, so sizeable gauge friction remains possible even when $|\xi|-\bar m$ is modest.

The displacement of the produced quanta away from the Hubble scale changes the curvature perturbation phenomenology. In the heavy, weak-backreaction regime we find $\mathcal P^{\rm id}_\zeta\simeq0.17 \,\mathcal C^2/\bar m^2$ for $|\xi|\simeq\bar m$. Gauge field production affects scalar perturbations before it affects the homogeneous inflaton: the additional friction becomes important at $\mathcal C\sim(2\pi|\xi|)^{-1}$, whereas background backreaction begins at $\mathcal C\gtrsim 0.4$. Incorporating this perturbative friction gives the benchmark scaling $\mathcal P^{\rm id}_\zeta\simeq0.08/\bar m^3$,  valid both in the mild and strong regimes. This implies that a strong backreaction regime with a sourced curvature perturbation spectrum below $10^{-9}$ should be possible, for gauge field masses of order a few hundred Hubble scales. 

The lattice simulations support the principal analytical ingredients. Before strong backreaction, the Whittaker solution and the modified Bessel approximation reproduce the evolution and amplitude of the gauge-field spectral peak. The curvature perturbation spectrum freezes after horizon crossing, and its plateau is lower in the  runs with larger gauge field mass, at comparable backreaction. In the strong backreacting regime an oscillatory behavior of the background is seen, similarly to the massless case~\cite{Cheng:2015oqa,Notari:2016npn}. In such a regime the analytical normalisation shows appreciable scatter and thus the simulations should  be regarded as evidence for the mass scaling and for the existence of the strong-backreaction regime, rather than as a precision calibration of a universal coefficient. Therefore, we conclude that, although the precise power-law dependences on $\bar m$ and $\xi$ are subject to order-one uncertainties, the suppression of vector-induced inhomogeneities at larger vector masses, for a fixed level of backreaction, is a robust finding of our analysis.

Several questions remain. Non-Gaussianity must be computed in the heavy gauge field regime to check if a realistic friction-dominated solution can be declared observationally viable, \emph{i.e.} with realistic CMB-scale perturbations. Note that in this respect Ref.~\cite{Caravano:2022epk} has shown that in the massless case non-gaussianity is suppressed  in the strong backreacting regime, and thus a similar suppression may be expected also in the massive case. The tensor spectrum is also expected to differ qualitatively from the massless case because its source is concentrated at subhorizon momenta. On the numerical side, larger-volume and higher-resolution simulations, explicit convergence tests, and a denser scan in $(\bar m,\mathcal C)$ are needed to improve the determination of the strong-regime normalisation of the curvature perturbation spectrum. 

Our lattice simulations have shown the presence of oscillations in the background in the strong backreaction regime, similarly to the results obtained in the massless case~\cite{Cheng:2015oqa,Notari:2016npn}. Observationally, it is relevant to check if such oscillations are imprinted on ${\cal P}_\zeta$ when $m/H$ is very large. That would however require much larger lattice simulations. 

Finally, whether the EFT bound on the coupling $\alpha$ can be satisfied depends on the ultraviolet origin of the vector mass and of the axial operator; concrete Higgs, St\"uckelberg, and medium-induced realisations should be analysed separately. These extensions will sharpen the observational and ultraviolet consistency tests, giving support to the central result established here: massive-gauge field production can replace the usual Hubble friction as the dominant breaking mechanism of the inflaton and can give rise to the observed amplitude of curvature perturbations. Massive gauge-field friction might therefore provide a concrete and dynamically viable alternative to the standard slow-roll paradigm, with distinctive scalar and tensor signatures. 

\vspace{5mm}
\noindent
\textbf{Note added}: While this manuscript was being finalised for submission, Ref.~\cite{Baker:2026xsd} appeared, which also investigates axion inflation coupled to a massive Abelian vector field and in particular analyses whether memory effects are still present for very large masses. As the two studies adopt different approaches, a direct comparison requires some care. Our work, by contrast, enters the strong backreaction regime with dedicated lattice simulations, demonstrating that the massive gauge field can maintain strong friction while reducing induced inhomogeneities at horizon crossing, providing key evidence that this strongly backreacting setup may constitute a phenomenologically viable model of inflation on CMB scales. Nevertheless, Ref.~\cite{Baker:2026xsd} appears to suggest that a gauge field with a mass of ${\cal O}(100)\,H$
may sustain a strong backreaction while remaining consistent with the CMB normalisation
of the scalar power spectrum, similarly to our findings.

\vspace{5mm}
\noindent
\textbf{Acknowledgements}

\vspace{4mm}

\noindent
The authors thank Ramkishor Sharma for assistance with the \texttt{Pencil Code} and acknowledge the use of NyX cluster at ICCUB University of Barcelona. R.~Z.~F. and J.~J.~T.~D. acknowledge the financial support provided by FCT-Fundação para a Ciência e Tecnologia (FCT), I.P., through the Strategic Funding UID/04650/2025 and UID/04564/2025 and national funds with DOI identifiers 10.54499/2023.11681.PEX, 10.54499/2024.00249.CERN funded by measure RE-C06-i06.m02-``Reinforcement of funding for International Partnerships in Science, Technology and Innovation'' of the Recovery and Resilience Plan–RRP, within the framework of the financing contract signed between the Recover Portugal Mission Structure (EMRP) and the Foundation for Science and Technology I.P. (FCT), as an intermediate beneficiary, as well as the advanced computing projects 2024.00249.CERN.F1. The authors also acknowledge the use of ChatGPT-5.6 Sol and Claude Opus 4.8 to improve the text, cross-check certain expressions, and provide coding assistance. 

\appendix

\section{\label{sec:cov-fe}Covariant Field Equations and Constraints}
The covariant equations following from Eq.~\eqref{eq:action} are
\begin{align}
\Box\phi-V_{,\phi}&=\frac{\alpha}{4f}F_{\mu\nu}\tilde F^{\mu\nu}~,\label{eq:scalar-covariant}\\
\nabla_\nu F^{\mu\nu}+m^2A^\mu&=-\frac{\alpha}{f}\tilde F^{\mu\nu}\nabla_\nu\phi~,\label{eq:vec-field-cov}\\
\m^2G_{\mu\nu}&=T^{(\phi)}_{\mu\nu}+T^{(A)}_{\mu\nu}~,
\end{align}
where
\begin{align}
T^{(\phi)}_{\mu\nu}&\equiv\nabla_\mu\phi\nabla_\nu\phi
-g_{\mu\nu}\left[\frac12(\nabla\phi)^2+V(\phi)\right],\\
T^{(A)}_{\mu\nu}&\equiv F_{\mu\rho}F_\nu{}^\rho-\frac14g_{\mu\nu}F_{\rho\sigma}F^{\rho\sigma}
+m^2\left(A_\mu A_\nu-\frac12g_{\mu\nu}A_\rho A^\rho\right).
\label{eq:stress-vector-standard}
\end{align}
The axial term does not contribute to the stress tensor because $\sqrt{-g}\,F_{\mu\nu}\tilde F^{\mu\nu}$ is metric independent. We use $\Box\equiv g^{\mu\nu}\nabla_\mu\nabla_\nu$ and the Levi--Civita connection. Equations~\eqref{eq:scalar-covariant} and \eqref{eq:vec-field-cov} reduce to Eqs.~\eqref{eq:eq-sf-full}--\eqref{eq:crossB} in the FLRW coordinates of Eq.~\eqref{eq:flrw-metric}.

In terms of the comoving electric and magnetic fields, the temporal and spatial vector equations are
\begin{align}
(\nabla^2-a^2m^2)A_0-\nabla\cdot\mathbf{A}'
&=-\frac{\alpha}{f}\nabla\phi\cdot(\nabla\times\mathbf A)~,
\label{eq:constraint-eq-A0}\\
\mathbf A''-(\nabla^2-a^2m^2)\mathbf A
+\nabla(\nabla\cdot\mathbf A-A_0')
&=\frac{\alpha}{f}\left[\phi'(\nabla\times\mathbf A)
-\nabla\phi\times(\mathbf A'-\nabla A_0)\right].
\end{align}
The Bianchi identity is simply
\begin{equation}
\nabla_\mu\tilde F^{\mu\nu}=0~.
\label{eq:Bianchi-dual}
\end{equation}
Taking the divergence of Eq.~\eqref{eq:vec-field-cov}, using Eq.~\eqref{eq:Bianchi-dual} and the symmetry of $\nabla_\mu\nabla_\nu\phi$, gives the Proca constraint
\begin{equation}
m^2\nabla_\mu A^\mu=0~.
\label{eq:constraint-eq-fix}
\end{equation}
The identity
\begin{equation}
\nabla_\mu\nabla_\nu F^{\mu\nu}=0
\label{eq:double-divergence-F-vanishes}
\end{equation}
follows from antisymmetry of $F^{\mu\nu}$ and symmetry of the Ricci tensor. For $m\ne0$, Eq.~\eqref{eq:constraint-eq-fix} becomes $A_0^{'}+2\mathcal H A_0=\nabla\cdot\mathbf A$, which is Eq.~\eqref{eq:constraint-incoord}. Differentiating the constraint and using Eq.~\eqref{eq:constraint-eq-A0} yields
\begin{equation}
A_0''+2\mathcal H A_0'+(2\mathcal H'+a^2m^2-\nabla^2)A_0
=\frac{\alpha}{f}\nabla\phi\cdot(\nabla\times\mathbf A)~.
\end{equation}
This second-order equation does not make $A_0$ an independent propagating mode; it is a consequence of the constraint. With the conventions of Sec.~\ref{sec2-massive-eqs},
\begin{align}
F_{\mu\nu}F^{\mu\nu}
&=-\frac{2}{a^4}\left(|\mathbf E|^2-|\mathbf B|^2\right),\\
F_{\mu\nu}\tilde F^{\mu\nu}
&=-\frac{4}{a^4}\,\mathbf E\cdot\mathbf B~.
\label{eq:FtildeF-EB-sign}
\end{align}
The minus sign in the second line follows from $\epsilon^{0123}=+1/\sqrt{-g}$ and is the sign required for consistency with Eq.~\eqref{eq:eq-sf-full}.

\section{Mode Equations with Scalar-Field Gradients \label{App:Decomposition into Transverse and Longitudinal Modes}}
For an inhomogeneous scalar field, products in the coordinate-space equations become momentum-space convolutions. We use the Fourier convention of Eq.~\eqref{eq:Fourier-expansion-Amu} and write
\begin{equation}
\label{eq:Fourier-mode-scalar-field-decomp}
\phi(\eta,\mathbf x)=\int\frac{\textrm d^3\mathbf k}{(2\pi)^{3/2}}
\varphi(\eta,\mathbf k)e^{i\mathbf k\cdot\mathbf x}~.
\end{equation}
Fourier transforming Gauss's law, Eq.~\eqref{eq:nablaE}, gives the exact constraint
\begin{equation}
\label{eq:constraint-A0-inhomogeneous}
\mathcal A_0(\eta,\mathbf k)
=-\frac{i\mathbf k\cdot\mathbfcal A'(\eta,\mathbf k)}{k^2+a^2m^2}
+\frac{\alpha}{f(k^2+a^2m^2)}
\int\frac{\textrm d^3\mathbf q}{(2\pi)^{3/2}} \varphi(\eta,\mathbf q)
(\mathbf k\cross\mathbf q)\cdot\mathbfcal A_T(\eta,\mathbf p)~.
\end{equation}
The convolution in the second term contains only the component transverse to $ \mathbf p\equiv\mathbf k-\mathbf q$, because $(\mathbf k\cross\mathbf q)\cdot\mathbf p=0$.

It is useful to retain $\mathcal A_0$ until after projecting the spatial equation. Defining
\begin{align}
\mathbf S(\eta,\mathbf k)
\equiv\frac{\alpha}{f}\int\frac{\textrm d^3\mathbf q}{(2\pi)^{3/2}}\Big[
i\varphi'(\eta,\mathbf q)\,\mathbf p\cross\mathbfcal A(\eta,\mathbf p)
-i\varphi(\eta,\mathbf q)\,\mathbf q\cross\mathbfcal A'(\eta,\mathbf p)+\varphi(\eta,\mathbf q)(\mathbf k\cross\mathbf q)\mathcal A_0(\eta,\mathbf p)
\Big],
\label{eq:inhomogeneous-vector-source}
\end{align}
the longitudinal and transverse projections of Eq.~\eqref{eq:crossB} are
\begin{align}
\mathbfcal A_L''(\eta,\mathbf k)+a^2m^2\mathbfcal A_L(\eta,\mathbf k)
-i\mathbf k\,\mathcal A_0'(\eta,\mathbf k)
&=P_L(\mathbf k)\mathbf S(\eta,\mathbf k)~,
\label{eq:longitudinal-inhomogeneoussf}\\
\mathbfcal A_T''(\eta,\mathbf k)+(k^2+a^2m^2)\mathbfcal A_T(\eta,\mathbf k)
&=P_T(\mathbf k)\mathbf S(\eta,\mathbf k)~,
\label{eq:transverse-inhomogeneoussf}
\end{align}
where we have defined $P_T^{ij}(\mathbf k)\equiv\delta^{ij}-P_L^{ij}(\mathbf k)$, and $P_L^{ij}(\mathbf{k})\equiv k^{i} k^{j}/k^2$.
Equations~\eqref{eq:constraint-A0-inhomogeneous}--\eqref{eq:transverse-inhomogeneoussf} form a closed system. Substituting the constraint into $\mathbf S$ makes the transverse--longitudinal mixing explicit and generates terms through order $(\alpha/f)^2$.

As a check, for a homogeneous scalar,
$\varphi(\eta,\mathbf q)=(2\pi)^{3/2}\bar\phi(\eta)\delta^{(3)}(\mathbf q)$, the convolution term in Eq.~\eqref{eq:constraint-A0-inhomogeneous} vanishes and
\begin{equation}
\mathbf S(\eta,\mathbf k)
=i\frac{\alpha}{f}\bar\phi'(\eta)\,\mathbf k\cross\mathbfcal A(\eta,\mathbf k)~.
\end{equation}
Using $\mathbf k\cross\bm\epsilon_\pm=\mp ik\bm\epsilon_\pm$ then reproduces Eqs.~\eqref{eq:oscillator-freq} and \eqref{eq:oscillator-freq-wplusminus}, while the longitudinal projection together with Eq.~\eqref{eq:constraint-A0} reproduces Eq.~\eqref{eq:longitudinal-mode-homogeneous}. 

\section{Standard Quantisation for Quantum Fields\label{App:Fields quantization}}
To proceed further in Secs.~\ref{sec4-AnalyticalSolutionsforWeakBackreaction} and \ref{sec5-scalarpowerspectrum-fullsec}, and in particular to derive the electric, magnetic, and scalar power spectra, we promote the Fourier-expanded fields $A_0(\eta,\mathbf{x})$, $\mathbf{A}(\eta,\mathbf{x})$, and $\phi(\eta,\mathbf{x})$ to quantum operators as
\begin{align}
    \label{eq:Fourier-quantum-commutators-A0}&\hat{A}_0(\eta,\mathbf{x}) = -i \int \frac{\textrm{d}^3\mathbf{k}}{(2\pi)^{3/2}}\frac{ke^{i\mathbf{k}\cdot \mathbf{x}}}{k^2+a^2m^2}\left[\mathcal{A}_{L}^{'}(\eta,k)\hat{a}_{\mathbf{k},L}+\mathcal{A}_{L}^{'*}(\eta,k)\hat{a}_{-\mathbf{k},L}^{\dagger}\right],\\
    \label{eq:Avec-field-operator}&\hat{\mathbf{A}}(\eta,\mathbf{x}) = \sum_{\lambda = L,\pm} \int \frac{\textrm{d}^3\mathbf{k}}{(2\pi)^{3/2}}\bm{\epsilon}_{\lambda}(\mathbf{k}) e^{i\mathbf{k}\cdot \mathbf{x}}\left[\mathcal{A}_{\lambda}(\eta,k)\hat{a}_{\mathbf{k},\lambda}+\mathcal{A}^{*}_{\lambda}(\eta,k) \hat{a}^{\dag}_{-\mathbf{k},\lambda}\right],\\
    \label{eq:scalar-field-operator-modes-expansion}&\hat{\phi}(\eta,\mathbf{x}) = \int \frac{\textrm{d}^3 \mathbf{k}}{(2\pi)^{3/2}}e^{i\mathbf{k}\cdot \mathbf{x}}\left[\varphi(\eta,k)\hat{b}_{\mathbf{k}}+\varphi^{*}(\eta,k)\hat{b}_{-\mathbf{k}}^{\dagger}\right],
\end{align}
where we used Eq.~\eqref{eq:constraint-A0} in order to get Eq.~\eqref{eq:Fourier-quantum-commutators-A0}, and the fact that $\mathbf{k}\cdot \bm{\epsilon}_L(\mathbf{k}) = k$. The operators 
 $\hat{a}$ and $\hat{a}^{\dagger}$, and $\hat{b}$ and $\hat{b}^{\dagger}$, are the two sets of annihilation and creation operators satisfying $\hat{a}_{\mathbf{k},\lambda} \ket{0}=0$ and $\bra{0}\hat{a}^{\dag}_{\mathbf{k},\lambda} = 0$, and $\hat{b}_{\mathbf{k}}\ket{0}=0$ and $\bra{0}\hat{b}^{\dagger}_{\mathbf{k}}=0$, when acting on the vacuum state, as well as the commutation relations ($\lambda=L,\pm$):
\begin{eqnarray}
    \label{eq:comm-rel-ann-cr}\left[\hat{a}_{\mathbf{k},\lambda},\hat{a}^{\dag}_{\mathbf{k}',\lambda'}\right] &=& \hat{I}\delta_{\lambda,\lambda'}\delta^{(3)}(\mathbf{k}-\mathbf{k}')~,\\
    \label{eq:comm-rel-b-operators}\left[\hat{b}_{\mathbf{k}},\hat{b}^{\dag}_{\mathbf{k}'}\right] &=& \hat{I}\delta^{(3)}(\mathbf{k}-\mathbf{k}')~,
\end{eqnarray}
with $\hat{I}$ the identity operator. The rest of the commutators of these operators vanish. Furthermore, `$\dagger$' is used to denote the adjoint operator, and `$*$' the complex conjugate of the mode function. In the decomposition of Eq.~\eqref{eq:Avec-field-operator}, the sum runs over the longitudinal mode $L$ and the two transverse modes $\pm$, defined in the basis of circular polarisation vectors obeying $\mathbf{k}\cross \bm{\epsilon}_{\pm}(\mathbf{k}) = \mp ik\bm{\epsilon}_{\pm}(\mathbf{k})$. In view of the Fourier decompositions in Eqs.~\eqref{eq:electric-field-decom} and \eqref{eq:magnetic-field-decom}, together with the operator expansions in Eqs.~\eqref{eq:Fourier-quantum-commutators-A0} and \eqref{eq:Avec-field-operator}, the electric and magnetic field operators can be written as
\begin{align}
    \nonumber&\hat{\mathbf{E}}(\eta,\mathbf{x}) = \int \frac{\textrm{d}^3\mathbf{k}}{(2\pi)^{3/2}}e^{i\mathbf{k}\cdot \mathbf{x}}\left\{\frac{k\mathbf{k}}{k^2+a^2m^2}\left[\mathcal{A}_L^{'}(\eta,k)\hat{a}_{\mathbf{k},L}+\mathcal{A}^{'*}_L(\eta,k)\hat{a}^{\dagger}_{-\mathbf{k},L}\right]\right.\\
    \label{eq:electric-operator}&\phantom{---------------}\left.-\sum_{\lambda=L,\pm}\bm{\epsilon}_{\lambda}(\mathbf{k})\left[\mathcal{A}_{\lambda}^{'}(\eta,k)\hat{a}_{\mathbf{k},\lambda}+\mathcal{A}^{'*}_{\lambda}(\eta,k)\hat{a}^{\dagger}_{-\mathbf{k},\lambda}\right]\right\},\\
     \label{eq:magnetic-operator}&\hat{\mathbf{B}}(\eta,\mathbf{x}) =\sum_{\lambda = \pm} \lambda \int \frac{\textrm{d}^3\mathbf{k}}{(2\pi)^{3/2}}k\bm{\epsilon}_{\lambda}(\mathbf{k})e^{i\mathbf{k}\cdot \mathbf{x}}\left[\mathcal{A}_{\lambda}(\eta,k)\hat{a}_{\mathbf{k},\lambda}+\mathcal{A}^{*}_{\lambda}(\eta,k)\hat{a}^{\dagger}_{-\mathbf{k},\lambda}\right],
\end{align}
and hence the expectation value of the symmetrised product, $\langle \hat{\mathbf{E}}\cdot \hat{\mathbf{B}}\rangle$, which replaces the spatial average, is found to be (see footnote~\ref{footnote:the1/2})
\begin{equation}
    \label{eq:backreacting-term-EB-2}\langle \hat{\mathbf{E}}\cdot \hat{\mathbf{B}} \rangle = -\frac{1}{2}\sum_{\lambda = \pm} \lambda \int \frac{\textrm{d}^3\mathbf{k}}{(2\pi)^{3}}k\frac{\partial}{\partial \eta}\left[|\mathcal{A}_{\lambda}(\eta,k)|^2\right],
\end{equation}
where $|\mathcal{A}_{\lambda}|^2 = \mathcal{A}_{\lambda}\mathcal{A}^{*}_{\lambda}$ is the modulus squared of the mode functions. As for the corresponding power spectra in Eqs.~\eqref{eq:PE-power} and \eqref{eq:PB-power}, these are defined as
\begin{equation}
    \label{eq:pieces-of-energy-density-gauge-fields-rhoA}\langle \hat{\mathbf{E}}^2\rangle \equiv \int \textrm{d} \ln k~\mathcal{P}_{E}(k)~, \qquad \langle \hat{\mathbf{B}}^2\rangle \equiv \int \textrm{d} \ln k~\mathcal{P}_{B}(k)~, 
\end{equation}
and the comoving energy density of $A_{\mu}$ in Eq.~\eqref{eq:total-gauge-energy-density} is obtained from
\begin{equation}
    \rho_A \equiv \frac{1}{2}\left[\langle \hat{\mathbf{E}}^2\rangle + \langle \hat{\mathbf{B}}^2\rangle+a^2m^2 \left(\langle \hat{A}_0^2\rangle +\langle \hat{\mathbf{A}}^2\rangle \right)\right].
\end{equation}
It should be mentioned that these last three definitions are independent of the regime of backreaction considered. On the other hand, Eqs.~\eqref{eq:PE-power}, \eqref{eq:PB-power} and \eqref{eq:total-gauge-energy-density} are only valid when the scalar-field gradients are neglected, as we employed Eq.~\eqref{eq:constraint-A0} in order to arrive at those.

\section{\label{sec:app-asymp}Asymptotic Analysis}
In this appendix, we supplement the discussion in Sec.~\ref{sec4-AnalyticalSolutionsforWeakBackreaction} by providing further useful approximations. In particular, the subhorizon limit taken to obtain Eq.~\eqref{eq:solution-to-Whittaker-eq-Aplusminus}, corresponding to $k|\eta|\rightarrow \infty$, yields the following asymptotic form of the Whittaker function (see Sec.~13.19 of Ref.~\cite{NIST:DLMF}): 
\begin{equation}
    \label{eq:asymptotics-subhorizon-Whittaker}W_{\pm i \xi,\mu}(2ik\eta) \simeq e^{-ik\eta}(2ik\eta)^{\pm i\xi}\sum_{n=0}^{\infty}(1/2+\mu\mp i\xi)_{n}(1/2-\mu\mp i \xi)_n\frac{(-2ik\eta)^{-n}}{n!}~.
\end{equation}
Here, $(\alpha)_n\equiv \alpha(\alpha+1)...(\alpha+n-1)=\Gamma(\alpha+n)/\Gamma(\alpha)$ denotes the Pochhammer symbol (the final equality holds provided that $\alpha \neq 0,-1,-2,...$). The above approximation is valid provided that $|\textrm{ph}(z)|\leq 3\pi/2-\delta$, where $\textrm{ph}(z)$ indicates the phase of a complex number $z$ taken on the principal branch, and $\delta>0$ is an arbitrary small constant. This subhorizon expression is not independent of the gauge-field mass unless $2k|\eta| \gg |1/2+\mu\mp i \xi||1/2-\mu\mp i\xi|$, in which case it reduces to $W_{\pm i\xi,\mu} \simeq e^{-ik\eta}(2ik\eta)^{\pm i\xi}$, as pointed out in footnote~\ref{footnotesub}.

\subsection{\label{sec:asymp-Tricomi-various}Asymptotics of the Tricomi Function}
In Sec.~\ref{app: Matching procedure}, we derive an approximate solution for the massive gauge-field mode functions in the regime $|\xi| > \bar{m}\gg 1$. In this appendix section, we present an alternative derivation based directly on the asymptotic expansion of the Tricomi function.

In the massless case, $\mu=1/2$, the Whittaker function can be recast in terms of Coulomb wave functions. When the mass does not vanish, this approach is no longer available, unless one assumes an imaginary mass, for which $\bar{m}^2$ takes discrete values, $\bar{m}^2 = -l(l+1)$, with $l=0,1,2,...$. One may then adopt an alternative procedure, relating the Whittaker function to the Tricomi confluent hypergeometric function $U(\alpha,\beta,z)$. Following Sec.~13.14 of Ref.~\cite{NIST:DLMF}, the relation reads 
\begin{equation}
    W_{\pm i \xi,\mu}(z) = e^{-z/2}z^{1/2+\mu}U(1/2+\mu\mp i \xi,1+2\mu,z)~.
\end{equation}
Using one of Kummer's transformations in Sec.~13.2 of the aforementioned reference, this expression can be rewritten as
\begin{equation}
    W_{\pm i \xi,\mu}(z) = e^{-z/2}z^{1/2-\mu}U(1/2-\mu\mp i \xi,1-2\mu,z)~,
\end{equation}
which, in this case, amounts to reversing the sign in front of $\mu$. We thus arrive at (see Eq.~\eqref{eq:solution-to-Whittaker-eq-Aplusminus})
\begin{equation}
    \label{eq:Aplusminus-U}\mathcal{A}_{\pm}=\frac{e^{-(z\pm \xi \pi)/2}}{\sqrt{2k}}z^{1/2-\mu}U(\alpha_{\pm},\beta,z)~,
\end{equation}
with $\alpha_{\pm}\equiv 1/2-\mu\mp i\xi$, and $\beta \equiv 1-2\mu$, Also, $z\equiv 2ik\eta$.

In the limit where $\alpha$ in $U(\alpha,\beta,z)$ is large compared with $\beta$ and $z$ (in our case $\alpha$ is the only argument of the Tricomi function that depends on $\xi$), the gauge mode functions can be approximated by     
\begin{equation}
   \label{eq:big-approx-Aplus}\mathcal{A}_{\pm}(\eta)\simeq \frac{2\sqrt{i\eta}e^{\mp\xi \pi/2}}{\alpha_{\pm}^{\mu}\Gamma(\alpha_{\pm})}\left[K_{1-\beta}\left(2\sqrt{z\alpha_{\pm}}\right)\sum^{\infty}_{n=0}\frac{p_n(z)}{\alpha_{\pm}^n}+\sqrt{\frac{z}{\alpha_{\pm}}}K_{-\beta}\left(2\sqrt{z\alpha_{\pm} }\right)\sum^{\infty}_{n=0}\frac{q_n(z)}{\alpha_{\pm}^n}\right], 
\end{equation} 
where $K_\nu(z)$ are the modified Bessel function of the second kind (see Sec.~13.8 of Ref.~\cite{NIST:DLMF}). We used the property $K_{-\nu}(z) = K_{\nu}(z)$. This approximation is valid so long as $|\textrm{ph}(\alpha_{\pm})| \leq \pi-\delta$, which holds for both polarisations, since $\alpha_{\pm}=1/2-\mu\mp i \xi$ is never real and negative, regardless of whether $\mu$ is taken to be real or purely imaginary. The functions $p_{n}(2ik\eta)$ and $q_{n}(2ik\eta)$ appearing in Eq.~\eqref{eq:big-approx-Aplus} are defined as
\begin{align}
    \label{eq:def-pn}&p_n(z) \equiv \sum_{s=0}^{n}\binom{n}{s}(s+2\mu)_{n-s}z^{s}c_{n+s}(z)~,\\
    \label{eq:def-qn}&q_n(z) \equiv \sum_{s=0}^{n}\binom{n}{s}(1+s+2\mu)_{n-s}z^{s} c_{n+s+1}(z)~,
\end{align}
or
\begin{align}
   \label{eq:series-explicit-pn}&p_n(i\tilde{z}) = (2\mu)_n c_n(i\tilde{z})+...+(i\tilde{z})^n c_{2n}(i\tilde{z})~,\\
    \label{eq:series-explicit-qn}&q_n(i\tilde{z}) = (1+2\mu)_n c_{n+1}(i\tilde{z}) +...+(i\tilde{z})^n c_{2n+1}(i\tilde{z})~,
\end{align}
with $\tilde{z}\equiv 2k\eta$. The coefficients $c_n$ satisfy a recurrence relation involving the Bernoulli numbers $B_s$,
\begin{equation}
    \label{eq:rec-rel-c}c_{n+1}(i\tilde{z}) = -\frac{1}{n+1}\sum_{s=0}^{n} \left(\mu_{s+1}+i\tilde{z}_{s+2}\right)c_{n-s}(i\tilde{z})~,
\end{equation}
for $n=0,1,2,...$, where
\begin{align}
    &\mu_s \equiv (1-2\mu)\frac{B_s}{s!}~,\\
    &\tilde{z}_s \equiv 2k\eta\frac{(s-1)B_s}{s!}~.
\end{align}
We have $c_0(z) = 1$ for all $z\in \mathbb{C}$. It is useful to recall that the Bernoulli numbers vanish for all odd indices greater than $1$, namely $B_{2n+1}=0$ for $n=1,2,...$. Consequently, $\mu_{2n+1}=\tilde{z}_{2n+1}=0$ for $n=1,2,...$.

One may now determine the leading terms of each $c_n(z)$. To this end, there is no need to distinguish whether $\mu_s$ in Eq.~\eqref{eq:rec-rel-c} is real or complex, that is, if $\mu$ is real or purely imaginary, respectively. In either case, one finds that the dominant term in the series for $c_n$, in the limit of large $k|\eta|$ or large $|\mu|$, is
\begin{equation}
    c_n(i\tilde{z}) \simeq (-1)^n\frac{\left(\mu_1+i\tilde{z}_2\right)^{n}}{n!}=\frac{\left(1/2-\mu-i\tilde{z}/12\right)^{n}}{n!}~,
\end{equation}
where we used the values of the Bernoulli numbers $B_1 = -\frac{1}{2}$ and $B_2 = \frac{1}{6}$, listed in Sec.~24.2 of Ref.~\cite{NIST:DLMF}. Substituting this into Eqs.~\eqref{eq:series-explicit-pn} and \eqref{eq:series-explicit-qn}, respectively, we find that, if the condition
\begin{equation}
    \label{eq:cond-massive-oneBessel}|z||\beta-z/6|^2 \ll 4|\alpha_{\pm}|
\end{equation}
holds, the first of the two series in Eq.~\eqref{eq:big-approx-Aplus} converges, whereas the second vanishes when multiplied by $\sqrt{z/\alpha_{\pm}}$. The same convergence is ensured for negligibly small $k|\eta|$. Whether $\mu$ is large or not is irrelevant, since we have already assumed that $\alpha_{\pm} = 1/2-\mu\mp i \xi$ is comparatively large relative to the other arguments of the Tricomi function, which implies that $|\xi| \gg \bar{m}$, as argued above Eq.~\eqref{eq:big-approx-Aplus}. Under these assumptions, the term containing the Bessel function $K_{-\beta}$ in the aforementioned equation becomes negligible, yielding
\begin{equation}
\label{eq:approx_Aplus_generic}\mathcal{A}_{\pm}(\eta) \simeq \frac{2\sqrt{i\eta}e^{\mp\xi\pi/2}K_{2\mu}\left(2\sqrt{2ik\eta(1/2-\mu\mp i\xi)}\right)}{(1/2-\mu\mp i\xi)^{\mu}\Gamma(1/2-\mu\mp i\xi)}~.
\end{equation}
In the limit $|\xi| \gg |\mu|$, one arrives at Eqs.~\eqref{eq:mod-sq-Aplus-largemass-kappa} and \eqref{eq:mod-sq-Aminus-largemass-kappa} with $\kappa \rightarrow 1$ (see Eq.~\eqref{eq:kappa-definition}). It should be noted that $\xi \eta >0$, since $\eta<0$ during inflation, and $\xi<0$ by our choice of sign for the scalar-field velocity. 

In practice, the condition in Eq.~\eqref{eq:cond-massive-oneBessel} is rather restrictive and is satisfied only on relatively large superhorizon scales ($-k \eta \ll 1$). However, as shown in Sec.~\ref{app: Matching procedure}, an alternative derivation that relies on weaker assumptions leads to the same result. Consequently, the large-$|\xi|$ solution can be extended beyond the regime defined by condition~\eqref{eq:cond-massive-oneBessel}, as also confirmed numerically.

In the massless case $\mu = 1/2$ (recall that $\mu \equiv \sqrt{\frac{1}{4}-\bar{m}^2}$), we have \cite{Barnaby:2011vw}
\begin{align}
    &\label{eq:massless-result-again}|\mathcal{A}_{+}(\eta)| \simeq \sqrt{\frac{-2\eta}{\pi}}e^{|\xi|\pi}\left|K_1\left(2\sqrt{2\xi k \eta}\right)\right|,\\
    &\label{eq:massless-result-again-minus}|\mathcal{A}_{-}(\eta)| \simeq \sqrt{\frac{-2\eta}{\pi}}\left|K_1\left(-2i\sqrt{2\xi k\eta}\right)\right|.
\end{align}
The large-argument asymptotic form of the Bessel functions, valid for $|2\sqrt{2\xi k\eta}|\gg1$, leads to the well-known expressions \eqref{eq:A+-massless-regime1} and \eqref{eq:A--massless-regime1}. In the massive case, however, if $|\mu| \gg 1$, the relevant condition for the leading-order term to dominate is $|\sqrt{2\xi k \eta}|\gg |\mu|$, as indicated below Eq.~\eqref{eq: matching procedure second regime}, and the asymptotic expansion of the Bessel functions must include additional $\mu$-dependent terms (see Secs.~10.17 and 10.40 of Ref.~\cite{NIST:DLMF}). In the heavy-gauge-field case, $\mu = i \tilde{\mu}$, and
\begin{align}
    \label{eq:large-argument-K-heavy}&K_{2i\tilde{\mu}}\left(2\sqrt{2\xi k \eta}\right) \simeq \frac{\sqrt{\pi}}{2}\frac{e^{-2\sqrt{2\xi k \eta}}}{(2\xi k \eta)^{1/4}}\left\{1+\sum^{\infty}_{n=1}\frac{1}{[-4\sqrt{2\xi k\eta}]^n n!}\prod^{n}_{s=1}\left[(s-1/2)^2+4\tilde{\mu}^2\right]\right\},\\
    \label{eq:large-argument-K-heavy-stable}&K_{2i\tilde{\mu}}\left(-2i\sqrt{2\xi k \eta}\right) \simeq \frac{\sqrt{i\pi}}{2}\frac{e^{2i\sqrt{2\xi k \eta}}}{(2\xi k \eta)^{1/4}}\left\{1+\sum^{\infty}_{n=1}\frac{1}{[4i\sqrt{2\xi k\eta}]^n n!}\prod^{n}_{s=1}\left[(s-1/2)^2+4\tilde{\mu}^2\right]\right\}.
\end{align}
Hence, Eq.~\eqref{eq:approx_Aplus_generic} may be written, respectively, as 
\begin{align}
    \label{eq:mod-sq-Aplus-largemass}&|\mathcal{A}_{+}(\eta)| \simeq \frac{1}{\sqrt{2k}}\left(\frac{k\eta}{2\xi}\right)^{1/4}e^{|\xi|\pi-2\sqrt{2\xi k \eta}}\left[1-\frac{1+16\tilde{\mu}^2}{2^4\sqrt{2\xi k \eta}}+\frac{(1+16\tilde{\mu}^2)(9+16\tilde{\mu}^2)}{2^9(\sqrt{2\xi k\eta})^2}-...\right],\\
    &|\mathcal{A}_{-}(\eta)| \simeq \frac{1}{\sqrt{2k}}\left(\frac{k\eta}{2\xi}\right)^{1/4}\left[1-\frac{1+16\tilde{\mu}^2}{2^5(\sqrt{2\xi k \eta})^2}+...\right]^{1/2}~.
\end{align}

Before moving on to the next section where we compute $\rho_A$ and $\langle \hat{\mathbf{E}}\cdot \hat{\mathbf{B}}\rangle$ in detail, we need to ascertain whether it is safe, regardless of how large $|\xi|$ is, to simplify the arguments of the Bessel functions in Eq.~\eqref{eq:approx_Aplus_generic} to the forms appearing in Eqs.~\eqref{eq:mod-sq-Aplus-largemass-kappa} and \eqref{eq:mod-sq-Aminus-largemass-kappa}, for the unstable and stable polarisation modes, respectively. As a matter of fact, a familiar example of a generally poor approximation is to neglect the real part of the argument of a gamma function, even when its imaginary part is comparatively larger (see Sec.~5.11 of Ref.~\cite{NIST:DLMF}).\footnote{Let $|\Gamma(x+iy)|^2$ denote the modulus squared of the gamma function, where $x$ and $y$ are the real and imaginary parts of its argument. Following the aforementioned reference, if one neglects $x$ in the argument, then $\sim |\Gamma(iy)|^2 \simeq (2\pi/|y|)e^{-\pi |y|}$. However, the more accurate asymptotic expression, obtained from Stirling's formula for large $|y|$, is $(2\pi/(|y|^{1-2x}))e^{-\pi|y|}$, and the missing power-law factor $|y|^{2x}$ can be significant even for moderate $|x|$.} To check if the approximations for the Bessel functions are valid (or to determine the additional conditions under which they are), we note that $K_{\nu}(z)$ is holomorphic in the complex plane except for a branch cut along the negative real axis. Since the domain under consideration does not encompass that region, we can Taylor expand the function in Eq.~\eqref{eq:approx_Aplus_generic} about $2\sqrt{\pm 2\xi k \eta}$, obtaining\footnote{\label{footnote-clarification-argument-stable}The argument of the Bessel function of the stable mode is $-2i\sqrt{2\xi k \eta}$, rather than $2i\sqrt{2\xi k \eta}$, because the limit inside the square root is taken to be $-2\xi k \eta-i0$, consistently with the principal branch, thereby approaching the negative real axis from below.}
\begin{align}
    \nonumber&K_{2i\tilde{\mu}}\left(2\sqrt{2ik\eta(1/2-i\tilde{\mu}-i\xi)}\right) \simeq K_{2i\tilde{\mu}}\left(2\sqrt{2\xi k \eta}\right)\\
    &\phantom{----------}+\sqrt{\frac{k\eta}{2\xi}}\left(\tilde{\mu}+\frac{i}{2}\right)\left[K_{1+2i\tilde{\mu}}\left(2\sqrt{2\xi k \eta}\right)+K_{1-2i\tilde{\mu}}\left(2\sqrt{2\xi k \eta}\right)\right],\\
    \nonumber&K_{2i\tilde{\mu}}\left(2\sqrt{2ik\eta(1/2-i\tilde{\mu}+i\xi)}\right) \simeq K_{2i\tilde{\mu}}\left(-2i\sqrt{2\xi k \eta}\right)\\
    &\phantom{-------}+i\sqrt{\frac{k\eta}{2\xi}}\left(\tilde{\mu}+\frac{i}{2}\right)\left[K_{1+2i\tilde{\mu}}\left(-2i\sqrt{2\xi k \eta}\right)+K_{1-2i\tilde{\mu}}\left(-2i\sqrt{2\xi k \eta}\right)\right],
\end{align}
where the recurrence relation for the Bessel function $K_{\nu}(z)$ given in Sec.~10.29 of Ref.~\cite{NIST:DLMF} has been used to compute $\textrm{d}K_{\nu}(z)/\textrm{d}z$. Therefore, the conditions for retaining only the zeroth-order term in the expansion are
\begin{equation}
    \sqrt{2\xi k \eta}\frac{\bar{m}}{2|\xi|}\left|\frac{K_{1+2i\tilde{\mu}}(z_0)+K_{1-2i\tilde{\mu}}(z_0)}{K_{2i\tilde{\mu}}(z_0)}\right| \ll 1~.
    \label{eq:bessel-taylor-criterion}
\end{equation}
Here
\begin{equation}
z_0=2\sqrt{2\xi k\eta}
\qquad\text{or}\qquad
z_0=-2i\sqrt{2\xi k\eta}~.
\end{equation}
Equation~\eqref{eq:bessel-taylor-criterion}, including the displayed Bessel-function ratio, is the criterion for truncating the Taylor expansion. It is satisfied for $|\xi|\gg\bar m$, away from $z_0=0$ and from zeros of $K_{2i\tilde\mu}(z_0)$. When $|\xi|$ is only comparable to $\bar m$, the argument replacement is the matching prescription encoded by $\kappa$, not a uniformly controlled Taylor expansion. The construction in Sec.~\ref{app: Matching procedure} and the numerical comparison in Fig.~\ref{fig:kA2-Bessel-Whittaker} validate that regime.

\subsection{\label{sec:app-detailed-EBandrho}Detailed Evaluation of Gauge-Field Densities}
Here, we provide the intermediate steps leading to the analytical expressions for the gauge-field energy density $\rho_A$ and the pseudo-scalar density $\langle \hat{\mathbf{E}}\cdot \hat{\mathbf{B}}\rangle$ in Eq.~\eqref{eq:main-result-rho-EB-very-heavy}. The calculation follows from substituting the Bessel approximation for the gauge-field mode functions into Eqs.~\eqref{eq:rhoA-approx-Aplus} and \eqref{eq:EB-approx-Aplus}, and evaluating the resulting integrals. 

We first change the variable of integration to $x\equiv 2\sqrt{2\xi \kappa k \eta}$ in those equations, and define \cite{Gradshteyn:1943cpj}
\begin{align}
    {\cal I}^\alpha_{\nu_1, \nu_2} \equiv &\int^{\infty}_0 \textrm{d}x~x^{\alpha} K_{\nu_1}(x) K_{\nu_2}(x)= \frac{2^{-2+\alpha} }{\Gamma(1+\alpha)}\Gamma\left(\frac{1+\alpha+\nu_1+\nu_2}{2}\right) \nonumber \\
    &\phantom{---}\times\Gamma\left(\frac{1+\alpha-\nu_1+\nu_2}{2}\right)\Gamma\left(\frac{1+\alpha+\nu_1-\nu_2}{2}\right)\Gamma\left(\frac{1+\alpha-\nu_1-\nu_2}{2}\right),
\end{align}
which holds provided that $\textrm{Re}(-\alpha)<1-|\textrm{Re}(\nu_1)|-|\textrm{Re}(\nu_2)|$. Then, Eqs.~\eqref{eq:rhoA-approx-Aplus} and \eqref{eq:EB-approx-Aplus} are found to become
\begin{align}
    \nonumber\label{eq:rhoA-integrals-massmu}& \rho_A \simeq \frac{e^{2|\xi|\pi}}{2^{11}\pi^3\kappa^3|\xi|^3\eta^4}\left[2(1+2\tilde{\mu}^2)\mathcal{I}^{5}_{2i\tilde{\mu},2i\tilde{\mu}}+\frac{1}{4} \mathcal{I}^{7}_{1+2i\tilde{\mu},1+2i\tilde{\mu}}+\frac{1}{4} \mathcal{I}^{7}_{1-2i\tilde{\mu},1-2i\tilde{\mu}} \right. \nonumber \\ & \left. \phantom{-------------} + \frac{1}{2} \mathcal{I}^{7}_{1+2i\tilde{\mu},1-2i\tilde{\mu}}-\mathcal{I}^{6}_{2i\tilde{\mu},1+2i\tilde{\mu}}-\mathcal{I}^{6}_{2i\tilde{\mu},1-2i\tilde{\mu}}+\frac{\mathcal{I}^{9}_{2i\tilde{\mu},2i\tilde{\mu}}}{16\kappa^2\xi^2} \right],\\
    \label{eq:EB-integrals-massmu} & \langle \hat{\mathbf{E}}\cdot \hat{\mathbf{B}}\rangle \simeq \frac{e^{2|\xi|\pi}}{2^{12}\pi^3(\kappa\xi\eta)^4}\left[\mathcal{I}^{7}_{2i\tilde{\mu},2i\tilde{\mu}}-\frac{1}{2}\mathcal{I}^{8}_{2i\tilde{\mu},1+2i\tilde{\mu}}-\frac{1}{2}\mathcal{I}^{8}_{2i\tilde{\mu},1-2i\tilde{\mu}}\right].
\end{align}
The integrals evaluate to
\begin{eqnarray}
    \label{eq:heavy-gauge-field-rhoA-analytical}\rho_A &\simeq& \frac{(1+\tilde{\mu}^2)(1+4\tilde{\mu}^2)\tilde{\mu}}{5\sinh(2\pi \tilde{\mu})}\left[\frac{11+8\tilde{\mu}^2}{21}+\frac{(4+\tilde{\mu}^2)(9+4\tilde{\mu}^2)}{63\kappa^2\xi^2}\right]\frac{e^{2|\xi|\pi}}{2^4\pi^2\kappa^3|\xi|^3\eta^4}~,\\
    \label{eq:heavy-gauge-field-EB-analytical}\langle \hat{\mathbf{E}}\cdot \hat{\mathbf{B}}\rangle &\simeq& -\frac{3(9+4\tilde{\mu}^2)(1+\tilde{\mu}^2)(1+4\tilde{\mu}^2)\tilde{\mu}}{35\sinh(2\pi\tilde{\mu})}\frac{e^{2|\xi|\pi}}{2^{6}\pi^2 \left(\kappa\xi\eta\right)^4}~,
\end{eqnarray}
where we have used properties of the gamma function (see Chap.~5 of Ref.~\cite{NIST:DLMF}) to compute their modulus squared. Assuming $\tilde{\mu}\simeq \bar{m} \gg 1$, the above expressions further simplify to Eq.~\eqref{eq:main-result-rho-EB-very-heavy}. Replacing $\tilde{\mu}$ with $\bar{m}$ introduces a relative error of $15.8\%$ for $\bar{m}=2$, which decreases to $8\%$ for $\bar{m}\simeq 8$. Overall, the approximation remains reasonably accurate even for comparatively small values of $\bar{m}$.  

\subsection{\label{sec:superhorizon-expressions}The Superhorizon Regime}
For completeness, we now estimate the superhorizon evolution of the gauge mode functions. For the asymptotic behaviour of the Tricomi function in the limit $k|\eta| \rightarrow 0$, we refer the reader to Sec.~13.2 of Ref.~\cite{NIST:DLMF}.

\subsubsection*{Massless Gauge Field}
For the massless gauge field, we have (see Eq.~\eqref{eq:Aplusminus-U}; recall that $z\equiv 2ik\eta$)
\begin{equation}
    \mathcal{A}_{\pm}(\eta) = \frac{e^{-(z\pm \xi \pi)/2}}{\sqrt{2k}}U(\mp i \xi,0,z)~,
\end{equation}
and this expression takes the following approximate form in the superhorizon limit $k|\eta|\rightarrow 0$:
\begin{equation}
    \mathcal{A}_{\pm} \simeq \frac{e^{\mp \xi \pi/2}}{\sqrt{2k}\Gamma(1\mp i \xi)}~.
\end{equation}
Notice that the amplitude of the gauge modes remains constant on superhorizon scales, as is the case for massless fields. Depending on the polarisation, the modulus is given by
\begin{align}
    \label{eq:mod-plus-A-superhorizon-massless}&|\mathcal{A}_{+}| \simeq \frac{e^{|\xi|\pi}}{2\sqrt{\pi|\xi|k}}~,\\
    \label{eq:mod-minus-A-superhorizon-massless}&|\mathcal{A}_{-}| \simeq \frac{1}{2\sqrt{\pi|\xi|k}}~,
\end{align}
for large $|\xi|$ (see Sec.~5.11 of Ref.~\cite{NIST:DLMF}). Thus, the unstable mode dominates over the mode that is not enhanced by the instability.

\subsubsection*{Light Gauge Field}
The massive case is somewhat more intricate, as three scenarios must be considered separately. The first corresponds to $\mu=0$, which is equivalent to $\bar{m}=1/2$, and then
\begin{equation}
    \mathcal{A}_{\pm}(\eta) = \frac{e^{-(z\pm \xi \pi)/2}}{\sqrt{2k}}z^{1/2}U(1/2\mp i \xi,1,z)~.
\end{equation}
The superhorizon limit $|z|\rightarrow 0$ yields\footnote{\label{footnote:indication-logarithm-phase}In this appendix, the distinction between `$\textrm{log}$' and `$\textrm{ln}$' is that the former denotes the complex logarithm, whereas the latter indicates the natural logarithm. Thus, for a complex number $z$
\begin{equation}
    \log(z) = \ln(|z|) +i\textrm{arg}(z)~,
\end{equation}
where $\textrm{arg}(z)$ is the angle measured from the positive real axis, which includes the different branches of the multivalued complex logarithm, namely $\textrm{arg}(z) = \textrm{ph}(z)+2n\pi$, with $n\in \mathbb{Z}$ and $-\pi < \textrm{ph}(z) \leq \pi$.}
\begin{equation}
    \label{eq:superhorizon-mu0-specialcase}\mathcal{A}_{\pm}(\eta)\simeq -\frac{e^{\mp \xi \pi/2}}{\sqrt{2k} \Gamma(1/2\mp i \xi)}z^{1/2}\left[\log(z) +\psi(1/2\mp i \xi)+2\gamma\right],
\end{equation}
where $\gamma\simeq 0.5772...$ is the Euler-Mascheroni constant, and $\psi(z)=\Gamma^{'}(z)/\Gamma(z)$ is the so-called `digamma function' \cite{NIST:DLMF}, with the prime denoting differentiation with respect to the argument of the gamma function. In the limit of large $|\xi|$, the asymptotic expansion of the digamma function is
\begin{equation}
    \psi(1/2\mp i \xi) \simeq \log(1/2\mp i \xi)-\frac{1}{1\mp 2i \xi}~.
\end{equation}
Consequently 
\begin{align}
    &\mathcal{A}_{+}(\eta) \simeq \frac{-e^{-\xi\pi/2}\sqrt{2ik\eta}}{\sqrt{2k} \Gamma(1/2- i \xi)}\left[\ln(2\xi k\eta)-\frac{1}{4\xi^2}+2\gamma+\frac{i}{2|\xi|}\right],\\
    &\mathcal{A}_{-}(\eta)\simeq \frac{-e^{\xi \pi/2}\sqrt{2ik\eta}}{\sqrt{2k}\Gamma(1/2+i\xi)}\left[\ln(2\xi k \eta)-\frac{1}{4\xi^2}+2\gamma-i\left( \pi+\frac{1}{2|\xi|}\right)\right],
\end{align}
or, at leading logarithmic order as $k|\eta|\to0$,
\begin{align}
\label{eq:mod-plus-A-superhorizon-murealzero-amplitude}
|\mathcal A_+(\eta)|&\simeq
\sqrt{\frac{-\eta}{2\pi}}e^{\pi|\xi|}\left|\ln\!\left(2|\xi|k|\eta|\right)+2\gamma\right|,\\
|\mathcal A_-(\eta)|&\simeq
\sqrt{\frac{-\eta}{2\pi}}\left|\ln\!\left(2|\xi|k|\eta|\right)+2\gamma\right|.
\end{align}
We again find $|\mathcal A_+|\simeq e^{\pi|\xi|}|\mathcal A_-|$ at leading logarithmic order, while both massive mode functions continue to evolve on superhorizon scales.

In the case of real $\mu$ corresponding neither to the massless framework nor to $\mu=0$, we obtain\footnote{\label{footnote-mu-approx-Tricomi}Notice that this formula applies only when $0<\textrm{Re}(\beta)<1$, with $\beta$ denoting the second argument of $U(\alpha,\beta,z)$ \cite{NIST:DLMF}. In our case, $\beta=1-2\mu$. For real $\mu$, this excludes $\mu=0$ and $\mu=1/2$. The case of purely imaginary $\mu$, considered next, is also excluded, since $\textrm{Re}(\beta) = 1$.}
\begin{equation}
    \mathcal{A}_{\pm}(\eta)\simeq \frac{e^{\mp \xi \pi/2}}{\sqrt{2k}}z^{1/2-\mu}\frac{\Gamma(2\mu)}{\Gamma(1/2+\mu\mp i \xi)}~,
\end{equation}
on superhorizon scales. Taking a large $|\xi|$ gives
\begin{align}
    \label{eq:mod-plus-A-superhorizon-murealnotzero-amplitude}&|\mathcal{A}_{+}(\eta)|\simeq \frac{e^{|\xi|\pi}\Gamma(2\mu)}{2\sqrt{\pi k }|\xi|^{\mu}}(-2k\eta)^{1/2-\mu}~,\\
    &|\mathcal{A}_{-}(\eta)|\simeq \frac{\Gamma(2\mu)}{2\sqrt{\pi k}|\xi|^{\mu}}(-2k\eta)^{1/2-\mu}~,
\end{align}
where $|\mathcal{A}_{+}|\simeq e^{|\xi|\pi}|\mathcal{A}_{-}|$ once again. Nevertheless, on sufficiently large scales, the amplitudes of the gauge mode functions decay, although the unstable mode remains dominant because of the exponential factor. As a consistency check, setting $\mu=1/2$ reproduces the expressions in Eqs.~\eqref{eq:mod-plus-A-superhorizon-massless} and \eqref{eq:mod-minus-A-superhorizon-massless}.

\subsubsection*{Heavy Gauge Field}
Finally, when $\mu$ is purely imaginary, the approximation above is no longer applicable, as discussed in footnote~\ref{footnote-mu-approx-Tricomi}. Instead, the superhorizon limiting form is \cite{NIST:DLMF}
\begin{equation}
    \label{eq:superhorizon-muimaginary}\mathcal{A}_{\pm}(\eta) \simeq \frac{e^{\mp \xi \pi/2}}{\sqrt{2k}}z^{1/2}\left[\frac{\Gamma(2i\tilde{\mu})z^{-i\tilde{\mu}}}{\Gamma(1/2+i\tilde{\mu}\mp i \xi)}+\frac{\Gamma(-2i\tilde{\mu})z^{i\tilde{\mu}}}{\Gamma(1/2-i\tilde{\mu}\mp i \xi)}\right].
\end{equation}
Because $z^{i\tilde{\mu}} = e^{\pi\tilde{\mu}}(-z)^{i\tilde{\mu}}$, with $z=2ik\eta$, the presence of both, the non-zero $\xi$ in the gamma function, and the exponential factor $e^{\pi \tilde{\mu}}$, prevents the second addend inside the brackets from being treated as the complex conjugate of the first, and vice versa. Consequently, there will be an oscillatory term when taking the modulus of that expression, but it will be damped as the $z^{1/2}$ factor goes to zero,
\begin{align}
    \nonumber&|\mathcal{A}_{\pm}(\eta)| \simeq e^{\pm |\xi|\pi}\sqrt{\frac{-\eta}{2\tilde{\mu}\sinh(2\pi\tilde{\mu})}}\left\{1+e^{\mp 2|\xi|\pi}\cosh(2\pi\tilde{\mu})\right.\\
    \label{eq:superhorizon-muimaginary1}&\left.\phantom{-------------}+e^{\mp |\xi|\pi}\sqrt{2\left[\cosh(2\pi|\xi|)+\cosh(2\pi \tilde{\mu})\right]}\cos(\Theta(\eta))\right\}^{1/2},
\end{align}
where `$\Theta$' denotes the time-dependent argument (see footnote~\ref{footnote:indication-logarithm-phase}) of the complex number $[\Gamma(2i\tilde{\mu}) z^{-i\tilde{\mu}}]^2/[\Gamma(1/2+i\tilde{\mu}\mp i \xi)\Gamma(1/2+i\tilde{\mu}\pm i \xi)]$, given by
\begin{align}
    \nonumber&\Theta(\eta) =-\pi-2\tilde{\mu}\left[\ln(2k|\eta|)+\gamma\right]+\sum_{s=1}^{\infty}\left[\frac{2\tilde{\mu}}{s}-2\textrm{ph}\left(1+\frac{2i\tilde{\mu}}{s}\right)+\textrm{ph}\left(1+\frac{1/2+i\tilde{\mu}\mp i \xi}{s}\right)\right.\\
    &\left.\phantom{--}+\textrm{ph}\left(1+\frac{1/2+i\tilde{\mu}\pm i \xi}{s}\right)\right]+\textrm{ph}\left(\frac{1}{2}+i\tilde{\mu}\mp i\xi\right)+\textrm{ph}\left(\frac{1}{2}+i\tilde{\mu}\pm i\xi\right)+2n\pi~,
\end{align}
with $n\in \mathbb{Z}$. The so-called `Weierstrass form' of the gamma function was employed \cite{NIST:DLMF}: 
\begin{equation}
    \log[\Gamma(z)] = -\log(z)-\gamma z +\sum_{n=1}^{\infty}\left[\frac{z}{n}-\log\left(1+\frac{z}{n}\right)\right].
\end{equation}
In deriving Eq.~\eqref{eq:superhorizon-muimaginary1}, we used the following relations to express the gamma functions in terms of hyperbolic cosines:\footnote{We emphasise that no approximations were made in going from Eq.~\eqref{eq:superhorizon-muimaginary} to Eq.~\eqref{eq:superhorizon-muimaginary1}.} 
\begin{eqnarray}
    |\Gamma(1/2-i\tilde{\mu}+i\xi)|&=&\sqrt{\frac{\pi}{\cosh[\pi(|\xi|+\tilde{\mu})]}}~,\\
    |\Gamma(1/2-i\tilde{\mu}-i\xi)|&=&\sqrt{\frac{\pi}{\cosh[\pi(|\xi|-\tilde{\mu})]}}~.
\end{eqnarray}
Expressed in terms of inverse tangents, $\Theta(\eta)$ reads
\begin{align}
    \nonumber&\Theta(\eta) = -\pi - 2\tilde{\mu}\left[\ln(2k|\eta|)+\gamma\right]+\sum_{s=0}^{\infty}\left[\arctan\left(\frac{\tilde{\mu}\mp \xi}{s+1/2}\right)+\arctan\left(\frac{\tilde{\mu}\pm \xi}{s+1/2}\right)\right]\\
    &\phantom{--------------------}+2\sum_{s=1}^{\infty}\left[\frac{\tilde{\mu}}{s}-\arctan\left(\frac{2\tilde{\mu}}{s}\right)\right]+2n\pi~.
\end{align}
As expected from the complex number appearing below Eq.~\eqref{eq:superhorizon-muimaginary1}, the resulting $\Theta(\eta)$ is the same for both polarisations, since $\Gamma(1/2+i\tilde{\mu}- i \xi) \Gamma(1/2+i\tilde{\mu}+ i \xi) = \Gamma(1/2+i\tilde{\mu}+i\xi)\Gamma(1/2+i\tilde{\mu}-i\xi)$. The logarithm is the only time-dependent contribution that grows without bound as $k|\eta|\to0$; all remaining terms in Eq.~\eqref{eq:superhorizon-muimaginary1} contribute a time-independent phase. Absorbing the explicit $-\pi$ and the constant terms into $\Theta_0$, we may write
\begin{equation}
\label{eq:superhorizon-limit-argument-periodic}
\Theta(\eta)=2\tilde\mu\ln(2k|\eta|)+\Theta_0
\quad (\mathrm{mod}\ 2\pi)~,
\end{equation}
where $\Theta_0$ depends on $\xi$, $\tilde\mu$, and the branch convention. It must be retained when the phase of the logarithmic oscillations, rather than only their envelope, is required.

One observes that the amplitudes of the gauge mode functions are not constant on superhorizon scales, as in the other two cases of a massive gauge field ($\mu=0$ and real $\mu$ with $0<\mu<1/2$). However, these amplitudes can be simplified further by assuming the realistic case of large $|\xi|$ and large $\tilde{\mu}$, with $|\xi|+\tilde{\mu}\gg 1$ and $|\xi|>\tilde{\mu}$. In this regime, the two amplitudes read
\begin{align}
    \label{eq:Aplusabs-superhor-xilargemu}&|\mathcal{A}_{+}(\eta)| \simeq \sqrt{\frac{-\eta\left[1-\cos(\Theta)\right]}{\tilde{\mu}}}e^{(|\xi|-\tilde{\mu})\pi}~,\\
    &|\mathcal{A}_{-}(\eta)| \simeq \sqrt{\frac{-\eta}{2\tilde{\mu}}}~.
\end{align}
As these expressions show, $|\mathcal{A}_{+}|$ is not exactly equal to $e^{|\xi|\pi}|\mathcal{A}_{-}|$, unlike in the preceding cases. Rather, the exponential factor includes $\tilde{\mu}$, and the prefactor differs slightly, since the cosine term is not exponentially suppressed for the `$+$' mode. When $|\xi|=\tilde\mu$, the exponential factor in Eq.~\eqref{eq:Aplusabs-superhor-xilargemu} disappears. This point lies just below the exact tachyonic threshold $|\xi|=\bar m=\sqrt{\tilde\mu^2+1/4}$; the relative difference is of order $\bar m^{-2}$ in the heavy limit. The two amplitudes then reduce to
\begin{align}
    \label{eq:ampl-xiequalsmutilde-exactly}&|\mathcal{A}_{+}(\eta)|\simeq \sqrt{\frac{-\eta}{\tilde{\mu}}\left[\frac{3}{2}-\sqrt{2}\cos(\Theta)\right]}~,\\
    &|\mathcal{A}_{-}(\eta)|\simeq \sqrt{\frac{-\eta}{\tilde{\mu}}}~.
\end{align}
This is reasonable because $|\xi| = \tilde{\mu}$ implies that (see Eq.~\eqref{eq:oscillator-freq-wplusminus})
\begin{equation}
    \label{eq:effective-freq-xiequalsmu}\omega^2_{\pm}\eta^2 \simeq \left|k|\eta|\mp |\xi|+\frac{i}{2}\right|^2,
\end{equation}
and therefore there is no instability ($\omega^2_{\pm}$ is not negative). As is clear from Eq.~\eqref{eq:effective-freq-xiequalsmu}, the amplitudes of the two polarisation modes do not match exactly when $|\xi|=\tilde{\mu}$, despite the absence of enhancement, because the effective frequency is different for each mode, indeed. 

To verify Eq.~\eqref{eq:Aplusabs-superhor-xilargemu} in particular, we may consider the superhorizon limit of Eq.~\eqref{eq:mod-sq-Aplus-largemass-kappa}, bearing in mind that this expression was derived under assumptions other than the superhorizon evolution itself. In the limit $k|\eta|\to0$, the small-argument form of Eq.~\eqref{eq:mod-sq-Aplus-largemass-kappa} gives (see Sec.~10.45 of Ref.~\cite{NIST:DLMF})
\begin{equation}
|\mathcal A_+(\eta)|\simeq
\sqrt{\frac{-2\eta}{\tilde\mu}}\,
 e^{\pi(|\xi|-\tilde\mu)}
\left|\sin\!\left[
\tilde\mu\ln\!\left(2|\xi|\kappa k|\eta|\right)
-\arg\Gamma(1+2i\tilde\mu)
\right]\right|.
\end{equation}
This reproduces the $\sqrt{-\eta}$ envelope, the factor $e^{\pi(|\xi|-\tilde\mu)}$, and the logarithmic oscillation frequency in Eq.~\eqref{eq:Aplusabs-superhor-xilargemu}. The constant phase should not be discarded when individual oscillations are compared.

\begin{figure}
    \centering
\includegraphics[width=1.00\linewidth]{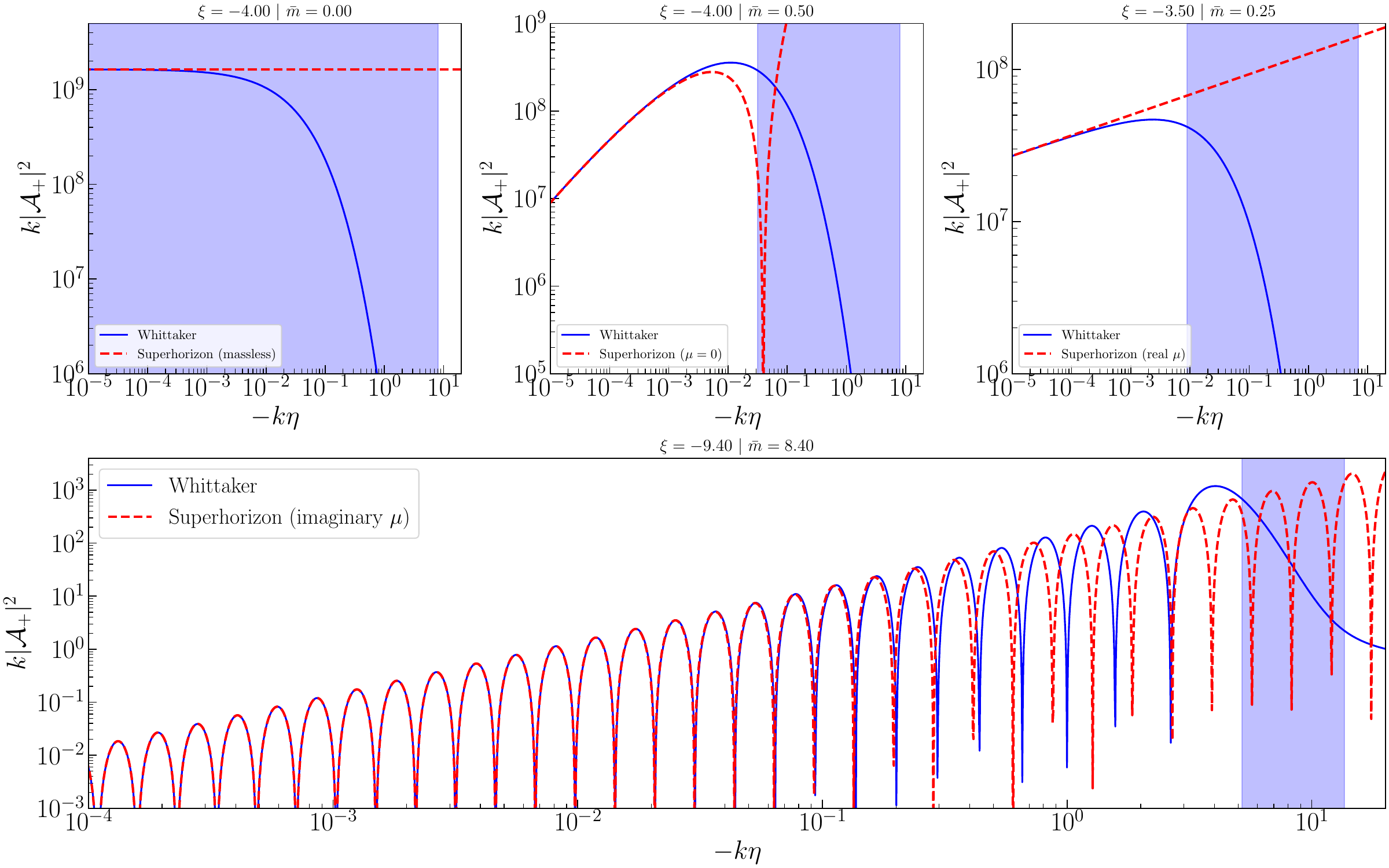}
    \caption{Comparison of $k|\mathcal{A}_{+}|^2$ obtained from the full Whittaker solution, Eq.~\eqref{eq:solution-to-Whittaker-eq-Aplusminus} (solid blue), and the superhorizon approximation (red dashed) for representative values of $(\xi,\bar m)$. The upper panels correspond to the massless case \eqref{eq:mod-plus-A-superhorizon-massless} and to two examples with real $\mu$, Eqs.~\eqref{eq:mod-plus-A-superhorizon-murealzero-amplitude} and \eqref{eq:mod-plus-A-superhorizon-murealnotzero-amplitude}, while the lower panel shows a representative case with imaginary $\mu$, Eq.~\eqref{eq:Aplusabs-superhor-xilargemu}. The shaded regions indicate the instability bands, given by Eq.~\eqref{eq:Instability Band 2}.}
    \label{fig:kA2-superhorizon-Whittaker}
\end{figure}

Figure~\ref{fig:kA2-superhorizon-Whittaker} compares the superhorizon approximation with the exact Whittaker solution for some values of $(\xi,\bar{m})$, illustrating the massless, light, and heavy gauge-field regimes. As expected, the agreement improves as the modes evolve deeper into the superhorizon regime, where the asymptotic expansion becomes increasingly accurate. While the agreement is less precise near horizon crossing, the approximation successfully reproduces the late-time behaviour of the mode functions in all cases considered.  

\section{Detailed Calculation of the Inverse-Decay Contribution\label{app:detailed-calculation-inverse-decay}}
The purpose of this appendix is to provide the detailed derivation of the inverse-decay contribution to the scalar power spectrum presented in Secs.~\ref{sec:scalarpowerspectrum-weak}--\ref{sec:scalarpowerspectrum-weak-CASE-HEAVY-FIELDS}. Starting from the sourced perturbation equation, we derive the corresponding integral representation and perform the manipulations leading to the semi-analytical expressions employed in the main text.

\subsection{\label{app:detailed-calculation-inverse-decay-POWER-SPECTRUM}Inverse-Decay Power Spectrum}
We present the intermediate steps leading to the semi-analytical spectrum in Eq.~\eqref{eq:the-ratio-semianalytical-intermediate}. As in the case of the gauge field, the scalar field is promoted to a quantum operator and expanded in mode functions (see Eq.~\eqref{eq:scalar-field-operator-modes-expansion}). These can be split into two contributions: the first is the (vacuum) solution of the homogeneous equation, namely, Eq.~\eqref{eq:sf-eq-pert-linear-Fourier} with no source term on the right-hand side, while the second is a particular solution of Eq.~\eqref{eq:sf-eq-pert-linear-Fourier} that captures the effect of the gauge fields. The former is denoted by $u_k=u(\eta,k)$, and the field operator in momentum space reads 
\begin{equation}
    \label{eq:exp-deltaphi-vac-operators}\hat{\delta \phi}_{\textrm{vac}}(\eta,\mathbf{k})=a^{-1}(\eta)\left[u(\eta,k)\hat{b}_{\mathbf{k}}+u^{*}(\eta,k)\hat{b}^{\dag}_{-\mathbf{k}}\right],
\end{equation}
with the annihilation and creation operators, $\hat{b}$ and $\hat{b}^{\dag}$, respectively, satisfying Eq.~\eqref{eq:comm-rel-b-operators}, whereas the rest of commutators vanish. The quantum operator associated with the particular solution, $\hat{\delta \phi}_{\textrm{id}}(\eta,\mathbf{k})$, where `id' refers to the inverse-decay process mentioned at the beginning of Sec.~\ref{sec:scalarpowerspectrum-weak}, depends on the source $\hat{J}(\eta,\mathbf{k})$ (see Eq.~\eqref{eq:source-Fouriermodes-sf-evolution}). Therefore, the relevant annihilation and creation operators are those of the gauge field, namely $\hat{a}$ and $\hat{a}^{\dag}$ (see Eqs.~\eqref{eq:Fourier-quantum-commutators-A0} and \eqref{eq:Avec-field-operator}). These commute with those in Eq.~\eqref{eq:exp-deltaphi-vac-operators}, implying that the two-point correlation function of $\hat{\zeta}(\eta,\mathbf{k})$ in Fourier space can be written as follows (the time dependence is not shown explicitly, since the correlators are evaluated at equal time):
\begin{equation}
    \label{eq:2-point-correlator-Rgaugeinvariant}\langle \hat{\zeta}(\mathbf{k})\hat{\zeta}(\mathbf{k}')\rangle = \left(\frac{\mathcal{H}}{\phi^{'}}\right)^2 \left[\langle\hat{\delta \phi}_{\textrm{vac}}(\mathbf{k})\hat{\delta \phi}_{\textrm{vac}}(\mathbf{k}')\rangle+\langle\hat{\delta \phi}_{\textrm{id}}(\mathbf{k})\hat{\delta \phi}_{\textrm{id}}(\mathbf{k}')\rangle\right],
\end{equation}
where (recall that $\eta<0$)
\begin{equation}
    \label{eq:deltaphi-invdecay-operator}\hat{\delta \phi}_{\textrm{id}}(\eta,\mathbf{k}) = a^{-1}(\eta)\int^{0}_{-\infty}\textrm{d}\tilde{\eta}~ G_k(\eta,\tilde{\eta}) \hat{J}(\tilde{\eta},\mathbf{k})~.
\end{equation}
$G_k$ is the retarded mode Green function for the mode $k$ and, for suitably normalised mode functions $u_k$,\footnote{\label{Wronskian-normalisation}Our convention is $u_k u_k^{*'}-u_k^*u_k'=i$. This Wronskian is time independent because the homogeneous equation associated with Eq.~\eqref{eq:sf-eq-pert-linear-Fourier} contains no first-derivative term.} it reads
\begin{equation}
    \label{eq:green-function-retarded-real}G_k(\eta,\tilde{\eta}) = i\theta(\eta-\tilde{\eta}) \left[u_k(\eta)u^{*}_k(\tilde{\eta})-u^{*}_k(\eta)u_k(\tilde{\eta})\right].
\end{equation}
$\theta(\eta-\tilde{\eta})$ denotes the Heaviside step function, which equals $1$ whenever $|\tilde{\eta}| > |\eta|$, and vanishes otherwise, thereby enforcing causality. One may verify that $G_{k}$ is real-valued.

To determine the full contribution to the two-point correlator in Eq.~\eqref{eq:2-point-correlator-Rgaugeinvariant}, we must first specify the evolution of the mode functions $u_k(\eta)$. These satisfy the following equation (cf. Eq.~\eqref{eq:sf-eq-pert-linear-Fourier}):
\begin{equation}
    u_k^{''} +\left(k^2+a^2m^2_{\phi}-\frac{a^{''}}{a}\right) u_k=0~.
\end{equation}
Upon introducing the change of variable $u_k \equiv \sqrt{-\eta} \chi_k$, we obtain the equation
\begin{equation}
    (-k\eta)^2 \frac{\textrm{d}^2\chi_k}{\textrm{d}(-k\eta)^2}+(-k\eta)\frac{\textrm{d}\chi_k}{\textrm{d}(-k\eta)}+\left[\left(-k\eta\right)^2-\nu^2(\eta)\right]\chi_k = 0~,
\end{equation}
where $\nu(\eta)$ is defined by
\begin{equation}
    \nu(\eta) \equiv \sqrt{\frac{1}{4}-\eta^2\left(a^2m^2_{\phi}-\frac{a^{''}}{a}\right)}~,
\end{equation}
with $m_{\phi} = m_{\phi}(\eta)$ and $a=a(\eta)$ in general. At first order in the slow-roll approximation, $\nu$ becomes constant and takes the form
\begin{equation}
    \label{eq:nu-sr-suppressed-terms}\nu \simeq \sqrt{\frac{9}{4}+3\epsilon -\bar{m}_{\phi}^2}~,
\end{equation}
where $\epsilon \equiv 1-\mathcal{H}^{'}/\mathcal{H}^2$ denotes the first Hubble-flow parameter, and $\bar{m}_{\phi}$ is the inflaton mass in units of the Hubble rate, analogously to $\bar{m}$ for the gauge field. Both quantities are constant at first order in the aforementioned expansion, and may be negligibly small if the background evolution is sufficiently close to de Sitter and, in the case of $m^2_{\phi}$, slow-roll regime is sustained by the slope of the potential. Because $\nu$ is then constant, the differential equation above reduces to the Bessel equation \cite{NIST:DLMF}, and the mode functions $u_k$ can therefore be written as a linear combination of the Hankel functions of the first and second kind, $H^{(1)}_{\nu}$ and $H^{(2)}_{\nu}$, respectively. Consequently 
\begin{equation}
    u_k(\eta) = \sqrt{-\eta} \left[\alpha_1 H_{\nu}^{(1)}(-k\eta) + \alpha_2 H_{\nu}^{(2)}(-k\eta)\right],
\end{equation}
where $\alpha_1$ and $\alpha_2$ are constants of integration. Imposing the Bunch-Davies vacuum on subhorizon scales, together with the proper normalisation of the mode functions (see footnote~\ref{Wronskian-normalisation}) and the asymptotic form of the Hankel functions for large argument (see Sec.~10.17 of Ref.~\cite{NIST:DLMF}), yields $\alpha_2 = 0$ and $\alpha_1 = \sqrt{\pi}e^{i\pi(\nu+1/2)/2}/2$. We may then write the standard result
\begin{equation}
    u_k(\eta) = \frac{\sqrt{-i\pi \eta}}{2}e^{i\pi\nu/2}H_{\nu}^{(1)}(-k\eta)~.
\end{equation}
The retarded Green function becomes
\begin{equation}
    G_k(\eta,\tilde{\eta}) = i\frac{\pi}{4}\sqrt{\eta\tilde{\eta}}\theta(\eta-\tilde{\eta}) \left[H_{\nu}^{(1)}(-k\eta)H_{\nu}^{(2)}(-k\tilde{\eta})-H_{\nu}^{(1)}(-k\tilde{\eta})H_{\nu}^{(2)}(-k\eta)\right],
\end{equation}
which is real, as noted below Eq.~\eqref{eq:green-function-retarded-real}, since $\left[H_{\nu}^{(1,2)}(x)\right]^{*} = H_{\nu}^{(2,1)}(x)$ when $x$ and $\nu$ are real. The two-point correlation function $\langle\hat{\delta \phi}_{\textrm{vac}}(\mathbf{k})\hat{\delta \phi}_{\textrm{vac}}(\mathbf{k}')\rangle$ is then given by (see Eq.~\eqref{eq:exp-deltaphi-vac-operators}; recall also that $\hat{b}_{\mathbf{k}}\ket{0}=0$ and $\bra{0}\hat{b}^{\dag}_{\mathbf{k}} = 0$, as well as the commutation relation in Eq.~\eqref{eq:comm-rel-b-operators})
\begin{equation}
    \label{eq:two-point-vac-part-deltaphi}\langle\hat{\delta \phi}_{\textrm{vac}}(\mathbf{k})\hat{\delta \phi}_{\textrm{vac}}(\mathbf{k}')\rangle = \frac{|u_k|^2}{a^2}\delta^{(3)}(\mathbf{k}+\mathbf{k}') = \frac{-\pi \eta}{4a^2}\left|H_{\nu}^{(1)}(-k\eta)\right|^2\delta^{(3)}(\mathbf{k}+\mathbf{k}')~.
\end{equation}
For the expressions above to take these forms, $\nu$ was assumed to be purely real, which is equivalent to small $\bar{m}^2_{\phi}$. 

For the vacuum contribution, the dimensionless power spectrum, $\mathcal{P}^{\textrm{vac}}_{\zeta}$, is defined by
\begin{equation}
    \label{eq:def-pvac-deltaphicorrelator}\langle \hat{\zeta}_{\textrm{vac}}(\mathbf{k})\hat{\zeta}_{\textrm{vac}}(\mathbf{k}')\rangle \equiv \left(\frac{\mathcal{H}}{\phi^{'}}\right)^2\langle\hat{\delta \phi}_{\textrm{vac}}(\mathbf{k})\hat{\delta \phi}_{\textrm{vac}}(\mathbf{k}')\rangle \equiv \frac{2\pi^2}{k^3}\mathcal{P}^{\textrm{vac}}_{\zeta}(k)\delta^{(3)}(\mathbf{k}+\mathbf{k}')~, 
\end{equation}
or (see Eq.~\eqref{eq:two-point-vac-part-deltaphi}) 
\begin{equation}
    \mathcal{P}^{\textrm{vac}}_{\zeta}(-k\eta)= \frac{H^2}{8\pi\dot \phi^2}\left(\frac{k}{a}\right)^2 (-k\eta) \left|H_{\nu}^{(1)}(-k\eta)\right|^2~. 
\end{equation}
The overdot in $\dot \phi$ denotes a derivative with respect to cosmic time $t$. Before we determine the end-of-inflation limit ($-k\eta\rightarrow 0$) of the full power spectrum in Eq.~\eqref{eq:2-point-correlator-Rgaugeinvariant}, we can already provide the corresponding limit of the expression above (see Sec.~10.7 of Ref.~\cite{NIST:DLMF}; recall also that $\nu>0$ and is real, as can be read off from Eq.~\eqref{eq:nu-sr-suppressed-terms}):
\begin{equation}
    \label{eq:PVAC-end-of-inflation-not-yet-dS}\mathcal{P}^{\textrm{vac}}_{\zeta}\simeq \frac{H^4|\Gamma(\nu)|^2}{\pi^3\dot \phi^2(1+\epsilon)^2}\left(\frac{-k\eta}{2}\right)^{3-2\nu}~.
\end{equation}
The reader should recall that $\epsilon$ is slow-roll suppressed. In the de Sitter case, $\epsilon= 0$ and $\nu=3/2$, and then $\mathcal{P}^{\textrm{vac}}_{\zeta} \simeq H^4/(4\pi^2 \dot \phi^2)$. Using Eq.~\eqref{eq:def-xi}, this can also be expressed in terms of $\xi$ as
\begin{equation}
    \label{eq:def-As-intermsof-xi}\mathcal{P}_{\zeta}^{\textrm{vac}} = \frac{\alpha^2}{16\pi^2\xi^2}\frac{H^2}{f^2}~.
\end{equation}

Turning to the contribution from inverse decay,
\begin{align}
\langle\hat{\delta\phi}_{\textrm{id}}(\mathbf k)
\hat{\delta\phi}_{\textrm{id}}(\mathbf k')\rangle
=a^{-2}(\eta)\int_{-\infty}^{0}\!\textrm d\tilde\eta\,
G_k(\eta,\tilde\eta)\int_{-\infty}^{0}\!\textrm d\tilde{\tilde\eta}\,
G_{k'}(\eta,\tilde{\tilde\eta})
\langle\hat J(\tilde\eta,\mathbf k)
\hat J(\tilde{\tilde\eta},\mathbf k')\rangle~.
\label{eq:correlator-id-deltaPHi-operator}
\end{align}
The next ingredient is the unequal-time connected source correlator. To evaluate it, we introduce
\begin{equation}
\hat{\mathcal J}(\eta,\mathbf k)
\equiv\frac{\alpha}{a(\eta)f}
\int\frac{\textrm d^3\mathbf q}{(2\pi)^{3/2}}
\hat{\mathbf E}(\eta,\mathbf q)\cdot
\hat{\mathbf B}(\eta,\mathbf k-\mathbf q)~.
\end{equation}
The source appearing in Eq.~\eqref{eq:source-Fouriermodes-sf-evolution} is its fluctuation,
\begin{align}
\label{eq:connected-source-operator}
\hat J(\eta,\mathbf k)
&\equiv \hat{\mathcal J}(\eta,\mathbf k)
-\langle\hat{\mathcal J}(\eta,\mathbf k)\rangle\nonumber\\
&=\frac{\alpha}{a(\eta)f}
\left[(\hat{\mathbf E}\cdot\hat{\mathbf B})(\eta,\mathbf k)
-(2\pi)^{3/2}\delta^{(3)}(\mathbf k)
\langle\hat{\mathbf E}\cdot\hat{\mathbf B}\rangle\right].
\end{align}
Here $\hat{\mathbf E}\cdot\hat{\mathbf B}$ denotes the symmetrised product, as explained in footnote~\ref{footnote:the1/2}. This makes the corresponding real-space composite operator Hermitian; its Fourier modes satisfy $\hat{\mathcal J}(\eta,\mathbf k)^\dagger=\hat{\mathcal J}(\eta,-\mathbf k)$, and similarly for $\hat J$. The subtraction removes the homogeneous contribution and is immaterial for nonzero external momentum, but it is essential for stating the connected correlator unambiguously.

Given Eqs.~\eqref{eq:Avec-field-operator}, \eqref{eq:electric-operator}, and \eqref{eq:magnetic-operator}, we readily find that the electric and magnetic field operators in Fourier space are given by (see also footnote~\ref{footnote:zeromass-electric})
\begin{align}
    &\hat{\mathbf{E}}(\eta,\mathbf{k}) = \left(\frac{k^2}{k^2+a^2m^2}-1\right)\hat{\mathbf{A}}_L^{'}(\eta,\mathbf{k}) -\sum_{\lambda=\pm}\hat{\mathbf{A}}_{\lambda}^{'}(\eta,\mathbf{k})~,\\
    &\hat{\mathbf{B}}(\eta,\mathbf{k}) = k\left[\hat{\mathbf{A}}_{+}(\eta,\mathbf{k})-\hat{\mathbf{A}}_{-}(\eta,\mathbf{k})\right],
\end{align}
and hence
\begin{align}
    \label{eq:expression-below}\hat{\mathbf{E}}(\eta,\mathbf{q}) \cdot \hat{\mathbf{B}}(\eta,\mathbf{k}-\mathbf{q}) \simeq -|\mathbf{k}-\mathbf{q}|\left[ \hat{\mathbf{A}}_{+}^{'}(\eta,\mathbf{q})\cdot \hat{\mathbf{A}}_{+}(\eta,\mathbf{k}-\mathbf{q})\right],
\end{align}
where the tachyonic growth of the `$+$' polarisation has already been assumed in order to simplify the calculation considerably.\footnote{Notice that, in contrast to the backreaction of the gauge field on the background scalar-field evolution, the longitudinal mode does affect the dynamics of the inflaton inhomogeneities because of the convolution, \textit{i.e.} the fact that $\bm{\epsilon}_L(\mathbf{q})\cdot \bm{\epsilon}_{\pm}(\mathbf{k}-\mathbf{q}) \neq 0$, which depends on the angle between the vectors $\mathbf{q}$ and $\mathbf{k}-\mathbf{q}$. However, this effect is negligible under the present assumptions, due to the amplified transverse polarisation and the essential fact that the homogeneous (background) scalar field does not affect the dynamics of the longitudinal mode.} Also, 
\begin{equation}
    \hat{\mathbf{A}}_{+}(\eta,\mathbf{k}) = \bm{\epsilon}_{+}(\mathbf{k}) \left[\mathcal{A}_{+}(\eta,k)\hat{a}_{\mathbf{k},+}+\mathcal{A}_{+}^{*}(\eta,k) \hat{a}^{\dag}_{-\mathbf{k},+}\right].
\end{equation}
Using that
\begin{equation}
    \bm{\epsilon}_{\pm}(\mathbf{q}) \cdot \bm{\epsilon}_{\pm}(\mathbf{k}-\mathbf{q}) = -\frac{1}{2}\left[1+\frac{|\mathbf{q}|^2-\mathbf{q}\cdot \mathbf{k}}{|\mathbf{q}||\mathbf{k}-\mathbf{q}|}\right],
\end{equation}
and $\bm{\epsilon}_{+}(-\mathbf{k}) = \bm{\epsilon}_{-}(\mathbf{k})$, we have 
\begin{align}
    \nonumber&\langle \hat{\mathcal J}(\tilde{\eta},\mathbf{k})\hat{\mathcal J}(\tilde{\tilde{\eta}},\mathbf{k}')\rangle \simeq \frac{\alpha^2\delta^{(3)}(\mathbf{k}+\mathbf{k}')}{4a(\tilde{\eta})a(\tilde{\tilde{\eta}})f^2}\int \frac{\textrm{d}^3\mathbf{q}}{(2\pi)^3}\left[1+\frac{|\mathbf{q}|^2-\mathbf{q}\cdot \mathbf{k}}{|\mathbf{q}||\mathbf{k}-\mathbf{q}|}\right]^2|\mathbf{k}-\mathbf{q}|\\
    \nonumber&\phantom{----------}\times\mathcal{A}_{+}^{'}(\tilde{\eta},|\mathbf{q}|)\mathcal{A}_{+}(\tilde{\eta},|\mathbf{k}-\mathbf{q}|)\left[|\mathbf{k}-\mathbf{q}|\mathcal{A}^{'*}_{+}(\tilde{\tilde{\eta}},|\mathbf{q}|)\mathcal{A}_{+}^{*}(\tilde{\tilde{\eta}},|\mathbf{k}-\mathbf{q}|)\right.\\
    \nonumber&\phantom{-------}\left.+|\mathbf{q}|\mathcal{A}_{+}^{*}(\tilde{\tilde{\eta}},|\mathbf{q}|)\mathcal{A}_{+}^{'*}(\tilde{\tilde{\eta}},|\mathbf{k}-\mathbf{q}|)\right]+\frac{\alpha^2\delta^{(3)}(\mathbf{k})\delta^{(3)}(\mathbf{k}')}{4a(\tilde{\eta})a(\tilde{\tilde{\eta}})f^2}\int\frac{\textrm{d}^3\mathbf{q}\textrm{d}^3\mathbf{p}}{(2\pi)^3}|\mathbf{q}||\mathbf{p}|\\
    \nonumber&\phantom{---------------}\times\left[\mathcal{A}^{'*}_{+}(\tilde{\eta},|\mathbf{q}|)\mathcal{A}_{+}(\tilde{\eta},|\mathbf{q}|)+\mathcal{A}_{+}^{'}(\tilde{\eta},|\mathbf{q}|)\mathcal{A}_{+}^{*}(\tilde{\eta},|\mathbf{q}|)\right]\\
    &\phantom{---------------}\times\left[\mathcal{A}^{'*}_{+}(\tilde{\tilde{\eta}},|\mathbf{p}|)\mathcal{A}_{+}(\tilde{\tilde{\eta}},|\mathbf{p}|)+\mathcal{A}_{+}^{'}(\tilde{\tilde{\eta}},|\mathbf{p}|)\mathcal{A}^{*}_{+}(\tilde{\tilde{\eta}},|\mathbf{p}|)\right],
\end{align}
where the commutation relation in Eq.~\eqref{eq:comm-rel-ann-cr}, together with the properties of the creation and annihilation operators, were also used. The unequal-time correlator need not be real; instead it obeys $C(\tilde\eta,\tilde{\tilde\eta})^*=C(\tilde{\tilde\eta},\tilde\eta)$, as required by Hermiticity. This relation is sufficient to make the final equal-time scalar two-point function real. The last term above is the disconnected zero-momentum contribution and is cancelled exactly by the subtraction in Eq.~\eqref{eq:connected-source-operator}. After making convenient changes of variables in the Fourier transform, such as $\mathbf{q}\rightarrow \mathbf{k}-\mathbf{q}$, the connected piece can be rewritten in the more compact form 
\begin{align}
    \nonumber&\langle \hat{J}(\tilde{\eta},\mathbf{k}) \hat{J}(\tilde{\tilde{\eta}},\mathbf{k}')\rangle \simeq \frac{\alpha^2 \delta^{(3)}(\mathbf{k}+\mathbf{k}')}{8a(\tilde{\eta})a(\tilde{\tilde{\eta}})f^2}\int \frac{\textrm{d}^3\mathbf{q}}{(2\pi)^3}\left[1+\frac{|\mathbf{q}|^2-\mathbf{q}\cdot \mathbf{k}}{|\mathbf{q}||\mathbf{k}-\mathbf{q}|}\right]^2\\
    \label{eq:correlator-twopoint-JJ}&\phantom{------------------}\times\mathcal{K}(\tilde{\eta};|\mathbf{q}|,|\mathbf{k}-\mathbf{q}|)\left[\mathcal{K}(\tilde{\tilde{\eta}};|\mathbf{q}|,|\mathbf{k}-\mathbf{q}|)\right]^{*}~, 
\end{align}
with
\begin{equation}
    \label{eq:def-BigA-firsttime}\mathcal{K}(\eta;q,p)\equiv q\,\mathcal{A}^{'}_{+}(\eta,p)\mathcal{A}_{+}(\eta,q)+p\,\mathcal{A}^{'}_{+}(\eta,q)\mathcal{A}_{+}(\eta,p)~.
\end{equation}

The two-point correlator in Eq.~\eqref{eq:correlator-twopoint-JJ} can then be substituted back into Eq.~\eqref{eq:correlator-id-deltaPHi-operator}, and one must evaluate the two integrals in order to estimate the contribution from inverse decay to the total power spectrum. Upon inserting the Green functions and the source correlator into Eq.~\eqref{eq:correlator-id-deltaPHi-operator}, we find
\begin{align}
    \nonumber&\langle\hat{\delta \phi}_{\textrm{id}}(\mathbf{k})\hat{\delta \phi}_{\textrm{id}}(\mathbf{k}')\rangle \simeq \frac{\alpha^2\pi^2\eta}{2^7 a^{2}(\eta)f^2}\delta^{(3)}(\mathbf{k}+\mathbf{k}')\int \frac{\textrm{d}^3\mathbf{q}}{(2\pi)^3}\left[1-\frac{\mathbf{q}\cdot (\mathbf{k}-\mathbf{q})}{|\mathbf{q}||\mathbf{k}-\mathbf{q}|}\right]^2\\
    &\times\int^{\eta}_{-\infty} \textrm{d}\tilde{\eta}~\frac{\sqrt{-\tilde{\eta}}}{a(\tilde{\eta})}\mathcal{H}^{\nu}_k(\eta,\tilde{\eta})\mathcal{K}(\tilde{\eta};|\mathbf{q}|,|\mathbf{k}-\mathbf{q}|)\int_{-\infty}^{\eta} \textrm{d}\tilde{\tilde{\eta}}~\frac{\sqrt{-\tilde{\tilde{\eta}}}}{a(\tilde{\tilde{\eta}})}\mathcal{H}^{\nu}_k(\eta,\tilde{\tilde{\eta}})\left[\mathcal{K}(\tilde{\tilde{\eta}};|\mathbf{q}|,|\mathbf{k}-\mathbf{q}|)\right]^{*}~,
\end{align}
where the Heaviside functions were taken into account in redefining the upper limits of integration, and the following compact notation was introduced:
\begin{equation}
    \mathcal{H}^{\nu}_k(\eta,\tilde{\eta}) \equiv H^{(1)}_{\nu}(-k\eta)H^{(2)}_{\nu}(-k\tilde{\eta})-H_{\nu}^{(1)}(-k\tilde{\eta}) H_{\nu}^{(2)}(-k\eta)~. 
\end{equation}
Because $\mathcal{H}^{\nu}_k(\eta,\tilde{\eta}) = -\left[\mathcal{H}^{\nu}_{k}(\eta,\tilde{\eta})\right]^{*}$, we may simply write
\begin{align}
    \nonumber&\langle\hat{\delta \phi}_{\textrm{id}}(\mathbf{k})\hat{\delta \phi}_{\textrm{id}}(\mathbf{k}')\rangle \simeq -\frac{\alpha^2\pi^2\eta}{2^7 a^{2}(\eta)f^2}\delta^{(3)}(\mathbf{k}+\mathbf{k}')\int \frac{\textrm{d}^3\mathbf{q}}{(2\pi)^3}\left[1-\frac{\mathbf{q}\cdot (\mathbf{k}-\mathbf{q})}{|\mathbf{q}||\mathbf{k}-\mathbf{q}|}\right]^2\\
    \label{eq:two-point-id-part-deltaphi}&\phantom{-----------------}\times\left|\int^{\eta}_{-\infty} \textrm{d}\tilde{\eta}~\frac{\sqrt{-\tilde{\eta}}}{a(\tilde{\eta})}\mathcal{H}^{\nu}_k(\eta,\tilde{\eta})\mathcal{K}(\tilde{\eta};|\mathbf{q}|,|\mathbf{k}-\mathbf{q}|)\right|^2.
\end{align}
The dimensionless power spectrum sourced by inverse decay, $\mathcal{P}^{\textrm{id}}_{\zeta}$, is then 
\begin{equation}
    \label{eq:def-pid-deltaphicorrelator}\langle \hat{\zeta}_{\textrm{id}}(\mathbf{k})\hat{\zeta}_{\textrm{id}}(\mathbf{k}')\rangle \equiv \left(\frac{\mathcal{H}}{\phi^{'}}\right)^2\langle\hat{\delta \phi}_{\textrm{id}}(\mathbf{k})\hat{\delta \phi}_{\textrm{id}}(\mathbf{k}')\rangle \equiv \frac{2\pi^2}{k^3}\mathcal{P}^{\textrm{id}}_{\zeta}(k)\delta^{(3)}(\mathbf{k}+\mathbf{k}')~, 
\end{equation}
or (see Eq.~\eqref{eq:two-point-id-part-deltaphi})
\begin{align}
    \nonumber&\mathcal{P}^{\textrm{id}}_{\zeta}(-k\eta) \equiv \frac{\alpha^2H^2}{2^8 f^2\dot \phi^2}\left(\frac{k}{a}\right)^2(-k\eta)\int \frac{\textrm{d}^3\mathbf{q}}{(2\pi)^3}\left[1-\frac{\mathbf{q}\cdot (\mathbf{k}-\mathbf{q})}{|\mathbf{q}||\mathbf{k}-\mathbf{q}|}\right]^2 \\
    &\phantom{-----------------}\times\left|\int^{\eta}_{-\infty} \textrm{d}\tilde{\eta}~\frac{\sqrt{-\tilde{\eta}}}{a(\tilde{\eta})}\mathcal{H}^{\nu}_k(\eta,\tilde{\eta})\mathcal{K}(\tilde{\eta};|\mathbf{q}|,|\mathbf{k}-\mathbf{q}|)\right|^2.
\end{align}
Evaluating at the end of inflation, one takes $\eta\to0^-$ while retaining the leading dependence on the observation time:
\begin{equation}
\mathcal H_k^\nu(\eta,\tilde\eta)
\xrightarrow[\eta\to0^-]{}
-\frac{2i}{\pi}\Gamma(\nu)
\left(\frac{2}{-k\eta}\right)^\nu
\operatorname{Re}H_\nu^{(1)}(-k\tilde\eta)~.
\end{equation}
Writing $\mathcal H_k^\nu(0,\tilde\eta)$ would be misleading because the displayed asymptotic form diverges as a power of $(-k\eta)^{-1}$ before it is combined with the prefactors in the power spectrum. It follows that
\begin{equation}
    \mathcal{P}^{\textrm{id}}_{\zeta}\simeq \frac{\alpha^2\pi}{8 f^2}\mathcal{P}^{\textrm{vac}}_{\zeta} \int \frac{\textrm{d}^3\mathbf{q}}{(2\pi)^3}\left[1-\frac{\mathbf{q}\cdot (\mathbf{k}-\mathbf{q})}{|\mathbf{q}||\mathbf{k}-\mathbf{q}|}\right]^2 \mathcal{S}(|\mathbf{q}|,|\mathbf{k}-\mathbf{q}|)~,
\end{equation}
where
\begin{equation}
    \label{eq:BigI-first-definition}\mathcal{S}(|\mathbf{q}|,|\mathbf{k}-\mathbf{q}|) \equiv \left|\int^{0}_{-\infty}\textrm{d}\tilde{\eta}\frac{\sqrt{-\tilde{\eta}}}{a(\tilde{\eta})}\textrm{Re}\left[H_{\nu}^{(1)}(-k\tilde{\eta})\right]\mathcal{K}(\tilde{\eta};|\mathbf{q}|,|\mathbf{k}-\mathbf{q}|)\right|^2.
\end{equation}
Since both the argument and the order of $H^{(1)}_{\nu}(-k\eta)$ are real and positive, its real part is the Bessel function of the first kind, $J_{\nu}(-k\eta)$ (see Sec.~10.4 of Ref.~\cite{NIST:DLMF}). In a quasi-de Sitter background, $\nu \simeq 3/2$, so that
\begin{equation} \label{eq: Hankel function quasi de Sitter}
    \textrm{Re}\left[H_{3/2}^{(1)}(-k\tilde{\eta})\right] = \sqrt{\frac{2}{\pi}}(-k\tilde{\eta})^{-3/2}\left[\sin(-k\tilde{\eta})+k\tilde{\eta}\cos(-k\tilde{\eta})\right],
\end{equation}
and $\mathcal{S}(|\mathbf{q}|,|\mathbf{k}-\mathbf{q}|)$ simplifies to
\begin{equation}
    \label{eq:BigS-q-p-expression}\mathcal{S}(|\mathbf{q}|,|\mathbf{k}-\mathbf{q}|) \simeq \frac{2H^2}{\pi k^3}\left|\int^{0}_{-\infty}\textrm{d}\tilde{\eta}~\left[\sin(-k\tilde{\eta})+k\tilde{\eta}\cos(-k\tilde{\eta})\right]\mathcal{K}(\tilde{\eta};|\mathbf{q}|,|\mathbf{k}-\mathbf{q}|)\right|^2.
\end{equation}
On the other hand, the three-dimensional momentum integral over $\mathbf{q}$ can be reduced to two integrals with measures $\textrm{d}q$ and $\textrm{d}p$, where $q\equiv |\mathbf{q}|$ and $p\equiv |\mathbf{k}-\mathbf{q}|$. Exploiting rotational invariance, we choose $\mathbf{k}$ to lie along the polar axis. It then follows that
\begin{equation}
    1-\frac{\mathbf{q}\cdot(\mathbf{k}-\mathbf{q})}{|\mathbf{q}||\mathbf{k}-\mathbf{q}|} = 1-\frac{k\cos \theta-q}{p}=\frac{(p+q)^2-k^2}{2qp}~,
\end{equation}
where $\theta$ denotes the polar angle. Furthermore, $\textrm{d}^3\mathbf{q} = -q^2\textrm{d}q\textrm{d}(\cos\theta)\textrm{d}\varphi=\frac{qp}{k}\textrm{d}q\textrm{d}p\textrm{d}\varphi$, with $\varphi\in[0,2\pi]$ the azimuthal angle. 

Putting everything together, we arrive at 
\begin{align}
    \nonumber&\frac{\mathcal{P}^{\textrm{id}}_{\zeta}}{\mathcal{P}^{\textrm{vac}}_{\zeta}}\simeq \frac{\alpha^2}{64\pi^2 k^4}\frac{H^2}{f^2}\int^{\infty}_0 \frac{\textrm{d}q}{q}\int^{k+q}_{|k-q|}\frac{\textrm{d}p}{p}\left[(q+p)^2-k^2\right]^2 \\
    \label{eq:PidPvac-ratio-beforeA}&\phantom{---------------}\times \left|\int^{0}_{-\infty}\textrm{d}\tilde{\eta}\left[\sin(-k\tilde{\eta})+k\tilde{\eta}\cos(-k\tilde{\eta})\right]\mathcal{K}(\tilde{\eta};q,p)\right|^2.
\end{align}
In view of Eq.~\eqref{eq:mod-sq-Aplus-largemass-kappa}, the source kernel $\mathcal{K}(\tilde{\eta};q,p)$, defined in Eq.~\eqref{eq:def-BigA-firsttime}, is given by 
\begin{align}
    \nonumber&\mathcal{K}(\tilde{\eta};q,p) \simeq -\frac{e^{2|\xi|\pi}}{\pi}\left\{qK_{2\mu}(2\sqrt{2 \xi \kappa p\tilde{\eta}})K_{2\mu}(2\sqrt{2 \xi \kappa q\tilde{\eta}})-q\sqrt{2 \xi\kappa p \tilde{\eta}}K_{2\mu}(2\sqrt{2 \xi\kappa q\tilde{\eta}})\right.\\
    &\left.\phantom{------------}\times\left[K_{1-2\mu}(2\sqrt{2 \xi\kappa p\tilde{\eta}})+K_{1+2\mu}(2\sqrt{2 \xi\kappa p\tilde{\eta}})\right]+(q \leftrightarrow p)\right\}, 
\end{align}
for both real and purely imaginary $\mu$.

Upon defining $x\equiv -k\tilde{\eta}$, we may change the integration variable in the inner integral in Eq.~\eqref{eq:PidPvac-ratio-beforeA} so that we arrive at Eq.~\eqref{eq:the-ratio-semianalytical-many-bessels1}, where $q_{*} \equiv q/k$ and $p_{*} \equiv p/k$ denote real, positive quantities, introduced for convenience, and we replaced the modulus squared with the square because the inner integral, as written above, is real, regardless of whether $\mu$ is real or purely imaginary. The gauge-field modes contribute most strongly within and just past the
instability band,
\begin{equation}
  |\xi|-\sqrt{\xi^{2}-\bar{m}^{2}}\;\lesssim\;-k\eta\;\lesssim\;
  |\xi|+\sqrt{\xi^{2}-\bar{m}^{2}}~.
\end{equation}
The integrals are dominated by the region $y_{q},y_{p}\gtrsim1$, where $y_{q}\equiv2\sqrt{2|\xi|\kappa\,q_{*}x}$ and $y_{p}\equiv2\sqrt{2|\xi|\kappa\,p_{*}x}$: at small $x$ the window function behaves as $\sin x-x\cos x\simeq x^{3}/3$, which suppresses the integrand faster than the mode functions can grow, so that no contribution arises from $y_{q},y_{p}\lesssim1$. In this region the large-argument form of the mode functions applies.

For massless and light gauge fields this already characterises the dominant contribution. For heavy fields the amplitudes peak at subhorizon scales near the turning point $y_{q},y_{p}\sim2\tilde{\mu}\sim\bar{m}$. Given that in that regime
\begin{equation}
  K_{2\mu}(y)\;\ll\;\frac{y}{2}\big[K_{1-2\mu}(y)+K_{1+2\mu}(y)\big]
  \label{eq:drop-K-against-derivative}
\end{equation}
up to corrections of $\mathcal{O}(\tilde{\mu}^{-1})$,  Eq.~\eqref{eq:the-ratio-semianalytical-many-bessels1} yields the intermediate form (cf.~Eq.~(3.27) of Ref.~\cite{Fumagalli:2023loc})
\begin{align}
&\frac{\mathcal P^{\textrm{id}}_\zeta}{\mathcal P^{\textrm{vac}}_\zeta}
\simeq\frac{\kappa}{2\pi^2}\mathcal{P}_{\zeta}^{\textrm{vac}}|\xi|^3e^{4\pi|\xi|}
\int_0^\infty\!\textrm d q_*
\int_{|1-q_*|}^{1+q_*}\!\textrm d p_*\,
\big[(q_*+p_*)^2-1\big]^2 \Bigg\{\int_0^\infty\!\textrm d x\,\sqrt{x}\,[\sin x-x\cos x]  \nonumber\\
&\quad\times 
\Bigg[\sqrt{q_*}\,K_{2\mu}\!\left(2\sqrt{2|\xi|\kappa q_*x}\right) \left\{
K_{1-2\mu}\!\left(2\sqrt{2|\xi|\kappa p_*x}\right)
+K_{1+2\mu}\!\left(2\sqrt{2|\xi|\kappa p_*x}\right)
\right\}
+(q_*\leftrightarrow p_*)\Bigg]\Bigg\}^2 .
\label{eq:the-ratio-semianalytical-intermediate}
\end{align}

The bracket in Eq.~\eqref{eq:the-ratio-semianalytical-intermediate} still involves modified Bessel functions of two distinct orders, $2\mu$ and $1\pm2\mu$. In the large-argument region $y_{q},y_{p}\gtrsim1$ that dominates the integrals, all three approach the common leading form $K_{\nu}(y)\to\sqrt{\pi/2y}\,e^{-y}$, independent of the order (see Sec.~10.40 of Ref.~\cite{NIST:DLMF}), which motivates the approximation
\begin{equation}
  K_{1-2\mu}(y)+K_{1+2\mu}(y)\;\simeq\;2\,K_{2\mu}(y)~.
  \label{eq:collapse-two-bessel}
\end{equation}
Using $K_{-\nu}=K_{\nu}$, the symmetrised terms then combine into $2(\sqrt{q_{*}}+\sqrt{p_{*}})K_{2\mu}(y_{q})K_{2\mu}(y_{p})$, which yields the compact two-Bessel expression in Eq.~\eqref{eq:the-ratio-semianalytical}.

The approximation~\eqref{eq:collapse-two-bessel} is exact only in the strict large-argument limit. For imaginary order $\mu=i\tilde{\mu}$ its accuracy is set by $2\tilde{\mu}/y$ rather than by $1/y$, so it degrades near the turning point $y\sim2\tilde{\mu}$ that the dominant modes probe in the heavy regime. Equation~\eqref{eq:the-ratio-semianalytical} is therefore a simplified estimate. 

\begin{figure}[t]
\centering
\includegraphics[width=1.00\linewidth]{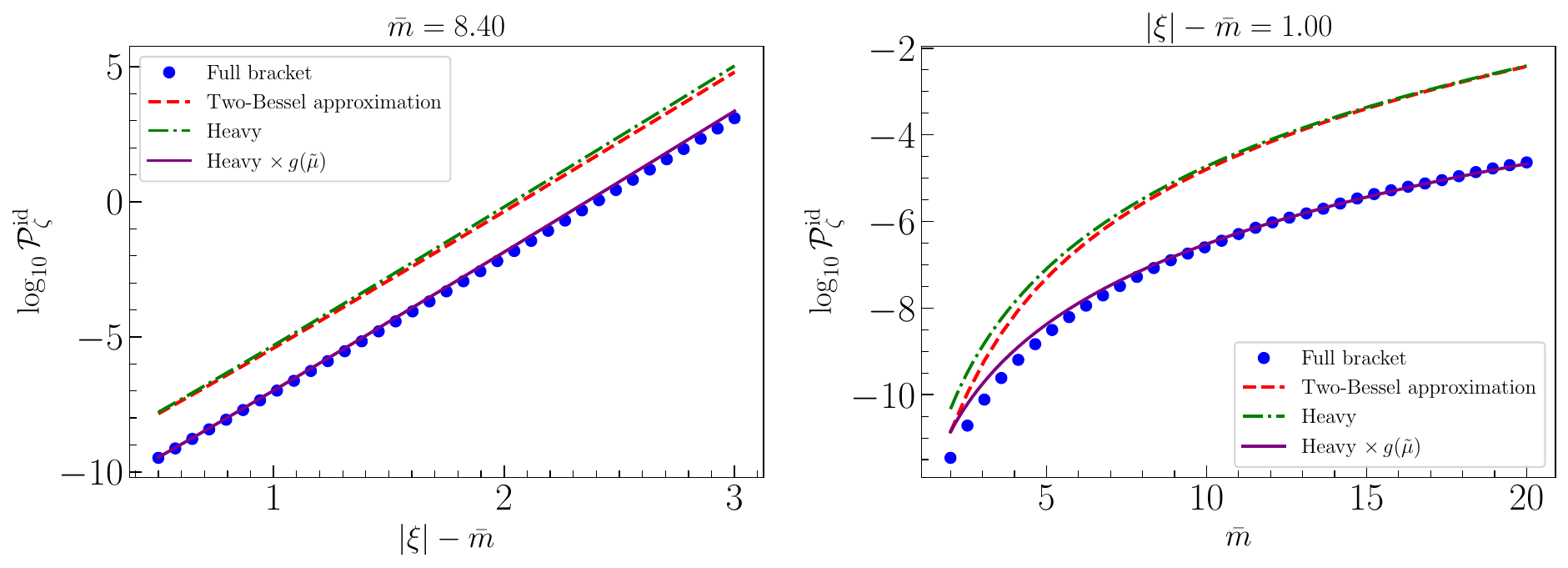}
\caption{Comparison of $\log_{10}\mathcal P_\zeta^{\rm id}$ obtained from the full Bessel expression, Eq.~\eqref{eq:the-ratio-semianalytical-many-bessels1} (blue circles), the reduced two-Bessel expression, Eq.~\eqref{eq:the-ratio-semianalytical} (red dashed), the heavy closed form, Eq.~\eqref{eq:loop-contribution-weak-backreaction} (green dash-dotted), and the one including the correction factor $g(\tilde \mu) = (\tilde \mu^{-2}+0.06\tilde \mu^{-1})$, Eq.~\eqref{eq:loop-contribution-weak-backreaction-correction} (purple solid). All curves are evaluated at fixed $\mathcal{P}^{\textrm{vac}}_\zeta = 2.1\times 10^{-9}$. The two-Bessel expression retains the main parameter dependence of the full bracket but it becomes less accurate once the mass increases for fixed $|\xi|-\bar{m}$.}
\label{fig:Pzeta_lattice_full_collapsed}
\end{figure}

As shown in Fig.~\ref{fig:Pzeta_lattice_full_collapsed}, the two-Bessel approximation \eqref{eq:the-ratio-semianalytical} differs from the full-bracket expression by up to one-to-two orders of magnitude. Despite this mismatch, the single pair of modified Bessel functions makes the loop integrals far more tractable, admitting a semi-analytical treatment. We therefore adopt it in what follows for our analytical estimates, supplementing it in the heavy gauge-field regime with a correction prefactor (cf.~Eq.~\eqref{eq: Power spectrum, inverse decay, with friction}) that improves the agreement substantially.

\subsection{Power Spectrum for Heavy Fields during Weak Backreaction \label{app:heavy-weak-spectrum}}
We now give the reduction that leads from Eq.~\eqref{eq:the-ratio-semianalytical-intermediate} to Eq.~\eqref{eq:loop-contribution-weak-backreaction}. For $\tilde\mu\gg1$ and small argument compared to the index, the Bessel envelope is Eq.~\eqref{eq:heavy-bessel-envelope}. With the definition
\begin{equation}
\Theta_s\equiv\tilde\mu\ln(2\kappa|\xi|s)-\arg\Gamma(1+2i\tilde\mu)~,
\end{equation}
and the momentum dependent UV cutoff $\Lambda(u)\equiv\frac{\tilde\mu^2}{c\kappa|\xi|u}~,$
where $u=\max(q_*,p_*)$. The time integral then contains
\begin{equation}
\int_0^{\Lambda(u)}\!\textrm d x\,F(x)
\sin[\tilde\mu\ln x+\Theta_{q_*}]
\sin[\tilde\mu\ln x+\Theta_{p_*}]~,
\qquad
F(x)\equiv\sqrt{x}\,[\sin x-x\cos x]~.
\label{eq:heavy-time-integral-app}
\end{equation}
Using the product-to-sum identity, the term with phase $2\tilde\mu\ln x+\Theta_{q_*}+\Theta_{p_*}$ is suppressed by $1/\tilde\mu$ after integration by parts. The leading term is therefore proportional to
\begin{equation}
\frac12\cos\!\left[\tilde\mu\ln\!\left(\frac{q_*}{p_*}\right)\right]
G[\Lambda(u)]~,
\end{equation}
where
\begin{align}
G(\Lambda)&\equiv\int_0^\Lambda\!\textrm d x\,F(x)=\frac54\sqrt{2\pi}\,C_{\rm F}\!\left(\sqrt{\frac{2\Lambda}{\pi}}\right)
-\frac52\sqrt{\Lambda}\cos\Lambda
-\Lambda^{3/2}\sin\Lambda~,
\label{eq:G-Lambda-correct}
\end{align}
and $C_{\rm F}(z)=\int_0^z\cos(\pi t^2/2)\,\textrm d t$ is the Fresnel cosine integral.

The remaining integrand is symmetric under $p_*\leftrightarrow q_*$. Restricting to $q_*>p_*$ and doubling the result, averaging the rapidly varying $\cos^2[\tilde\mu\ln(q_*/p_*)]$ to $1/2$, and expanding the $p_*$ integral for $q_*\gg1$ gives the leading term $32q_*^5$. One then obtains
\begin{equation}
\frac{\mathcal P^{\rm id}_\zeta}{\mathcal P^{\rm vac}_\zeta}
\simeq\frac{32c_\ell}{c^6}\mathcal{P}_{\zeta}^{\textrm{vac}}
\frac{\tilde\mu^{10}}{\kappa^5|\xi|^3}
 e^{4\pi(|\xi|-\tilde\mu)}~,
\end{equation}
where
\begin{equation}
c_\ell\equiv\int_0^\infty\frac{\textrm d\Lambda}{\Lambda^7}
G(\Lambda)^2 \simeq0.04~.
\label{eq:definition-cl}
\end{equation}
For $\Lambda\ll1$, $G(\Lambda)\propto\Lambda^{9/2}$ and the integrand scales as $\Lambda^2$; for $\Lambda\gg1$ it scales as $\Lambda^{-4}$. The integral is therefore finite and dominated by $\Lambda=\mathcal O(1)$, which justifies treating $c$ as an order-one matching constant. Direct comparison with the numerical four-dimensional integral gives $c\simeq1.27$, with the correction factor $g(\tilde \mu) = (\tilde \mu^{-2}+0.06\tilde \mu^{-1})$, for the parameter range used in Fig.~\ref{fig:PidrPvac_heavy_fixedxim}.

\begin{table}[t]
\centering
\small
\begin{tabular}{cccccc}
\toprule
$\bar m_i=m/H_i$ & sampled $|\xi|-\bar m$ & $\alpha\m/f$ & $10^6H_i/\m$ & $\mathcal C_{\rm rep}$ & $\mathcal P_\zeta^{\rm id}$\\
\midrule
$4.75$  & $[1.16,2.37]$ & $750$  & $8.68$  & $0.07$  & $1.5\times10^{-3}$\\
$4.75$  & $[1.30,2.40]$ & $750$  & $8.68$  & $0.10$  & $6.0\times10^{-4}$\\
$4.75$  & $[1.60,2.00]$ & $1000$ & $8.68$  & $0.60$  & $6.0\times10^{-3}$\\
$4.75$  & $[0.60,2.20]$ & $1350$ & $8.68$  & $1.20$  & $7.0\times10^{-3}$\\
$8.25$  & $[0.98,0.98]$ & $930$  & $8.68$  & $0.004$ & $1.5\times10^{-7}$\\
$8.25$  & $[0.93,1.00]$ & $930$  & $8.68$  & $0.009$ & $5.0\times10^{-7}$\\
$8.25$  & $[0.87,0.88]$ & $1400$ & $13.02$ & $0.020$ & $1.8\times10^{-6}$\\
$8.25$  & $[1.03,1.10]$ & $950$  & $8.68$  & $0.022$ & $2.0\times10^{-6}$\\
$8.25$  & $[0.90,1.50]$ & $1000$ & $8.68$  & $0.12$  & $3.0\times10^{-5}$\\
$9.50$  & $[0.40,1.30]$ & $1200$ & $8.68$  & $0.15$  & $1.4\times10^{-4}$\\
$9.50$  & $[0.83,1.46]$ & $1500$ & $8.68$  & $0.45$  & $1.5\times10^{-3}$\\
$11.00$ & $[0.85,0.87]$ & $900$  & $6.51$  & $0.008$ & $1.0\times10^{-7}$\\
$11.00$ & $[0.80,1.65]$ & $2100$ & $8.68$  & $0.80$  & $2.0\times10^{-4}$\\
$16.89$ & $[0.90,1.20]$ & $2300$ & $8.68$  & $0.35$  & $3.0\times10^{-5}$\\
\bottomrule
\end{tabular}
\caption{Parameter points used in the analytical--lattice comparison. Here $H_i\equiv \sqrt{V(\phi_0)/(3 \m^2)}$ The interval in the second column is the range sampled during the analysis window, and $\mathcal C_{\rm rep}$ is a representative value in that window rather than a time average. The tabulated spectra are approximate plateau amplitudes read from the simulation outputs.}
\label{tab:parameter_points}
\end{table}

\bibliography{apssamp}


\end{document}